\documentclass[aps,physrev,twocolumn,superscriptaddress]{revtex4-1}

\usepackage[pdftex]{graphicx}
\usepackage{dcolumn}
\usepackage{bm}
\usepackage{amsmath, amssymb,amsfonts}
\usepackage{mathrsfs}
\usepackage{type1cm}
\usepackage[colorlinks,linkcolor=blue,citecolor=blue]{hyperref}

\begin{document}

\title{Fluctuations of squeezing fields beyond the Tomonaga--Luttinger liquid paradigm}

\author{Kazuma Nagao}
\email{knagao@physnet.uni-hamburg.de}
\affiliation{%
Zentrum f\"ur Optische Quantentechnologien and Institut f\"ur Laserphysik, Universit\"at Hamburg, 22761 Hamburg, Germany\\
}%
\affiliation{%
The Hamburg Center for Ultrafast Imaging, Luruper Chaussee 149, 22761 Hamburg, Germany
}%
\author{Ludwig Mathey}
\affiliation{%
Zentrum f\"ur Optische Quantentechnologien and Institut f\"ur Laserphysik, Universit\"at Hamburg, 22761 Hamburg, Germany\\
}%
\affiliation{%
The Hamburg Center for Ultrafast Imaging, Luruper Chaussee 149, 22761 Hamburg, Germany
}%

\date{\today}

\begin{abstract}

The concept of Tomonaga--Luttinger liquids (TLL) on the basis of the free-boson models is ubiquitous in theoretical descriptions of low-energy properties in one-dimensional quantum systems.
In this work, we develop a squeezed-field path-integral description for gapless one-dimensional systems beyond the free-boson picture of the TLL paradigm.
In the squeezed-field description, the parameter of the Bogoliubov transformation for the TL Hamiltonian becomes a dynamical squeezing field, and its fluctuations give rise to corrections to the free-boson results.
We derive an effective nonlinear Lagrangian describing the dispersion relation of the squeezing field, and interactions between the excitations of the TLL and the squeezing modes.
Using the effective Lagrangian, we analyze the imaginary-time correlation function of a vertex operator in the non-interacting limit.
We show that a side-band branch emerges due to the fluctuation of the squeezing field, in addition to the standard branch of the free-boson model of the TLL paradigm.
Furthermore, we perturbatively analyze the spectral function of the density fluctuations for an ultracold Bose gas in one dimension.
We evaluate the renormalized values of the phase velocities and spectral weights of the TLL and side-band branches due to the interaction between the TLL and the squeezing modes.
At zero temperature, the renormalized dispersion relations are linear in the momentum, but at nonzero temperatures, these acquire a nonlinear dependence on the momentum due to the thermal population of the excitation branches.
 
\end{abstract}

\maketitle

\section{Introduction}

Restricted dimensionality in many-body systems gives rise to phenomena associated with strong fluctuations and strong correlations between particles \cite{nagaosa1999quantum,giamarchi2004quantum,fradkin2013field}.
Among these systems of low dimensionality, one dimensional (1D) quantum systems, e.g., quantum wires such as carbon nanotubes \cite{avouris2008carbon,ishii2003direct,shi2015observation,sato2019strong,wang2020nonlinear}, 1D liquid $^{3}{\rm He}$ and $^{4}{\rm He}$ \cite{wada2001helium,taniguchi2005possible,savard2011hydrodynamics,bertaina2016one}, and ultracold gases trapped in a 1D potential \cite{cazalilla2011one,wen2013fermi,capponi2016phases,stoferle2004transition,kinoshita2006a,citro2007evidence,widera2008quantum,haller2010pinning,hu2011detecting,gring2012relaxation,cheneau2012light,danshita2013universal,fabbri2015dynamical,kunimi2017thermally,yang2017quantum,ozaki2020semiclassical}, have offered a unique platform for exploring in and out of equilibrium many-body phenomena that cannot be captured within naive mean-field treatments due to significant quantum and thermal fluctuations.

A unifying property observed in generic gapless boson and fermion systems in 1D is that the low-lying excitations of the spectrum are dominated by collective bosonic excitations \cite{tomonaga1950remarks,luttinger1963exactly,mattis1965exact,cazalilla2004bosonizing}.
This property can be universally described by Tomonaga--Luttinger liquid (TLL) theory that is established by linearization of the microscopic band structure and bosonization relations \cite{nagaosa1999quantum,giamarchi2004quantum,fradkin2013field,gogolin2004bosonization,haldane1981effective,furusaki1994kondo,cazalilla2004bosonizing,mathey2004luttinger,mathey2007commensurate,tokuno2008dynamics,kitagawa2010ramsey,eggel2011dynamical,ruhman2015topological,okamoto2016hierarchical,ashida2016quantum,matveev2017second,matveev2018hybrid,imamura2019from}.
In TLL theory, a harmonic fluid Hamiltonian, referred to as the TL Hamiltonian, arises as a low-energy effective model of a microscopic system. Important examples include 1D spinless fermions, dilute Bose gases in 1D, and various spin one-half chains \cite{giamarchi2004quantum,gogolin2004bosonization}.
The central dynamical fields of the TL formalism are the collective variables ${\hat \theta}(x)$ and ${\hat \phi}(x)$, which correspond to the phase and density fluctuations of a single-component system \cite{haldane1981effective}.
In terms of these fields, the TL Hamiltonian is written as \cite{giamarchi2004quantum,haldane1981effective}   
\begin{align}
{\hat H}_{0} = \frac{\hbar v_0}{2\pi}\int dx \left[ K (\partial_{x}{\hat \theta})^2 + \frac{1}{K} (\partial_x {\hat \phi})^2 \right]. \label{eq: tll model}
\end{align}
Here $v_0$ is the zero-temperature sound velocity and $K$ is the TL parameter controlling the correlations of the system. 
The canonical variables satisfy the commutation relation $[{\hat \phi}(x),\partial_{x'}{\hat \theta}(x')]=i\pi \delta(x-x')$ \cite{giamarchi2004quantum}.
The TL Hamiltonian also emerges as a renormalization group fixed point in the scaling limit of quantum critical systems, and defines the concept of gapless TLL phases \cite{sachdev2011quantum}.
In the standard treatment of TLL theory, in order to evaluate physical quantities, Eq.~(\ref{eq: tll model}) is diagonalized by using a Bogoliubov transformation for ${\hat \phi}$ and ${\hat \theta}$ \cite{haldane1981effective,giamarchi2004quantum}.
Then, within the harmonic-fluid approximation, the system is decomposed into free bosons ${\hat H}_0 \rightarrow \sum_{k \neq 0}\hbar \omega_{k} {\hat b}^{\dagger}_{k}{\hat b}_{k}$, where $\omega_{k}=v_0|{k}|$ is the linear dispersion.
Due to this exact solvability, key quantities, such as the local magnetization or spatiotemporal correlation functions, can be evaluated analytically.

\begin{figure}
\begin{center}
\includegraphics[width=80mm]{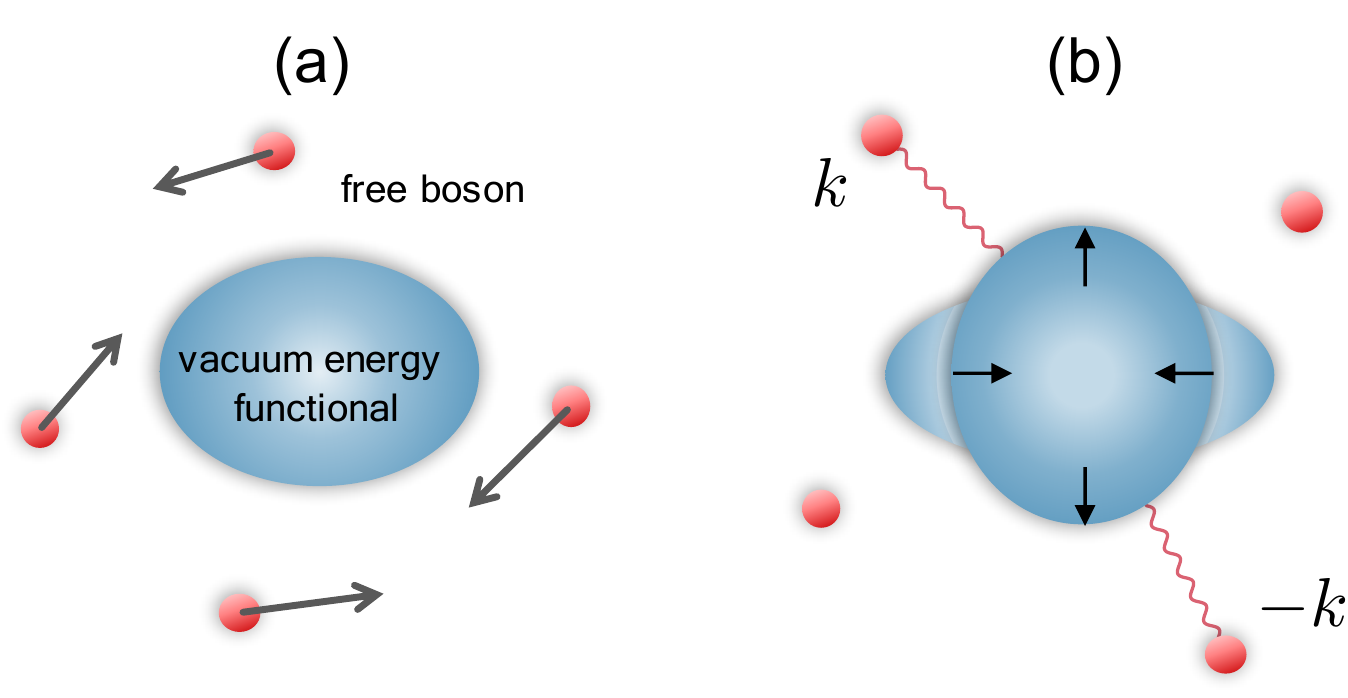}
\vspace{-4mm}
\caption{
Schematic illustration of the squeezed-field description for the TL Hamiltonian.
(a) In the standard TLL description, the vacuum energy functional, whose value is determined from an equilibrium Bogoliubov-transformation angle, is static and elementary excitations of the vacuum are non-interacting bosons.
(b) In the squeezed-field description, crudely speaking, the vacuum energy functional is allowed to breathe around an equilibrium state.
As a consequence, the bosons are scattered by fluctuations of the energy functional, giving rise to an effective coupling between $k$ and $-k$ modes.
}
\vspace{-8mm}
\label{fig: topfigure}
\end{center}
\end{figure}

In the last decade, the low-energy excitation properties associated with TLL theory have been explored in well-controlled experiments \cite{barak2010interacting,fabbri2015dynamical,wang2020nonlinear}.
Since the free boson model ignores correlations among different modes and modifications of the ground state and excitation properties caused by them, it is insufficient to describe the full spectral properties of the experimental systems.
It is therefore imperative to develop a useful framework beyond the free-boson descriptions. 
Studies that utilize refermionization, and kinematic approaches have been reported in Refs. \cite{pustilnik2006dynamic,imambekov2009universal,schmidt2010spin,imambekov2012one,matveev2013decay,protopopov2014relaxation} and \cite{apostolov2013thermal,buchhold2015nonequilibrium,samanta2019thermal}, respectively.
 
In this paper, we develop a squeezed-field path-integral formalism for gapless 1D quantum systems as an extended framework beyond the free-boson picture of the TLL paradigm.
The squeezed-field method has been recently introduced in Ref.~\cite{seifie2019squeezed} and applied to a weakly-interacting dilute Bose gas in 3D.
The underlying idea of our formalism for the TL Hamiltonian is schematically illustrated in Fig.~\ref{fig: topfigure}. 
In the squeezed-field method for the 3D case, the squeeze parameter of the Bogoliubov transformation for the quadratic Hamiltonian of the Bogoliubov approximation can fluctuate around its equilibrium value, leading to a new type of collective mode, namely, the squeezing mode \cite{seifie2019squeezed}.
Such fluctuations of the squeeze parameter can be interpreted as dynamics of the quantum depletion, which is not included in the standard Bogoliubov approximation.
In this work, we generalize the squeezed-field description to bosonized 1D systems and derive an effective Lagrangian formulating the coupling among the TLL bosonic excitation modes and the squeezing modes.
In order to illustrate the consequences of the squeezed-field description, we analyze a prototypical imaginary-time correlation function using path integrals with the effective Lagrangian. 
We also apply our formalism to an ultracold Bose gas in 1D and discuss the modifications to the spectral properties created by the coupling between the TLL and the squeezing modes.

This paper is organized as follows:
In Sec.~\ref{sec: effective models}, we introduce and explain our formalism of a squeeze-field description of a 1D system described by the TL Hamiltonian in the low-energy limit.
We derive an effective Lagrangian, which describes the dispersion of the squeezing modes and their couplings to the TLL modes.
In Sec.~\ref{sec: noninteracting}, as an instructive example, an application of the effective Lagrangian to a prototypical correlation function is presented in a simplified limit, where the coupling between the TLL and the squeezing modes is neglected.
In Sec.~\ref{sec: model}, we introduce a bosonization of a continuous model for a 1D ultracold Bose gas, which will be treated in Sec.~\ref{sec: spectral function}.
In Sec.~\ref{sec: spectral function}, we investigate a spectral function for the 1D gas by using the squeezed-field effective Lagrangian.
In particular, we discuss what effects can emerge in the spectral function due to the couplings between the TLL and squeezing modes for zero and nonzero temperatures.
We present a discussion of the results of Sec.~\ref{sec: spectral function} in Sec.~\ref{sec: discussions}.
In the final section, Sec.~\ref{sec: conclusion}, we summarize the main results of this paper.

\section{Effective model for the squeezing fields in 1D} \label{sec: effective models}

\subsection{Basic setups}

To explain our formalism, we write the Bogoliubov transformation for ${\hat \phi}$ and ${\hat \theta}$ in bosonized form \cite{giamarchi2004quantum,haldane1981effective}:
\begin{align}
{\hat \phi}(x) &= -\frac{i\pi}{L}\sum_{k\neq 0}\left(\frac{L|k|}{2\pi}\right)^{\frac{1}{2}}\frac{e^{-\frac{{\tilde \alpha}|k|}{2}-ikx}}{k} \left[ f_{k}{\hat b}^{\dagger}_{k} + f^*_{-k}{\hat b}_{-k} \right], \nonumber \\
{\hat \theta}(x) &= \frac{i\pi}{L}\sum_{k\neq 0}\left(\frac{L|k|}{2\pi}\right)^{\frac{1}{2}}\frac{e^{-\frac{{\tilde \alpha}|k|}{2}-ikx}}{|k|} \left[ g_{k}{\hat b}^{\dagger}_{k} - g^*_{-k}{\hat b}_{-k} \right], \nonumber
\end{align}
where ${\tilde \alpha}$ is a cutoff and $L$ is the system size. 
We note that the zero modes in the expansion \cite{haldane1981effective} are not relevant in our analysis.
The Bogoliubov transformation is generated by the two-mode squeeze operator ${\hat S}_{k}(\eta_{k})=e^{\eta_{k}{\hat b}_{k}^{\dagger}{\hat b}_{-k}^{\dagger}-\eta^*_{k}{\hat b}_{k}{\hat b}_{-k}}$, where $\eta_{k}=|\eta_{k}|e^{i\vartheta_{k}} \in \mathbb{C}$. 
The coefficients $f_{k}$ and $g_{k}$, that appear in the fields ${\hat \phi}$ and ${\hat \theta}$, are given by $f_{k} = u_{k} + v_{k}$ and $g_{k}=u_{k} - v_{k}$ with the Bogoliubov coefficients $u_{k}={\rm cosh}(|\eta_{k}|)$ and $v_{k}=e^{i\vartheta_{k}}{\rm sinh}(|\eta_{k}|)$, respectively. 
In this paper, we focus on the case of $K>1$. 
These fields ${\hat \phi}(x)$ and ${\hat \theta}(x)$ reduce to the fields ${\hat \phi}_0(x)$ and ${\hat \theta}_0(x)$ of Eq.~(\ref{eq: tll model}) by choosing $f_k = g_k = 1$, which corresponds to $K=1$, as described in Appendix~\ref{app: energy_functional}. 
The fields ${\hat \phi}(x)$ and ${\hat \theta}(x)$ are related to the fields ${\hat \phi}_0(x)$ and ${\hat \theta}_0(x)$ via the squeeze operation ${\hat S}^{-1} {\hat \phi}_0 {\hat S} = {\hat \phi}$ and ${\hat S}^{-1} {\hat \theta}_0 {\hat S} = {\hat \theta}$, where ${\hat S}=\prod_{k}{\hat S}_{k}$.

In the standard treatment of TLL theory, the parameters $\eta_{k}$ are chosen as a constant value $\eta_{k}=\eta^0 = \frac{1}{2}{\rm ln}K$~\cite{haldane1981effective}, which diagonalizes the TL Hamiltonian. 
However, in the squeezed-field description \cite{seifie2019squeezed}, $\eta_{k}$ is a dynamical field, and allowed to fluctuate around the equilibrium value $\eta_{k}=\eta^0$.
In what follows, according to the previous work \cite{seifie2019squeezed}, we will refer to $\eta_{k}$ as the squeezing field. 

To formulate the dynamics of $\eta_{k}$ explicitly, we consider the thermodynamic partition function ${\mathcal Z}={\rm Tr}e^{-\beta{\hat H}}$. 
${\hat H}$ is a microscopic Hamiltonian expressed in terms of ${\hat \phi}$ and ${\hat \theta}$.
The low-energy limit includes the TL Hamiltonian ${\hat H}_0$.
We insert the completeness relation of two-mode squeezed-coherent states, i.e., ${\hat 1}_{k\neq0} = {\cal N}\int {\cal D}(b,\eta) |\Psi\rangle\langle\Psi|$ where $|\Psi\rangle = \prod_{k>0}{\hat S}(\eta_{k}){\hat D}(b_{k}){\hat D}(b_{-k})|0\rangle$, ${\hat D}(b_{\pm k})=e^{b_{\pm k}{\hat b}^{\dagger}_{\pm k}-b^*_{\pm k}{\hat b}_{\pm k}}$ is the displacement operator of the coherent state, and ${\cal N}$ is a normalization constant \cite{seifie2019squeezed}, into infinitesimal slices of the imaginary time axis of the inverse temperature $\beta=(k_{\rm B}T)^{-1}$.
Then, we obtain a path-integral representation of the partition function ${\cal Z} = \int [{\cal D}(b,\eta)]e^{-\frac{1}{\hbar}\int^{\hbar\beta}_{0} d\tau {\cal L}(b,\eta)}$.
The corresponding Lagrangian is given by 
\begin{align}
{\cal L}(b,\eta)=\langle \Psi| \hbar\partial_{\tau} + {\hat H}_{0}|\Psi\rangle + {\cal L}_{\rm int}(\{b_{k},b_{-k},\eta_{k}\}). \label{eq: lagrangian}
\end{align}
The explicit evaluation of the dynamical term $\langle \Psi | \partial_{\tau} | \Psi \rangle$ is given in Appendix \ref{app: dynamical}.
In a saddle-point approximation, ${\cal L}(b,\eta)$ describes the mean-field trajectory of the squeezing field $\eta_{k}(\tau)$ as well as that of the bosonic modes $b_{k}(\tau)$ of TLLs.
Fluctuations around the saddle-point path provide quantum corrections to the classical limit \cite{altland2010condensed}.
Due to the periodicity of the trace in ${\cal Z}$, the squeezing field should fulfill periodic boundary conditions: $\eta_{k}(\hbar\beta) = \eta_{k}(0)$.
The last term ${\cal L}_{\rm int}$ in Eq.~(\ref{eq: lagrangian}) corresponds to nonlinear interaction terms of ${\hat \phi}$ and ${\hat \theta}$ fields, which appear as corrections to the TL Hamiltonian ${\hat H}_0$ \cite{imambekov2012one,haldane1981effective}. 

We mention that a class of the squeezed-coherent states have also been used as variational wave functions utilizing Dirac--Frenkel theory or the time-dependent variational principle (TDVP) for bosonic systems \cite{tsue1991time,guaita2019gaussian}.
We note that a real-time action derived after a Wick rotation of Eq.~(\ref{eq: lagrangian}) is equivalent to the definition of the TDVP action with respect to the squeezed-coherent state $|\Psi\rangle$.
In this sense, the squeezed-field description may be regarded as a {\it quantization} of TDVP theory, which stands for a classical system whose time evolution is a deterministic trajectory of a point in the phase space of $b_{k}$ and $\eta_{k}$.
We expect that the squeezing field mediates nonlinear interactions of $k$ and $-k$ modes, which are not captured within the free-boson description of the TL Hamiltonian.

\subsection{Fluctuation expansion} \label{sec: fluctuation expansion}

We derive an effective Lagrangian from Eq.~(\ref{eq: lagrangian}) by expanding in the fluctuations of the squeezing field.
In this paper, we will focus on the first term in Eq.~(\ref{eq: lagrangian}), and ignore the term ${\cal L}_{\rm int}$, which is of higher order.
At this order of approximation, the coherent state parameters $b_k$ and $b_{-k}$, and the squeeze parameter $\eta_k$ are coupled, while there is no coupling to other momentum modes. 
Therefore, the dynamics separate into three-mode systems.
Our main focus is to clarify how the free-boson description is modified when the dynamics of $\eta_{k}$ is included, as described by the effective Lagrangian.

We write $\delta \eta_{k} = \eta_{k}-\eta^{0}$ as a small deviation of $\eta_{k}$ around $\eta_{k}=\eta^{0}$.
Then, by expanding the Lagrangian ${\cal L}$ in $\delta \eta_{k}$ up to a leading order, we obtain a nonlinear Lagrangian of the form ${\cal L}_{\rm eff} = {\cal L}_0 + {\cal L}_3$, with ${\cal L}_0 = {\cal L}_{0,b} + {\cal L}_{0, \xi}$ and
\begin{align}
{\cal L}_{0,b} &= \sum_{k \neq 0} b^*_{k} (\hbar\partial_\tau + \hbar\omega_{k}) b_{k}, \\
{\cal L}_{0,\xi} &= \sum_{k > 0}{\tilde \xi}^*_{k}(\hbar\partial_\tau + \hbar \omega_{\xi,k} ){\tilde \xi}_{k},
\end{align}
and 
\begin{align}
{\cal L}_{3} &= \sum_{k>0}w_{1}(|b_{k}|^2+|b_{-k}|^2)\left(\hbar\partial_{\tau}{\tilde \xi}_{k}-\hbar\partial_{\tau}{\tilde \xi}^*_{k} \right) \nonumber \\
&\; + \sum_{k>0}b_{k}b_{-k}(-\hbar\partial_{\tau}+\hbar\omega_{\xi,k}){\tilde \xi}^*_{k}  \nonumber \\ 
&\; + \sum_{k>0}b^*_{k}b^*_{-k}(\hbar\partial_{\tau}+\hbar\omega_{\xi,k}){\tilde \xi}_{k}, \label{eq: lagrangian int}
\end{align}
where $\xi_{k} = (r^{-1}\delta \eta^{R}_{k}+ir \delta \eta^{I}_{k})/\sqrt{2}$, $r=[(K-1/K)/(2{\rm ln}K)]^{1/2}$, ${\tilde \xi}_{k} = \sqrt{2r^2} \xi_{k}$, and $w_{1}=2^{-1}(K^{1/2}-K^{-1/2})^2/(K-K^{-1})$.
Here $\delta \eta^{R}_{k} = {\rm Re}\left[ \delta \eta_{k} \right]$ and $\delta \eta^{I}_{k} = {\rm Im}\left[\delta \eta_{k}\right]$.
The derivation of Eq.~(\ref{eq: lagrangian int}) is presented in Appendix \ref{app: fluctuations}.
Figure~\ref{fig: coupling} shows $r$ and $w_{1}$ as functions of $K$.

The effective Lagrangian ${\cal L}_{\rm eff}$ has two quadratic contributions.
In particular, the term ${\cal L}_{0,b}$ describes a Lagrangian of the non-interacting TLL, while the term ${\cal L}_{0, \xi}$ describes the squeezing mode. 
The Lagrangian ${\cal L}_{0,b}$ contains the linear dispersion $\hbar \omega_{k} = \hbar v_0 |k|$ of the TLL bogolons, which are the $b_{k}$ variables.
The squeezing field $\xi_{k}$ has a dispersion $\hbar\omega_{\xi,k}=2\hbar\omega_{k}$, i.e. twice the TLL dispersion.
As shown in the derivation of the effective Lagrangian, the Hamiltonian of the quadratic squeezing-field Lagrangian, i.e., $\sum_{k > 0}\hbar \omega_{\xi,k}{\tilde \xi}^*_{k}{\tilde \xi}_{k}$ arises in the fluctuation expansion of the vacuum energy functional in $E=\langle \Psi | {\hat H}_0 | \Psi \rangle$, see also Appendices~\ref{app: energy_functional} and~\ref{app: fluctuations}.
We note that the ratio $\omega_{\xi,k}/\omega_{k}=2$ agrees with that of the Bogoliubov and squeezing-field dispersions in a 3D weakly-interacting Bose gas in the long wavelength limit \cite{seifie2019squeezed}.
We also note that the Langrangian ${\cal L}$ reduces to the free-boson model of TLL theory if we replace the squeezing field $\eta_{k}$ by its equilibrium value $\eta^0$, which corresponds to setting all variables $\xi_{k}$ to zero. 
Therefore the free-boson model is the mean-field approximation of the model ${\cal L}_{\rm eff}$, in the sense that the squeezing field $\eta_{k}$ is replaced by its expectation value. 

\begin{figure}
\begin{center}
\includegraphics[width=80mm]{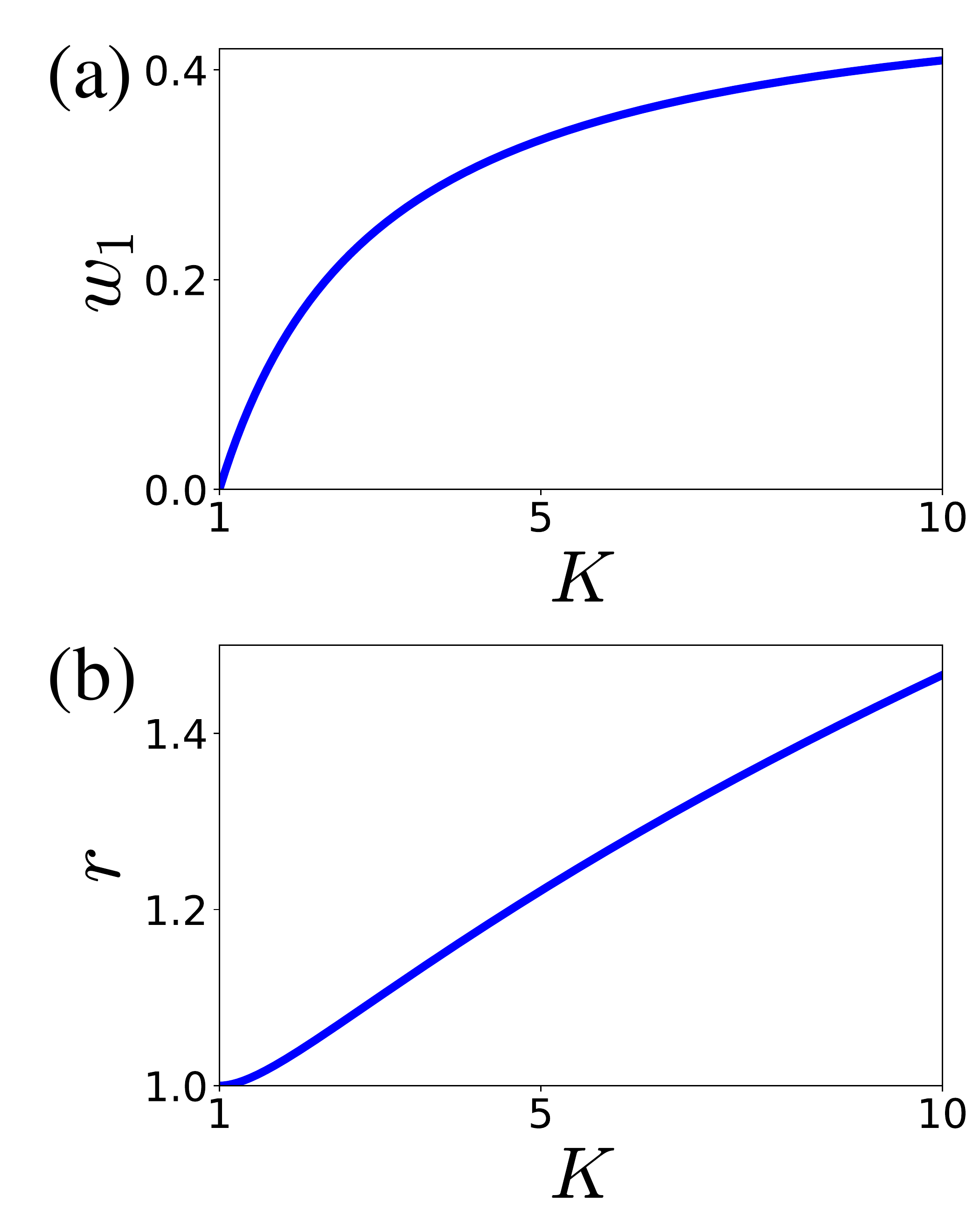}
\vspace{-4mm}
\caption{
$K$ dependence of the constants (a) $w_{1}$ and (b) $r$ of the effective Lagrangian ${\cal L}_{\rm eff}$.
}
\vspace{-8mm}
\label{fig: coupling}
\end{center}
\end{figure}

The nonlinear part of the effective Lagrangian ${\cal L}_{3}$ contains two types of vertices describing the interaction among the TLL and the squeezing modes.
The vertex $w_{1}b^*_{\pm k}b_{\pm k}\partial_{\tau}{\tilde \xi}_{k}$ describes processes, in which the fluctuation of the squeezing field is perturbed by the occupation of the TLL modes at $\pm k$.
This term contains the time derivative of the squeezing field and is generated from the fluctuation expansion of the dynamical term $\langle \Psi | \hbar \partial_{\tau} | \Psi \rangle$.
We note that the interaction strength $w_{1}$ is a monotonically increasing function of $K$, see Fig.~\ref{fig: coupling}.

The other vertex $b^*_{k}b^*_{-k}(\hbar\partial_{\tau}+\hbar\omega_{\xi,k}){\tilde \xi}_{k}$ describes the creation of a momentum-conserving pair of the TLL modes from a fluctuation of the squeezing field.
The prefactor of this scattering process is the inverse of the free propagator of the squeezing field, $G^{-1}_{\xi,k,\pm}(\tau,\tau') = (\pm \hbar\partial_{\tau}+\hbar\omega_{\xi,k})_{\tau,\tau'}$.
This inverse propagator is cancelled by free propagators of the squeezing field that appear in a perturbative expansion of multipoint correlation functions.
These cancellations directly influence the lower order poles of the correlators, see also Appendix~\ref{app: perturbation}.

Before closing this section, we define the Matsubara frequency expansion for the fields:
\begin{align}
b_{k}(\tau) = \frac{1}{(\beta\hbar)^{1/2}}\sum_{n \in {\mathbb Z}}e^{-i\omega_{n}\tau}b_{k}(\omega_{n}), \\
{\tilde \xi}_{k}(\tau) = \frac{1}{(\beta\hbar)^{1/2}}\sum_{n \in {\mathbb Z}}e^{-i\omega_{n}\tau}{\tilde \xi}_{k}(\omega_{n}).
\end{align}
Due to the periodicity of the fields, the Matsubara frequency $\omega_{n}$ is bosonic, i.e., $\omega_{n}=2\pi n/(\beta\hbar)$ ($n \in {\mathbb Z}$) \cite{altland2010condensed}.
Then, the non-interacting part of the total effective action reads 
\begin{align}
{\cal S}_{2} 
&= \int^{\hbar\beta}_{0}d\tau {\cal L}_{0} \nonumber \\
&= \sum_{n\in {\mathbb Z}}\sum_{k \neq 0}(-i\hbar\omega_{n}+\hbar\omega_{k})b^*_{k}(\omega_{n})b_{k}(\omega_{n}) \nonumber \\
&\;\;\;\; + \sum_{n\in {\mathbb Z}}\sum_{k > 0}(-i\hbar\omega_{n}+\hbar\omega_{\xi,k}){\tilde \xi}^*_{k}(\omega_{n}){\tilde \xi}_{k}(\omega_{n}).
\end{align}
The frequency representation of ${\cal S}_{3} = \int^{\hbar\beta}_{0}d\tau {\cal L}_{3}$ is shown in Appendix~\ref{app: perturbation}.

\section{Non-interacting correlation function at zero temperature} \label{sec: noninteracting}

In the remainder of this paper, we apply our formalism of squeezed fields to concrete physical quantities. 
As an instructive application, we analyze the imaginary time correlation function of the field ${\hat \phi}(x)$ in the non-interacting limit.
In this limit, the system is described by the effective Lagrangian ${\cal L}_{0}={\cal L}_{0,b}+{\cal L}_{0,\xi}$. 
It is regarded as a system of decoupled harmonic oscillators of two species.
We assume $T=0$ in this section.
Nonzero temperatures and nonlinear contributions will be discussed in Sec.~\ref{sec: spectral function}.

We consider the ground-state correlation function for a vertex operator of ${\hat \phi}$, in particular ${\hat {\cal V}} = e^{i{\hat \phi}(x,\tau)}$ \cite{fradkin2013field}.
This correlation function is a gauge-invariant expectation value for measuring correlations between two distinct points in space and time, which is widely discussed in standard TLL theory \cite{giamarchi2004quantum}.
We define  
\begin{align}
C_{\phi\phi}(x,\tau)
&=\langle T_{\tau}e^{i{\hat \phi}(x,\tau)} e^{-i{\hat \phi}(0,0)} \rangle_{0} \nonumber \\
&= e^{-\frac{1}{2}\langle [\phi(x,\tau)-\phi(0,0)]^2 \rangle_{0}}, \label{eq: vertex_correlation}
\end{align}
where $\langle \cdots \rangle_{0}$ is the vacuum expectation value at $T=0$.
$T_{\tau}$ is the time ordering operation along the $\tau$ axis and ${\hat \phi}(x,\tau)$ is the imaginary-time Heisenberg operator of ${\hat \phi}(x)$.
In the squeezed-field description, as indicated in the previous section, the vacuum average is interpreted as a functional average over the phase space of $b_{k}(\tau)$ and ${\tilde \xi}_{k}(\tau)$ variables.

In the non-interacting limit, there is no correlation between $b_{k}$ and ${\tilde \xi}_{k}$ in the functional average.
Hence, the two-point correlation function in the exponent of $C_{\phi\phi}(x,\tau)$, i.e., $\langle \phi(x,\tau)\phi(0,0) \rangle_{0}$ reduces to the form
\begin{align}
\langle \phi(x,\tau)\phi(0,0) \rangle_{0} &= \frac{\pi}{2L}\sum_{k\neq 0}\frac{1}{|k|}e^{-{\tilde \alpha}|k|+ikx} \\
&\;\;\;\;\;\;\; \times \langle f^*_{k}(\tau)f_{k}(0)\rangle_{0} \langle b_{k}(\tau)b^*_{k}(0)\rangle_{0}, \nonumber
\end{align}
where $\tau > 0$.
We note that the average $\langle f^*_{k}(\tau)f_{k}(0)\rangle_{0}$ reduces to a constant in the conventional TLL description: $\langle f^*_{k}(\tau)f_{k}(0)\rangle_{0} \rightarrow  |f_{k}(\eta^{0})|^2 = K$.
We expand the coefficient $f_{k}(\tau)$ to first order in $\delta\eta_{k}(\tau)$ as $f_{k}(\tau) \approx f_{k}(\eta^0) + \frac{\partial f_{k}(\eta^0)}{\partial \eta^{R}_{k}}\delta \eta^{R}_{k}(\tau) + \frac{\partial f_{k}(\eta^0)}{\partial \eta^{I}_{k}}\delta \eta^{I}_{k}(\tau) + \cdots$.
We perform the Gaussian integrations of $b_{k}$ and ${\tilde \xi}_{k}$ to evaluate $\langle \phi(x,\tau)\phi(0,0) \rangle_{0}$ while keeping the leading order contributions to obtain
\begin{align}
\langle \phi(x,\tau)\phi(0,0) \rangle_{0} &\approx \frac{K \pi}{2L}\sum_{k\neq 0}\frac{1}{|k|}e^{-{\tilde \alpha}|k|+ikx}e^{-\omega_{k}\tau} \label{eq: phi_phi_expansion} \\
& + \frac{\varGamma \pi}{2L}\sum_{k\neq 0}\frac{1}{|k|}e^{-{\tilde \alpha}|k|+ikx} e^{-(\omega_{k}+\omega_{\xi,k})\tau}, \nonumber
\end{align}
where we find that $\varGamma = \frac{K(K+3)(K-1)}{4(K+1)^2}$.
The fluctuation-correction term in the second line implies that the squeezing mode with dispersion $\omega_{\xi,k}=2v_0|k|$ produces an additional side-band dispersion as $\omega_{k}+\omega_{\xi,k}=3v_0|k|$ in the correlation function $C_{\phi\phi}$, in addition to the TLL dispersion $\omega_{k}=v_0|k|$.

To obtain an analytical form of $C_{\phi\phi}$, we take the continuum limit of the momentum sum, such that $\sum_{k>0}\rightarrow\int^{\infty}_{0} \frac{dk}{2\pi}$.
Evaluating the integration, which is parallel to that in Ref.~\cite{giamarchi2004quantum}, we arrive at the expression 
\begin{align}
C_{\phi\phi}(x,\tau) \approx \left[ \frac{{\tilde \alpha}^2}{x^2+({\tilde \alpha} + v_0\tau)^2} \right]^{\frac{\kappa_0}{2}}\left[ \frac{{\tilde \alpha}^2}{x^2+({\tilde \alpha} + 3v_0\tau)^2} \right]^{\frac{\kappa_1}{2}}.
\end{align}
Thus, at the leading order expansion, the correlation function of the vertex operator is a product composed of two propagators.
The first propagator, which is characterized by a critical exponent $\kappa_{0}=K/2$ and the sound velocity $v_{0}$, derives from the first term in Eq.~(\ref{eq: phi_phi_expansion}).
It is the standard result of TLL theory \cite{giamarchi2004quantum}. 

\begin{figure}
\begin{center}
\includegraphics[width=80mm]{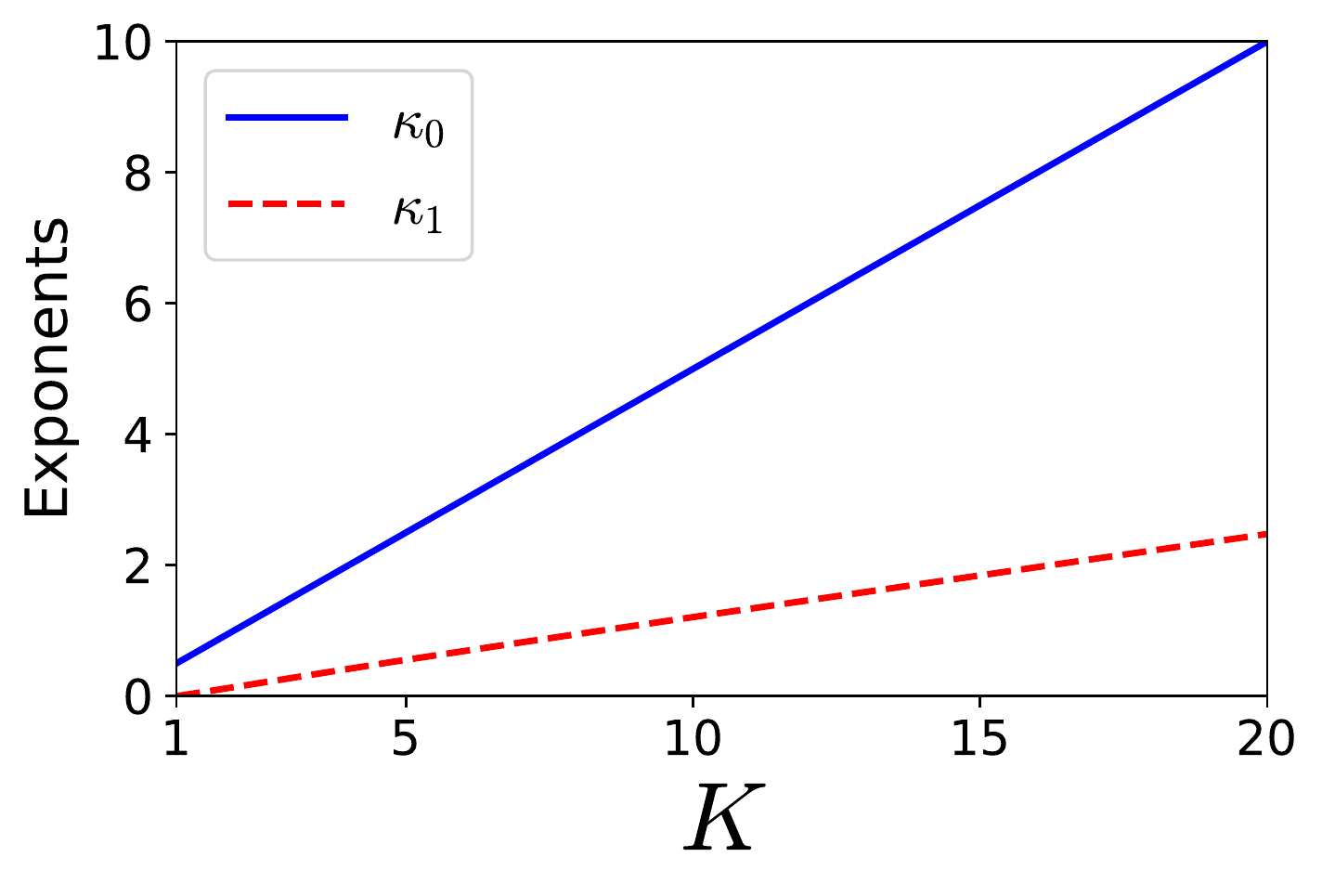}
\vspace{-4mm}
\caption{
Ground-state critical exponents in the correlation function $C_{\phi\phi}(x,\tau)$ in the non-interacting limit.
The exponent $\kappa_{1}$ characterizes the decay of the correlation of the squeezing-field contribution.
The exponent $\kappa_{0}$ is the standard exponent of TLL theory.
}
\label{fig: exponents}
\end{center}
\end{figure}

However, the second propagator, which reflects the fluctuations of the squeezing field, contains a propagation velocity of $3v_0$ and a critical exponent $\kappa_{1}$ given by
\begin{align}
\kappa_1 = \frac{1}{2}\varGamma = \frac{K(K+3)(K-1)}{8(K+1)^2}.
\end{align}
We note that $\kappa_1 \rightarrow K/8$ as $K \rightarrow \infty$ and that $\kappa_{1}$ is always bounded by $\kappa_{0}$ from above, for $K>1$.
In Fig.~\ref{fig: exponents}, $\kappa_{0}$ and $\kappa_{1}$ are displayed as a function of the TL parameter $K$.
As we show in Sec.~\ref{sec: spectral function}, the emergence of the side-band branch is observed in a spectral function of a 1D Bose gas system.
Furthermore, we point out that a similar side-band branch emerges in the single-particle Green's function for a dilute Bose gas in 3D in the squeezed-field path integral approach \cite{seifie2019squeezed}.

\section{Model for a 1D Bose gas} \label{sec: model}

In this section we summarize the phenomenological bosonization of bosons in 1D \cite{haldane1981effective,giamarchi2004quantum,cazalilla2011one} as a preparational step for Sec.~\ref{sec: spectral function}.

We consider a Bose gas, described by the microscopic Hamiltonian
\begin{align}
{\hat H}_{\rm B} = &-\frac{\hbar^2}{2m}\int dx {\hat \psi}^{\dagger}(x)\partial^2_{x}{\hat \psi}(x) \nonumber \\
&\;\;\;\;\;\;\;\;\;\;\;\;\;\;\;\;\; + \frac{1}{2}\iint dxdy {\hat \rho}(x)v(x-y){\hat \rho}(y), \label{eq: hamiltonian_boson}
\end{align}
where ${\hat \rho}(x) = {\hat \psi}^{\dagger}(x){\hat \psi}(x)$ is an atomic density at position $x$, and $m$ is the mass.
The Bose fields, ${\hat \psi}^{\dagger}(x)$ and ${\hat \psi}(x)$, obey the canonical commutation relation $[{\hat \psi}(x),{\hat \psi}^{\dagger}(x')]=\delta(x-x')$.
The first term of this Hamiltonian is the kinetic energy of the gas while the second term gives a density-density interaction with a short-range two-body potential $v(x-y)$. 
Throughout this work, we consider a homogeneous system, and exclude inhomogeneous features, such as trap effects. 
In particular, this implies that the system features Galilean invariance. 

In the bosonization \cite{giamarchi2004quantum,cazalilla2011one}, we write the creation operator in phase-density representation, i.e. as ${\hat \psi}^{\dagger}(x)=\sqrt{{\hat \rho}(x)}e^{-i{\hat \theta}(x)}$.
Furthermore, we represent the density ${\hat \rho}(x)$ in terms of a field ${\hat \phi}(x)$
\begin{align}
{\hat \rho}(x) = \left(\rho_0 - \frac{1}{\pi}\partial_{x}{\hat \phi}(x)\right)\sum_{l}e^{i2 l [\pi \rho_0 x - {\hat \phi(x)}]}. \label{eq: transform_density}
\end{align}
The field ${\hat \phi}(x)$ is the second canonical field of the TLL formalism.
The index $l$ in Eq.~(\ref{eq: transform_density}) is an integer and $\rho_0$ is the equilibrium density of the system.
Inserting the expression into Eq.~(\ref{eq: hamiltonian_boson}) and neglecting nonlinear terms, the Hamiltonian reduces to the TL Hamiltonian (\ref{eq: tll model}).
For a more complete discussion, see Refs.~\cite{haldane1981effective,giamarchi2004quantum,cazalilla2011one}.

We note that the Galilean invariance of the system implies the relation $v_0 K = \hbar \pi \rho_0 / m$ \cite{haldane1981effective,cazalilla2011one}, which will be used in Sec.~\ref{sec: spectral function}.

\section{Spectral function in the 1D dilute Bose gas} \label{sec: spectral function}

\begin{figure}
\begin{center}
\includegraphics[width=85mm]{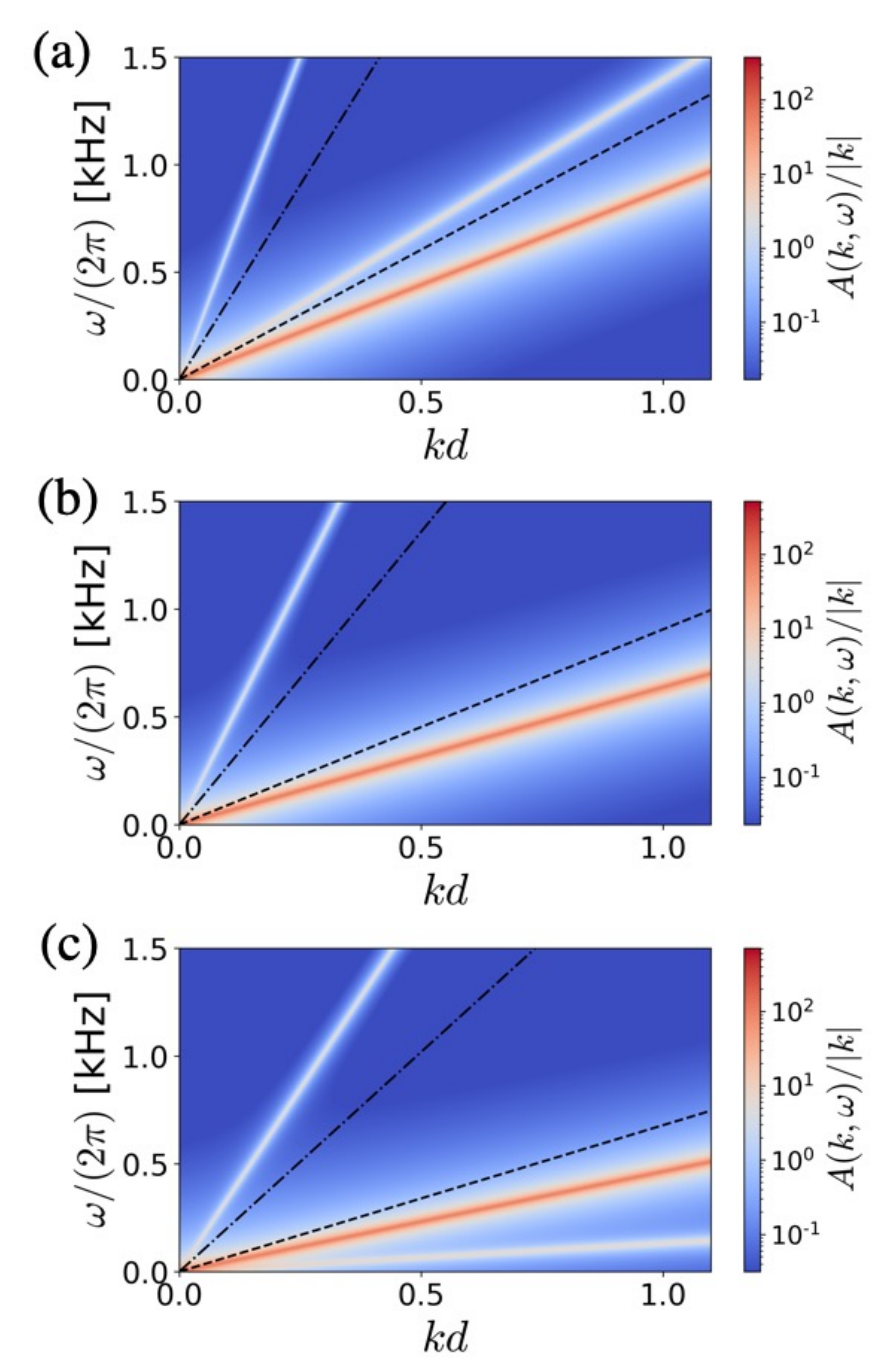}
\vspace{-4mm}
\caption{
Spectral function $A(k,\omega)$ of the 1D Bose gas at $T=0$ in the presence of the interactions between the TLL and squeezing modes.
We have set the parameters as (a) $K=9$, (b) $K=12$, (c) $K=16$, $d=1\mu \text{m}$, and $\rho_0=3.0 \times 10^{7} \text{m}^{-1}$.
The atomic species is ${}^{87}$Rb.
To display $A(k,\omega)$, we have replaced its delta function with Lorentzian functions with a width $\varepsilon = 0.01$ kHz.
We determine the velocity $v_0$ via $v_0 K = \hbar \pi \rho_0 / m$.
The dashed and dashed-dotted lines are the non-interacting TLL dispersion $\omega=\omega_{k}=v_0|k|$ and the non-interacting side-band dispersion of the squeezing field $\omega=\omega_{k}+\omega_{\xi,k}=3v_0|k|$. 
(a) The Lorentzian peaks from bottom to top are positioned at $\omega={\tilde v}_{1}|k|$, $\omega={\tilde v}_{3}|k|$, and $\omega=v_{5}|k|$, respectively.
$\tilde{v}_1$, $\tilde{v}_3$, and ${v}_5$ are given in the text. 
(b) For $K=12$, the values of ${\tilde v}_{1}$ and ${\tilde v}_{3}$ are almost the same.
(c) The renormalized velocity of the side-band branch ${\tilde v}_{3}$ is lowered below the TLL branch with velocity ${\tilde v}_{1}$.
}
\label{fig: zero_temperature}
\end{center}
\end{figure}

To characterize the dynamical properties of the Bose-gas Hamiltonian (\ref{eq: hamiltonian_boson}), we determine the spectral function $A(k,\omega)$ of the density fluctuations.
This quantity is defined as the imaginary part of the spatiotemporal Fourier transform of the retarded density-density correlation function $C^{\rm R}(x,t) = -i\Theta(t)\langle [{\hat \rho}(x,t),{\hat \rho}(0,0)] \rangle$ \cite{altland2010condensed}, i.e.,
\begin{align}
A(k,\omega) = -2\text{Im}\left[\int^{\infty}_{-\infty} dt \int dx e^{ikx+i\omega t} C^{\rm R}(x,t)\right]. 
\end{align}
$\Theta(t)$ is the unit step function ensuring causality.
We apply the squeezed field path integral formalism to $A(k, \omega)$, to obtain properties beyond standard TLL theory.
We incorporate the cubic interactions of the effective Lagrangian perturbatively.
We will discuss the features emerging in $A(k,\omega)$ due to the squeezed field description and compare them with TLL theory.
While our primary physical example is a Bose liquid in 1D with repulsive interactions, as similar discussion applies to a spinless Fermi liquid in 1D with attractive interactions.

To determine the spectral function $A(k,\omega)$, we calculate the imaginary-time correlation function corresponding to $C^{\rm R}(x,t)$.
From an analytical expression of the imaginary-time correlation function, we obtain the spectral function by performing an analytical continuation of the Matsubara-expansion coefficient as $\chi''(k,i\omega_{n}) = \int^{\hbar\beta}_{0} d\tau e^{i\omega_{n}\tau}\chi'(k,\tau)$ from $i\omega_{n}$ to $\omega+i 0^{+}$ \cite{altland2010condensed}: $A(k,\omega)=-2\text{Im}\left[\chi''(k,i\omega_{n}\rightarrow \omega+i0^{+})\right]$.
The long-wavelength limit of this quantity corresponds to the $l=0$ harmonic of the Haldane representation shown in Eq.~(\ref{eq: transform_density}), which gives $\chi(x,\tau)=-\frac{1}{\pi^2}\langle \partial_{x}\phi(x,\tau) \partial_x\phi(0,0) \rangle$.
We ignore the higher-harmonic contributions due to $l \neq 0$, which give corrections at multiples of a Fermi point $k_{\rm F}$ \cite{giamarchi2004quantum}.
We substitute the mode expansion of $\phi(x,\tau)$ in the correlation function and take the Fourier transform of $\chi(x,\tau)$ from $x$- to $k$-space to obtain
\begin{align}
\chi'(k,\tau)
&=\text{F.T.}_{x \rightarrow k}\left[ \chi(x,\tau)\right] \nonumber \\
&= -\frac{|k|}{2\pi}\langle X_{k}(\tau)X_{-k}(0) \rangle,
\end{align}
where $X_{k}(\tau)=f_{k}[\eta_{k}(\tau)]b^*_{k}(\tau)+f_{k}^*[\eta_{k}(\tau)]b_{-k}(\tau)$.
We note that the mode-expansion coefficient $f_{k}(\eta_{k})$ varies with $\eta_{k}(\tau)$ in the squeezed-field approach, which induces correlations between $b_{k}$ and $f_{k}$.

Next, we evaluate the spectral function $A(k,\omega)$. 
We adopt Feynman's diagrammatic perturbation theory for the Euclidean path integrals \cite{altland2010condensed}.
First, as discussed in Sec.~\ref{sec: noninteracting}, the correlation function $\chi'(k,\tau)$ is expanded in fluctuations of $\eta_{k}(\tau)$ up to a certain order.
Then, that expanded expression is reduced to a finite set of multipoint correlation functions of the field variables such as the 2-point function $\langle b_{k}(\tau)b^*_{k}(0)\rangle$, a 3-point function $\langle {\tilde \xi}^*_{k}(\tau) b^*_{k}(\tau)b_{k}(0)\rangle$, and 4-point functions such as $\langle {\tilde \xi}_{k}(\tau) b^*_{k}(\tau){\tilde \xi}^*_{k}(0)b^*_{-k}(0)\rangle$.
As the next step, we expand each correlation function with respect to the perturbative term of the effective Lagrangian, and draw the possible Feynman graphs, which correspond to the zeroth-order contribution and perturbative corrections due to the cubic vertices.
To elucidate the lower-order contributions of $\chi'(k,\tau)$, we carry out the perturbative analysis to the second order in the cubic vertices.
A more detailed discussion is given in Appendix \ref{app: perturbation}.

\begin{figure*}
\includegraphics[width=180mm]{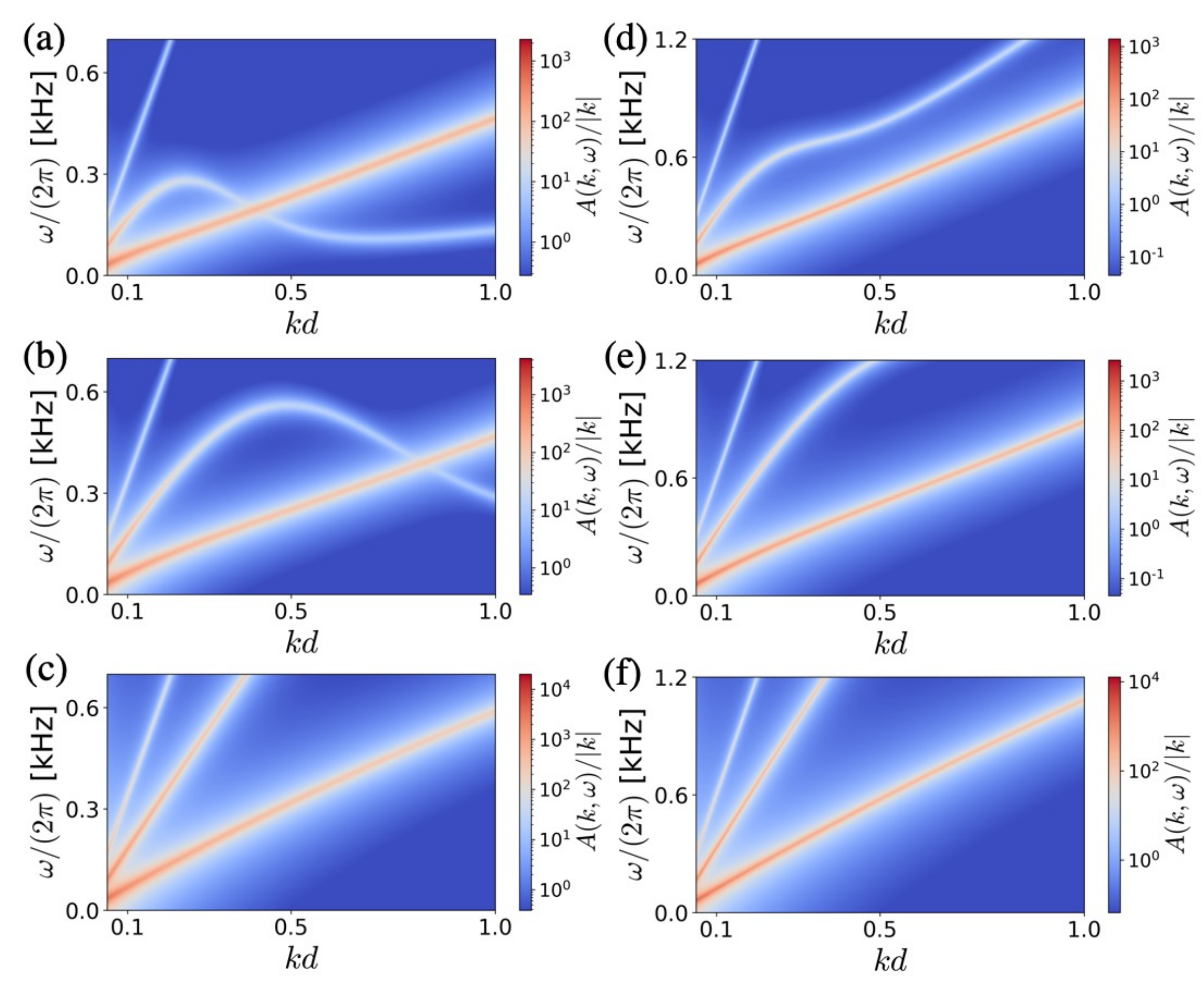}
\vspace{-4mm}
\caption{
Spectral function $A(k,\omega)$ at nonzero temperature.
The panels correspond to (a) $K=16$, $k_{\rm B}T=0.1E_0$, (b) $K=16$, $k_{\rm B}T=0.2E_0$, (c) $K=16$, $k_{\rm B}T=1.0E_0$, (d) $K=9$, $k_{\rm B}T=0.1E_0$, (e) $K=9$, $k_{\rm B}T=0.2E_0$, and (f) $K=9$, $k_{\rm B}T=1.0E_0$, respectively.
Other parameters are the same in Fig.~\ref{fig: zero_temperature}.
We note that each origin of this plot is at $kd=0.05$ to avoid the singularity of the Bose distribution.
}
\label{fig: finite_temperature}
\end{figure*}

First, we consider the non-interacting limit once again as studied in Sec.~\ref{sec: noninteracting}.
In this limit, the spectral function reads
\begin{align}
|k|^{-1}A(k,\omega) 
\approx {\cal A}_0&\left[\delta(\omega+v_{0}|k|) - \delta(\omega-v_{0}|k|) \right] \nonumber \\
+ {\cal B}_0&\left[\delta(\omega+3v_{0}|k|) - \delta(\omega-3v_{0}|k|)\right].
\end{align}
The spectral weights ${\cal A}_0$ and ${\cal B}_0$ are functions of $k$, $T$, and $K$.
The two-point propagators $\langle b^*_{k}(\tau)b_{k}(0) \rangle$ and $\langle b_{k}(\tau)b^*_{k}(0) \rangle$ yield the TLL dispersion with velocity $v_0$, whereas the four-point propagators, i.e., $\langle b^*_{k}(\tau){\tilde \xi}_{k}(\tau)b_{k}(0){\tilde \xi}^*_{k}(0) \rangle$, $\langle b^*_{k}(\tau){\tilde \xi}^*_{k}(\tau)b_{k}(0){\tilde \xi}_{k}(0) \rangle$, $\langle b_{k}(\tau){\tilde \xi}^*_{k}(\tau)b^*_{k}(0){\tilde \xi}_{k}(0) \rangle$, and $\langle b_{k}(\tau){\tilde \xi}_{k}(\tau)b^*_{k}(0){\tilde \xi}^*_{k}(0) \rangle$, create the side-band dispersion with $3v_0$.
The appearance of both dispersions in the spectral function $A(k,\omega)$ is consistent with the results of Sec.~\ref{sec: noninteracting}.
Furthermore, at zero temperature, ${\cal A}_0$ and ${\cal B}_0$ are proportional to $\kappa_0$ and $\kappa_1$, respectively.
Hence, the weight of the TLL dispersion is more dominant in the spectral function than that of the side-band dispersion.
At this level of approximation, the weight and velocity of the TLL branch are the same as in standard TLL theory.

Next, we include the cubic interactions of the TLL and the squeezing modes.
The perturbative analysis is presented in Appendix \ref{app: perturbation}.
We find that the weights and velocities of the TLL and the side-band dispersions are renormalized due to the interactions, and that an additional linear branch appears with velocity $v_{5}=5v_0$ in the spectral function:
\begin{align}
|k|^{-1}A(k,\omega) 
\approx {\cal A}&\left[\delta(\omega+{\tilde v}_{1}|k|) - \delta(\omega-{\tilde v}_{1}|k|) \right] \nonumber \\
+ {\cal B}&\left[\delta(\omega+{\tilde v}_{3}|k|) - \delta(\omega-{\tilde v}_{3}|k|)\right] \label{eq: spectral_function_second_order} \\
+ {\cal C}&\left[ \delta(\omega+v_{5}|k|) - \delta(\omega-v_{5}|k|) \right]. \nonumber
\end{align}
The weights of the peaks, ${\cal A}$, ${\cal B}$, and ${\cal C}$, are functions of $k$, $T$, and $K$.
The expressions are given in Appendix \ref{app: perturbation}.
We note that ${\cal A}$ and ${\cal B}$ approach ${\cal A}_0$ are ${\cal B}_0$ in the non-interacting limit, whereas ${\cal C}$ approaches zero.
The renormalized phase velocities ${\tilde v}_{1}$ and ${\tilde v}_{3}$ are given by 
\begin{align}
{\tilde v}_{1} = v_{0}\left(1-\frac{2{\cal A}'}{\cal A}\right),\;\; {\tilde v}_{3} = 3v_{0}\left(1-\frac{2{\cal B}'}{\cal B}\right), \label{eq: vel}
\end{align}
where ${\cal A}'$ and ${\cal B}'$ depend on $k$, $T$, and $K$, and are given in Appendix \ref{app: perturbation}.

In Fig.~\ref{fig: zero_temperature}, we plot the spectral function of Eq.~(\ref{eq: spectral_function_second_order}) in the $k$-$\omega$ plane at zero temperature.
At zero temperature, the constants ${\cal A}$, ${\cal B}$, ${\cal C}$, ${\cal A}'$, and ${\cal B}'$ depend only on $K$.
Therefore, the dispersions are still linear as in the non-interacting approximation.
As indicated in Fig.~\ref{fig: zero_temperature}, the peaks at $\omega = \omega_{k}$ and $\omega = \omega_{k}+\omega_{\xi,k}$ of the non-interacting limit are lowered by the couplings.
For $K < 12$, the renormalized side-band dispersion has a higher velocity than the renormalized TLL dispersion [Fig.~\ref{fig: zero_temperature}(a)]. 
For values of $K$ larger than a threshold value around $K \sim 12$ [Fig.~\ref{fig: zero_temperature}(b)], the side-band dispersion has a velocity smaller than the velocity of the TLL dispersion [Fig.~\ref{fig: zero_temperature}(c)].
Furthermore, in all panels of Fig.~\ref{fig: zero_temperature}, the additional branch with $5v_0$ has a velocity larger than the other branches for $K>1$.
In all cases, the weight of the TLL branch is dominant compared to the other branches.

Next, we present how the spectral function displayed in Eq.~(\ref{eq: spectral_function_second_order}) is modified at nonzero temperature.
In Fig.~\ref{fig: finite_temperature}, we plot the spectral function (\ref{eq: spectral_function_second_order}) for several temperatures.
The temperature unit is $E_0 = \hbar v_0/d$, where $v_0$ is determined via $v_0 K = \hbar \pi \rho_0 / m$, as before.
As visible in Fig.~\ref{fig: finite_temperature}(a), at $k_{\rm B}T  = 0.1E_0$, the TLL-mode dispersion with $K=16$ is almost linear in $k$ similar to the $T=0$ result.
In contrast, in the same figure, the side-band dispersion associated with ${\tilde v}_{3}$ significantly varies with $k$, in a non-linear fashion.
For smaller momenta, in particular for $kd \lesssim 0.41$, the side-band dispersion is larger than the TLL dispersion, i.e., ${\tilde v}_{1}(k,T)|k|<{\tilde v}_{3}(k,T)|k|$. 
For larger momenta, i.e. $kd \gtrsim 0.41$, this hierarchy is reversed and the two velocities closer to the ones at $T=0$.
Then, the two dispersions intersect at around $kd \approx 0.41$.
As demonstrated in detail in Appendix~\ref{app: perturbation}, the momentum dependence of the renormalized dispersions is due to the Bose distribution functions in the perturbative corrections. 
Taking this result into account, the increase of the velocity of the side-band branch can be attributed to the thermal population of the excitation branches at lower momenta.
As the temperature is raised, the thermal population of the excitation branches extends to higher momenta and the intersection point moves to the larger momenta, see Fig.~\ref{fig: finite_temperature}(b).
For higher temperatures, the excitation modes are populated over a wide momentum range and the side-band dispersion approaches a linear behavior, see Fig.~\ref{fig: finite_temperature}(c), that differs from the linear behavior at $T=0$.

\begin{figure}
\includegraphics[width=86mm]{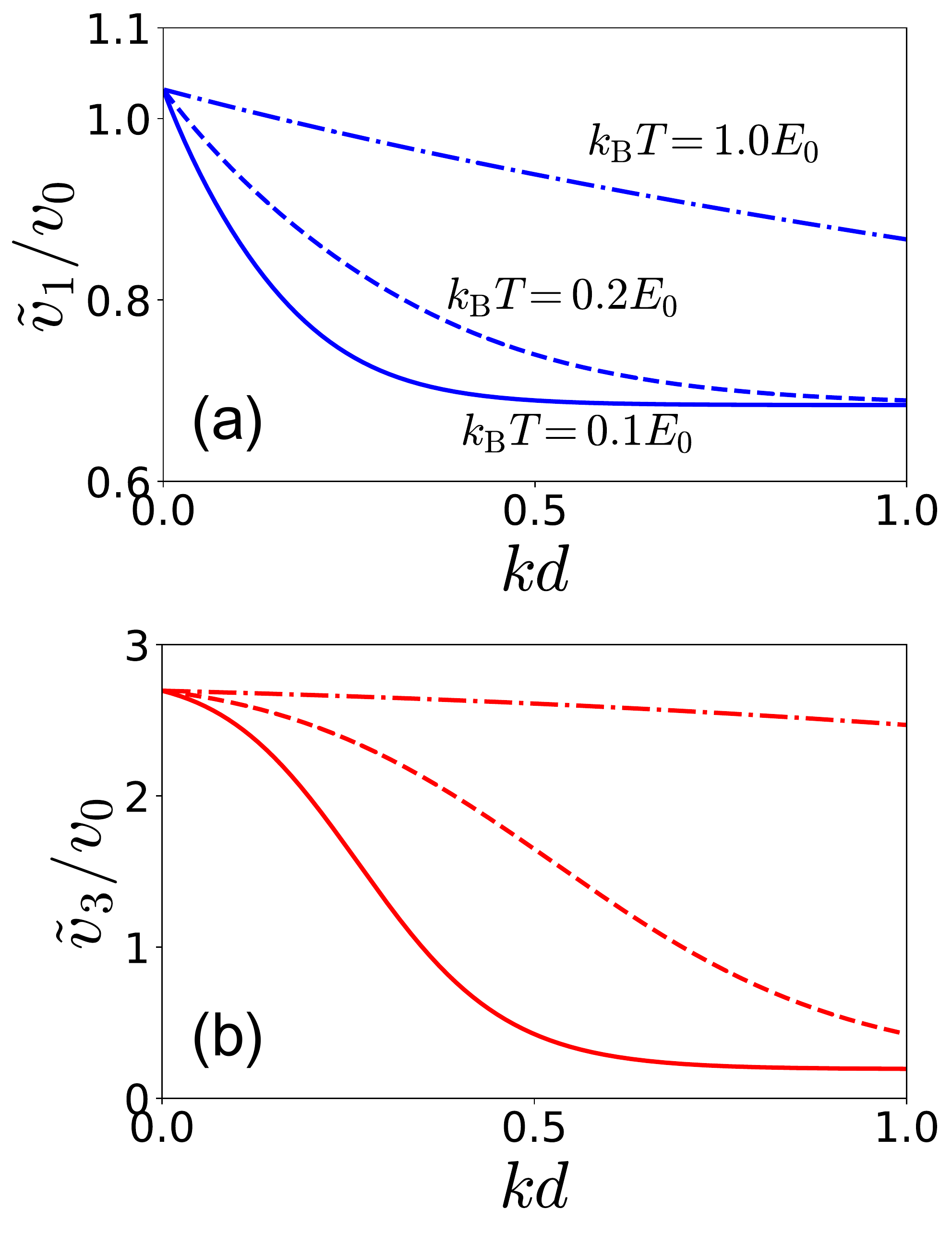}
\vspace{-4mm}
\caption{
Momentum dependence of the renormalized velocities (a) ${\tilde v}_{1}$ and (b) ${\tilde v}_{3}$ for different temperatures.
The velocities are scaled by the zero-temperature sound velocity $v_{0}$.
The TL parameter is set to $K=16$.
The solid, dashed, and dashed-dotted lines are $k_{\rm B}T=0.1E_0$, $k_{\rm B}T=0.2E_0$, and $k_{\rm B}T=1.0E_0$, respectively.
}
\label{fig: velocity}
\end{figure}

To elaborate on the temperature dependence of the excitation branches, we depict ${\tilde v}_{1}$ and ${\tilde v}_{3}$ as a function of $k$ in Fig.~\ref{fig: velocity} for fixed temperatures.
For $k_{\rm B}T = 0.1 E_0$ in Fig.~\ref{fig: finite_temperature}(a), ${\tilde v}_{3}$ is larger than ${\tilde v}_{1}$ in the range $kd < 0.41$. 
$v_1$ and $v_3$ approximately agree with each other around $k d \approx 0.41$. 
With increasing $k$, ${\tilde v}_{3}$ decreases more by a larger amount than ${\tilde v}_{1}$.
For momenta $kd > 0.5$, the two curves approach constants.
With increasing temperature, the curves fall off less rapidly.
This is particularly visible in Fig.~\ref{fig: finite_temperature}(c), for $k_{\rm B}T = 1.0 E_0$.

If the TL parameter is sufficiently small, such as $K=9$, we observe that the side-band dispersion does not display a significant momentum dependence compared to large $K$.
In Fig.~\ref{fig: finite_temperature}(d), in which we set $k_{\rm B}T = 0.1E_0$ corresponding to Fig.~\ref{fig: finite_temperature}(a), the side-band dispersion has a maximum in the range of $kd < 0.5$ due to the thermal populations of the excitation branches, however, the sideband dispersion has a larger frequency than the TLL dispersion.
This feature does not change for any temperature [Fig.~\ref{fig: finite_temperature}(e-f)]. 

Finally, the dispersion with $5v_0|k|$ is not affected by the thermal fluctuations because it is not corrected within second-order perturbation theory.
Higher order correction will modify this result.

\section{Discussion} \label{sec: discussions}

Ultracold Bose gases in 1D have been realized experimentally, and their excitation properties have been studied in Refs.~\cite{fabbri2015dynamical,yang2017quantum}.
In particular, in Ref.~\cite{fabbri2015dynamical}, the dynamical structure factor $S(k,\omega)=\int^{\infty}_{-\infty} dt \int dx e^{-ikx + i\omega t}\langle {\hat \rho}(x,t){\hat \rho}(0,0) \rangle$ has been measured for a gas of $^{87}{\rm Rb}$ by utilizing Bragg spectroscopy techniques.
This quantity is proportional to the spectral function $A(k,\omega)$ within linear response theory, see e.g.~\cite{altland2010condensed}.
Therefore our predictions of Sec.~\ref{sec: spectral function} can be studied experimentally in currently realizable setups.

However, we note that our analysis for the spectral function does not take into account the interactions between different momentum shells, which lead to broadening of the spectral peaks and damping of the excitations \cite{imambekov2012one}.
For the understanding of these additional modifications of our predictions, a perturbative framework will be developed elsewhere. 
The purpose of this extension of the analysis presented here, is to address  the question whether the side-band branch can be detected as a distinguishable peak.
Both experimental and numerical insight on this property would be of significant guidance.
For instance, the numerical insight could derive from quantitative simulations for a concrete realization, e.g., on the basis of time-evolving block decimation (TEBD) methods with matrix product states (MPS) \cite{schollwock2011the,vidal2007classical}.    

\section{Conclusions} \label{sec: conclusion}

In conclusion, we have developed a squeezed-field description of gapless quantum systems in one dimension (1D), which are described by the Tomonaga--Luttinger Hamiltonian in the low-energy limit.
We have derived an effective nonlinear Lagrangian to describe the coupling between the Tomonaga--Luttinger liquid (TLL) and the squeezing modes as an extended model beyond the free-boson picture of TLL theory. 

As a physically relevant application of this description, we have analyzed the imaginary-time correlation function of a vertex operator in order to illustrate the consequences of the fluctuation of the squeezing field in the non-interacting limit.
For this limit, the correlation function reduces to a product of the two propagators:
The first propagator reproduces the result of standard TLL theory with a linear dispersion $\omega_{k}=v_0|k|$ with the sound velocity $v_0$, and the critical exponent $\kappa_0 = K/2$.
However, the second propagator arises due to the fluctuations of the squeezing field, and exhibits a side-band dispersion at $3\omega_{k}=3v_0|k|$ with a propagation velocity $3v_0$.
Moreover, we determined the ground-state critical exponent of the second propagator as a function of the TL parameter $K$.
 
Furthermore, we have analyzed the spectral function of the density fluctuations for the 1D Bose gas perturbatively, on the basis of the effective Lagrangian. 
We have determined the renormalized values of the velocities and spectral weights of the two branches within the second-order expansion of the cubic vertices between the TLL and the squeezing modes.
At zero temperature, the renormalized velocity of the TLL branch is smaller (larger) than the velocity of the side-band branch for $1 < K \lesssim 12$ ($K \gtrsim 12$), but the dispersion relations remain linear in the momentum.
Nonzero-temperature contributions yield a nonlinear $k$-dependence of the dispersion relations.
The properties of the nonlinear dependence can be attributed to the thermal population of the excitations at lower momenta.
We also find that the interaction between the TLL and the squeezing modes produces an additional branch, whose velocity is larger than the renormalized velocities of the TLL dispersion and the side-band dispersion.
The property of this additional branch is not affected by the thermal excitations within the level of approximation.

Our formulation that was developed for single component 1D systems can be readily generalized to multicomponent 1D systems, such as 1D electrons in materials \cite{fradkin2013field} and SU(N) ultracold gases in a 1D potential \cite{capponi2016phases}.
We will embark on such generalizations in future studies.

Furthermore, our study provides an understanding of collective excitations beyond the Bogoliubov framework. 
This suggests the broad applicability of the arguments that we have presented here, to any system commonly described within the Bogoliubov approximation. 
In fact, the squeezed-field formalism advances the usually static Bogoliubov transformations to a dynamical field in the effective Lagrangians. 
The excitations of this field are the non-Bogoliubov excitations of the system, which we have presented here in the context of TLLs, but emphasize that the line of reasoning presented is widely applicable. 

\begin{acknowledgments}

We thank Ilias Seifie and Masahito Ueda for fruitful discussions and useful comments on this project. 
This work is supported by the Cluster of Excellence `Advanced Imaging of Matter' of the Deutsche Forschungsgemeinschaft (DFG) EXC 2056 - project ID 390715994.

\end{acknowledgments}

\appendix 

\section{The dynamical term of the Lagrangian} \label{app: dynamical}

The dynamical term of the imaginary-time Lagrangian is evaluated \cite{seifie2019squeezed} as 
\begin{align}
\langle \Psi | \partial_{\tau} | \Psi \rangle &= \frac{1}{2}\sum_{k\neq 0}\left( b^*_{k}\partial_{\tau}b_{k} - b_{k}\partial_{\tau}b^*_{k} \right) \nonumber \\
&\;\;\;\;\;\;\;\;\;\;\; + \sum_{k>0}(A^*_{k}\partial_{\tau}\eta_{k}-A_{k}\partial_{\tau}\eta^*_{k}). \label{eq: dynamical_term}
\end{align}
The functional $A_{k}$ is given by 
\begin{align}
A_{k} &= \frac{1}{2|\eta_{k}|^2}\left[ |v_{k}|^2 \eta_{k} (1+|b_{k}|^2 + |b_{-k}|^2) \right.  \\
&\left. + (|\eta_{k}|^2 + u_{k}v^*_{k}\eta_{k})b_{k}b_{-k} - (\eta_{k}^2 - u_{k}v_{k}\eta_{k})b^*_{k}b^*_{-k} \right]. \nonumber 
\end{align}
In the previous work \cite{seifie2019squeezed}, a dynamical term corresponding to Eq.~(\ref{eq: dynamical_term}) has been derived by expressing an overlap $\langle \Psi(\tau_{n+1})|\Psi(\tau_{n})\rangle$ in terms of functional integrals and directly evaluating it by means of the Gaussian integration techniques \cite{altland2010condensed}.
In this appendix, we present an alternative way to arrive at the same result, in which there is no need to represent the overlap in terms of functional integrals, therefore, one can obtain Eq.~(\ref{eq: dynamical_term}) without carrying out lengthy analyses of functional integration.

To begin with, we introduce a unitary operator ${\hat U}(b,\eta)=\prod_{k > 0} {\hat S}_{k}(\eta_{k}) {\hat D}_{k}(b_{k}) {\hat D}_{-k}(b_{-k})$. 
This operator generates the squeezed-coherent state involving $k$ and $-k$ modes, i.e., 
\begin{align}
|\Psi\rangle = {\hat U}|0\rangle=\left\{\prod_{k > 0} {\hat S}_{k}(\eta_{k}) {\hat D}_{k}(b_{k}) {\hat D}_{-k}(b_{-k})\right\}|0\rangle.
\end{align}
Using ${\hat U}$, we define a connection operator denoted by ${\hat U}^{\dagger} d {\hat U}$, where $d$ represents an exterior derivative of a function in the phase-space variables $(b_k,b_k^*,r_{k},\vartheta_{k})$.
It can be expressed as 
\begin{align}
{\hat U}^{\dagger} d {\hat U} 
&= {\hat U}^{\dagger} \sum_{k\neq 0} \left( d b_{k} \frac{\partial}{\partial b_{k}} +  d b^*_{k} \frac{\partial}{\partial b^*_{k}} \right) {\hat U} \nonumber \\
&+ {\hat U}^{\dagger} \sum_{k>0} \left( d r_{k}\frac{\partial}{\partial r_{k}} + d \vartheta_{k}\frac{\partial}{\partial \vartheta_{k}} \right) {\hat U}. \label{eq: connection}
\end{align}
We note that $\langle \Psi | \partial_{\tau} | \Psi \rangle = \langle 0 | {\hat U}^{\dagger} \partial_{\tau} {\hat U} | 0 \rangle$.
Hence, the dynamical term of the path-integral Lagrangian is equivalent to the vacuum average of ${\hat U}^{\dagger} d {\hat U}$ divided by $d\tau$. 

In Eq.~(\ref{eq: connection}), the terms with $db_{k}$ and $db^*_{k}$ are written as
\begin{align}
{\hat U}^{\dagger}db_{k} \frac{\partial}{\partial b_{k}}{\hat U} &= d b_{k}{\hat D}^{\dagger}_{k}(b_{k})\frac{\partial {\hat D}_{k}(b_{k})}{\partial b_{k}}, \nonumber \\
{\hat U}^{\dagger} db^*_{k} \frac{\partial}{\partial b^*_{k}} {\hat U} &= d b^*_{k}{\hat D}^{\dagger}_{k}(b_{k})\frac{\partial {\hat D}_{k}(b_{k})}{\partial b^*_{k}}.  
\end{align}
Utilizing the disentangling formula ${\hat D}(\alpha+\beta)={\hat D}(\alpha){\hat D}(\beta)e^{-(\alpha\beta^* - \alpha^*\beta)/2}$ \cite{walls2007quantum}, one can prove the relations 
\begin{align}
{\hat D}^{\dagger}_{k}\frac{\partial {\hat D}_{k}}{\partial b_{k}} = \left( {\hat b}^{\dagger}_{k} + \frac{b^*_{k}}{2} \right),\;{\hat D}^{\dagger}_{k}\frac{\partial {\hat D}_{k}}{\partial b^*_{k}} = -\left( {\hat b}_{k} + \frac{b_{k}}{2} \right). \label{eq: connection_b} 
\end{align}
Equation (\ref{eq: connection_b}) implies that the vacuum expectation value of ${\hat U}^{\dagger} d {\hat U}$ contains the dynamical term of the bosonic coherent-state path integrals \cite{altland2010condensed}:
\begin{align}
\left. \langle 0 | {\hat U}^{\dagger} d {\hat U} | 0 \rangle \right|_{d \eta = d \eta^* = 0} 
&= \langle 0 | {\hat U}^{\dagger} \sum_{k\neq 0} \left( d b_{k} \partial_{b_{k}} +  d b^*_{k} \partial_{b^*_{k}} \right) {\hat U} | 0 \rangle \nonumber \\
&= \frac{1}{2} \sum_{k\neq 0}(b^*_{k}db_{k} - b_{k}db^*_{k}). \label{eq: dynamical_term_coherent}
\end{align}

Likewise, one can calculate the dynamical term of the squeezing field in Eq.~(\ref{eq: dynamical_term}).
The first step is to write a connection operator for the squeeze operator ${\hat S}_{k}$ in a momentum shell 
\begin{align}
{\hat S}^{\dagger}_{k}(\eta_{k})d{\hat S}_{k}(\eta_{k}) = {\hat S}^{\dagger}_{k}\left(dr_{k}\frac{\partial}{\partial r_{k}} + d\vartheta_{k}\frac{\partial}{\partial \vartheta_{k}}\right){\hat S}_{k}.
\end{align}
We note that the squeeze operator has a polar representation, i.e. ${\hat S}_{k}(r_k,\vartheta_k)=e^{r_{k}(e^{i\vartheta_k}{\hat b}^{\dagger}_{k}{\hat b}^{\dagger}_{-k}-e^{-i\vartheta_k}{\hat b}_{k}{\hat b}_{-k})}$.
Hence, the first derivative of ${\hat S}_{k}(r_k,\vartheta_k)$ in $r_{k}$ reads
\begin{align}
\frac{\partial}{\partial r_{k}}{\hat S}_{k} = {\hat S}_{k}{\hat \Sigma}_{k}(\vartheta_k).
\end{align}
We have defined ${\hat \Sigma}_{k}(\vartheta_k)=e^{i\vartheta_k}{\hat b}^{\dagger}_{k}{\hat b}^{\dagger}_{-k}-e^{-i\vartheta_k}{\hat b}_{k}{\hat b}_{-k}=-{\hat \Sigma}^{\dagger}_{k}(\vartheta_k)$, which is an anti-Hermitian operator.

To evaluate $\partial_{\vartheta_k}{\hat S}_{k}$, we use the relation 
\begin{align}
e^{i \theta {\hat n}_{k}} {\hat b}_{k} e^{-i \theta {\hat n}_{k}} = e^{-i \theta}{\hat b}_{k}, \;\; e^{i \theta {\hat n}_{k}} {\hat b}^{\dagger}_{k} e^{-i \theta {\hat n}_{k}} = e^{i \theta}{\hat b}^{\dagger}_{k}.
\end{align}
This is a unitary transformation generated by the mode occupation ${\hat n}_{k}={\hat b}^{\dagger}_{k}{\hat b}_{k}$, and adds a phase factor $e^{-i\theta}$ or $e^{i\theta}$ to the annihilation or creation operator. 
$\partial_{\vartheta_{k}}{\hat S}_{k}$ is calculated as follows:
\begin{align}
\frac{\partial}{\partial \vartheta_{k}}{\hat S}_{k} &= \frac{\partial}{\partial \vartheta_{k}}\left(e^{r_{k} {\hat \Sigma}(\vartheta_{k})}\right)  \nonumber \\
&= \frac{\partial}{\partial \vartheta_{k}}\left(e^{\frac{i}{2}\vartheta_{k}{\hat {\cal N}}} e^{r_{k} {\hat \Sigma}(0)} e^{-\frac{i}{2}\vartheta_{k}{\hat {\cal N}}}\right) \nonumber \\
&= \frac{i}{2}e^{\frac{i \vartheta_{k}}{2}{\hat {\cal N}}} \left[{\hat {\cal N}},e^{ r_{k} {\hat \Sigma}(0)}\right] e^{-\frac{i \vartheta_{k}}{2}{\hat {\cal N}}}, 
\end{align}
where ${\hat {\cal N}}={\hat n}_{k}+{\hat n}_{-k}$. 
Hence, ${\hat S}^{\dagger}_{k}d{\hat S}_{k}$ can be written as
\begin{align}
{\hat S}^{\dagger}_{k}d{\hat S}_{k}
&= \left(e^{i\vartheta_{k}}{\hat b}^{\dagger}_{k}{\hat b}^{\dagger}_{-k}-e^{-i\vartheta_{k}}{\hat b}_{k}{\hat b}_{-k}\right)dr_{k} \nonumber \\
&\;\;\;\; + \frac{i}{2}{\hat S}^{\dagger}_{k}({\hat b}^{\dagger}_{k}{\hat b}_{k}+{\hat b}^{\dagger}_{-k}{\hat b}_{-k}){\hat S}_{k}d\vartheta_{k} \nonumber \\
&\;\;\;\; - \frac{i}{2}({\hat b}^{\dagger}_{k}{\hat b}_{k}+{\hat b}^{\dagger}_{-k}{\hat b}_{-k})d\vartheta_{k}.
\end{align}
The dynamical term of the squeezing field is given as the expectation value of ${\hat S}^{\dagger}_{k}d{\hat S}_{k}$ with respect to the coherent state $|b_{k},b_{-k}\rangle = {\hat D}_{k}(b_{k}){\hat D}_{-k}(b_{-k})|0\rangle$. 
After direct calculations, we arrive at 
\begin{widetext}
\begin{align}
\langle b_{k},b_{-k}| {\hat S}^{\dagger}_{k}d{\hat S}_{k}|b_{k},b_{-k}\rangle &=  \left(e^{i\vartheta_{k}}b^*_{k}b^*_{-k}-e^{-i\vartheta_{k}}b_{k}b_{-k}\right)dr_{k} - \frac{i}{2}(|b_{k}|^2+|b_{-k}|^2)d\vartheta_{k} \nonumber \\
&\;\;\;\; + \frac{i}{2}\left[u^2_{k}|b_{k}|^2+|v_{k}|^2(|b_{-k}|^2+1)+u_{k}v_{k}b^*_{k}b^*_{-k}+u_{k}v^*_{k}b_{k}b_{-k} \right]d\vartheta_{k} \nonumber \\
&\;\;\;\; + \frac{i}{2}\left[u^2_{k}|b_{-k}|^2+|v_{k}|^2(|b_{k}|^2+1)+u_{k}v_{k}b^*_{k}b^*_{-k}+u_{k}v^*_{k}b_{k}b_{-k} \right]d\vartheta_{k} \nonumber \\
&= -\frac{1}{2 r_{k}^2}\left[ |v_{k}|^2\eta_{k}(1+|b_{k}|^2+|b_{-k}|^2) + (r^2_{k}+u_{k}v^*_{k}\eta_{k})b_{k}b_{-k} - (\eta_k^2-u_{k}v_{k}\eta_{k})b^*_{k}b^*_{-k} \right]d\eta^*_{k} \nonumber \\
&\;\;\;\; + \frac{1}{2 r_{k}^2}\left[ |v_{k}|^2\eta^*_{k}(1+|b_{k}|^2+|b_{-k}|^2) + (r^2_{k}+u_{k}v_{k}\eta^*_{k})b^*_{k}b^*_{-k} - ((\eta^*_k)^2-u_{k}v^*_{k}\eta^*_{k})b_{k}b_{-k} \right]d\eta_{k}. \nonumber
\end{align}
\end{widetext}
Combining this result and Eq.~(\ref{eq: dynamical_term_coherent}), we obtain Eq.~(\ref{eq: dynamical_term}).

\section{The squeezed-field energy functional for the TL Hamiltonian} \label{app: energy_functional}

In this appendix, we derive the energy functional $E=\langle \Psi | {\hat H}_0 | \Psi \rangle$ of the squeezed field path integral for the TL Hamiltonian. 
First, we write the following mode expansion:
\begin{align}
{\hat \phi}_{0}(x) &= -\frac{i\pi}{L}\sum_{k\neq 0}\left(\frac{L|k|}{2\pi}\right)^{\frac{1}{2}}\frac{e^{-\frac{{\tilde \alpha}|k|}{2}-ikx}}{k} \left[ {\hat b}^{\dagger}_{k} + {\hat b}_{-k} \right], \nonumber \\
{\hat \theta}_{0}(x) &= \frac{i\pi}{L}\sum_{k\neq 0}\left(\frac{L|k|}{2\pi}\right)^{\frac{1}{2}}\frac{e^{-\frac{{\tilde \alpha}|k|}{2}-ikx}}{|k|} \left[ {\hat b}^{\dagger}_{k} - {\hat b}_{-k} \right]. \label{eq: mode expansion free}
\end{align}
This transformation diagonalizes the Hamiltonian ${\hat H}_0$ at $K=1$ in the boson Fock space \cite{giamarchi2004quantum}. 
Plugging this into ${\hat H}_{\rm 0}$ with $K > 1$, we obtain an off-diagonal quadratic boson model 
\begin{align}
{\hat H}_{\rm 0} &= \frac{\hbar v_0}{4}\sum_{k \neq 0}|k|e^{-{\tilde \alpha}|k|}\left(\frac{1}{K}-K\right){\hat b}^{\dagger}_{k}{\hat b}^{\dagger}_{-k} \nonumber \\
&+ \frac{\hbar v_0}{4}\sum_{k \neq 0}|k|e^{-{\tilde \alpha}|k|}\left(\frac{1}{K}-K\right){\hat b}_{-k}{\hat b}_{k} \nonumber \\
&+ \frac{\hbar v_0}{4}\sum_{k \neq 0}|k|e^{-{\tilde \alpha}|k|}\left(\frac{1}{K}+K\right){\hat b}^{\dagger}_{k}{\hat b}_{k} \nonumber \\
&+ \frac{\hbar v_0}{4}\sum_{k \neq 0}|k|e^{-{\tilde \alpha}|k|}\left(\frac{1}{K}+K\right){\hat b}_{-k}{\hat b}^{\dagger}_{-k}. \label{eq: Hamiltonian non-diagonal}
\end{align}

As the second step, we apply the squeeze transformation to the off-diagonal Hamiltonian.
The squeeze transform of the bosons in Eq.~(\ref{eq: Hamiltonian non-diagonal}) is given by 
\begin{align}
{\hat S}^{-1}_{k}(\eta_{k}) {\hat b}_{k} {\hat S}_{k}(\eta_{k}) &= u_{k}{\hat b}_{k} + v_{k}{\hat b}^{\dagger}_{-k}, \\
{\hat S}^{-1}_{k}(\eta_{k}) {\hat b}^{\dagger}_{k} {\hat S}_{k}(\eta_{k}) &= u_{k}{\hat b}^{\dagger}_{k} + v^*_{k}{\hat b}_{-k}.
\end{align}
Hence, ${\hat b}_{k}\pm{\hat b}^{\dagger}_{-k}$ read
\begin{align}
{\hat S}^{-1}_{k}({\hat b}_{k}+{\hat b}^{\dagger}_{-k}){\hat S}_{k} &= (u_{k} + v_{k}){\hat b}^{\dagger}_{k} + (u_{k} + v^*_{k}){\hat b}_{-k}, \nonumber \\
{\hat S}^{-1}_{k}({\hat b}_{k}-{\hat b}^{\dagger}_{-k}){\hat S}_{k} &= (u_{k} - v_{k}){\hat b}^{\dagger}_{k} - (u_{k} - v^*_{k}){\hat b}_{-k}. \nonumber
\end{align}
The coefficients in the righthand side are given by
\begin{align}
u_{k} = {\rm cosh}(r_{k}),\;\;\; v_{k} = e^{i\vartheta_{k}}{\rm sinh}(r_{k}). \nonumber
\end{align}
Comparing the above expressions to the mode expansion presented in the main text, we find 
\begin{align}
f_{k} = u_{k} + v_{k},\;\;\;\; g_{k} = u_{k} - v_{k}. \nonumber
\end{align}
If we put $\vartheta_{k}=0$, these reduce to $f_{k}=e^{r_{k}}$ and $g_{k}=e^{-r_{k}}$ \cite{haldane1981effective}.

Utilizing the above relations, the Hamiltonian transforms into
\begin{align}
{\hat S}^{-1}{\hat H}_{\rm 0}{\hat S}
&= \frac{\hbar v_0}{4}\sum_{k \neq 0}|k|e^{-{\tilde \alpha}|k|}V_{++}(\eta_{k}){\hat b}^{\dagger}_{k}{\hat b}^{\dagger}_{-k} \nonumber \\
&+ \frac{\hbar v_0}{4}\sum_{k \neq 0}|k|e^{-{\tilde \alpha}|k|}V^*_{++}(\eta_{k}){\hat b}_{k}{\hat b}_{-k} \nonumber \\
&+ \frac{\hbar v_0}{4}\sum_{k \neq 0}|k|e^{-{\tilde \alpha}|k|}V_{+-}(\eta_{k}){\hat b}^{\dagger}_{k}{\hat b}_{k}  \\
&+ \frac{\hbar v_0}{4}\sum_{k \neq 0}|k|e^{-{\tilde \alpha}|k|}V_{+-}(\eta_{k}){\hat b}_{-k}{\hat b}^{\dagger}_{-k}, \nonumber
\end{align}
where 
\begin{align}
V_{++} &= \left(\frac{1}{K}-K\right)(u_{k}^2+v_{k}^2) + 2\left(\frac{1}{K}+K\right)u_{k}v_{k}, \nonumber \\
V_{+-} &= \left(\frac{1}{K}+K\right)(u_{k}^2+|v_{k}|^2) + \left(\frac{1}{K}-K\right)u_{k}(v_{k}+v^*_{k}). \nonumber
\end{align}
We note that, if we substitute the following values into $u_{k}$ and $v_{k}$ subject to $K>1$, the off-diagonal coefficient $V_{++}$ vanishes, and the Hamiltonian is diagonalized as ${\hat H}_0 = \sum_{k \neq 0}\hbar \omega_{k} {\hat b}^{\dagger}_{k}{\hat b}_{k}$:
\begin{align}
u^{0}_{k} = \frac{1}{2}\left( \sqrt{K} + \frac{1}{\sqrt{K}} \right),\;\;\;\; v^{0}_{k} = \frac{1}{2}\left( \sqrt{K} - \frac{1}{\sqrt{K}} \right).
\end{align}
The energy functional $E$ is defined as an expectation value of the squeezed Hamiltonian ${\hat S}^{-1}{\hat H}_0{\hat S}$ with respect to the coherent state:
\begin{align}
E &= \frac{1}{4}\sum_{k\neq 0}\hbar v_0|k|e^{-{\tilde \alpha} |k|}V_{+-}(\eta_{k}) \nonumber \\
&+ \frac{1}{4}\sum_{k\neq 0}\hbar v_0|k|e^{-{\tilde \alpha} |k|}V_{+-}(\eta_{k})(|b_{k}|^2+|b_{-k}|^2) \\
&+ \frac{1}{4}\sum_{k\neq 0}\hbar v_0|k|e^{-{\tilde \alpha} |k|}(V_{++}(\eta_{k})b^*_{k}b^*_{-k}+{\rm C.c.}). \nonumber
\end{align} 
We note that the vacuum-energy functional described in the first line, which is independent of $b_{k}$ and $b^*_{k}$, emerges as the Hamiltonian is rewritten into the normal-ordered form.
In the leading-order fluctuation expansion, it gives the dispersion of the squeezing field.

\section{Fluctuation expansion of the Lagrangian} \label{app: fluctuations}

We here give a detailed discussion about the fluctuation expansion of the squeezed-field Lagrangian. 

We first consider the dynamical term of the Lagrangian $\langle \Psi | \partial_{\tau} | \Psi \rangle$.
Suppose that $A_{k}$ is expanded in $\delta \eta^{R}_{k}$ and $\delta \eta^{I}_{k}$ up to the first order, i.e., $A_{k} \approx A^0_{k} + \left. \frac{\partial A_{k}}{\partial \eta^{R}_{k}} \right|_{\eta_{k}=\eta_{k}^0} \delta\eta^{R}_{k} +  \left. \frac{\partial A_{k}}{\partial \eta^{I}_{k}}\right|_{\eta_{k}=\eta_{k}^{0}} \delta\eta^{I}_{k}$.
If one uses the $\xi$ basis, in which $\delta\eta^{R}_{k}$ and $\delta\eta^{I}_{k}$ are expressed as
\begin{align}
\delta \eta^{R}_{k} = \frac{r}{\sqrt{2}}(\xi_k + \xi^*_{k}),\;\;\; \delta \eta^{I}_{k} = \frac{1}{i\sqrt{2} r}(\xi_k - \xi^*_{k}), \label{eq: transformation eta to xi}
\end{align}
the dynamical term of the squeezing field is written as
\begin{align}
\sum_{k>0}(A^*_{k}\partial_{\tau}\eta_{k}-{\rm C.c.}) 
&\approx 2{\tilde A}\sum_{k>0}\xi^*_{k}\partial_{\tau}\xi_{k}  \nonumber \\
&+2V_{1}' \sum_{k > 0}|b_{k}|^2(\partial_{\tau}\xi_{k}-\partial_{\tau}\xi^*_{k}) \nonumber \\
&+2V_{1}' \sum_{k > 0}|b_{-k}|^2(\partial_{\tau}\xi_{k}-\partial_{\tau}\xi^*_{k}) \nonumber \\
&-\sqrt{2{\tilde A}}\sum_{k > 0}b_{k}b_{-k}\partial_{\tau}\xi_{k}^*  \nonumber \\
&+\sqrt{2{\tilde A}}\sum_{k > 0}b^*_{k}b^*_{-k}\partial_{\tau}\xi_{k}, 
\end{align}
where $V_{1}'= \frac{1}{\sqrt{2}r}\frac{1}{4{\rm ln}K}\left( \sqrt{K} - \frac{1}{\sqrt{K}} \right)^2$, ${\tilde A} = \frac{1}{2{\rm ln}K}\left(K-\frac{1}{K}\right)$, and $\delta \eta_{k} = \delta \eta_{-k}$.
We note that the linear terms are zero because the boundary condition ensures $\int^{\hbar \beta}_{0} d\tau \partial_{\tau} \delta \eta_{k}(\tau) = \delta \eta_{k}(\hbar \beta) - \delta \eta_{k}(0) = 0$. 
With a rescaled field ${\tilde \xi}_{k} = \sqrt{2{\tilde A}}\xi_{k}$, we obtain the dynamical term of the interacting Lagrangian in Eq.~(\ref{eq: lagrangian int}). 
The vertex coefficient $w_{1}$ in Eq.~(\ref{eq: lagrangian int}) is related to $V_{1}'$ such that 
\begin{align}
w_{1} = \frac{2V_{1}'}{\sqrt{2{\tilde A}}} = \frac{1}{2}\frac{\left(\sqrt{K}-\frac{1}{\sqrt{K}}\right)^2}{K-\frac{1}{K}}.
\end{align}

To obtain the contributions from the Hamiltonian part in the Lagrangian, we expand the energy functional $E=\langle \Psi | {\hat H}_0 |\Psi\rangle$ in fluctuations of $\eta_{k}$ to the second order $O(\delta \eta_{k}^2)$. 
The differential coefficients of $V_{+-}$ at $\delta \eta_{k}=0$ read
\begin{align}
(\partial_{R}V_{+-})_0&=(\partial_{I}V_{+-})_0 = 0, \nonumber \\
(\partial_{R}\partial_{I}V_{+-})_0 &= 0, \nonumber \\
(\partial_{R}^{2}V_{+-})_0 &= 8, \\
(\partial_{I}^{2}V_{+-})_0 &= 2\left(K-\frac{1}{K}\right)^{2}\frac{1}{({\rm ln}K)^2}. \nonumber
\end{align}
Here $\partial_{R/I}=\partial/\partial \eta^{R/I}_{k}$ was introduced for brevity. 
Likewise, the first and second differential coefficients of $V_{++}$ read
\begin{align}
(\partial_{R}V_{++})_0 &= 4, \nonumber \\
(\partial_{I}V_{++})_0 &= \frac{i}{{\rm ln}\sqrt{K}}\left(K-\frac{1}{K}\right), \nonumber \\
(\partial_{R}\partial_{I}V_{++})_0 &= i\left[ \frac{4}{\eta^{0}} - \left(K-\frac{1}{K}\right)\frac{1}{(\eta^{0})^2} \right], \\
(\partial_{R}^2V_{++})_0 &= 0, \nonumber \\
(\partial_{I}^2V_{++})_0 &= \frac{4}{\eta^{0}} + \frac{1}{2}\left(K-\frac{1}{K}\right)\left(\sqrt{K}-\frac{1}{\sqrt{K}}\right)^2\frac{1}{(\eta^{0})^2}.  \nonumber
\end{align}
Combining these results, we obtain the Hamiltonian included in the effective Lagrangian:
\begin{align}
E &\approx \sum_{k \neq 0}\hbar \omega_{k} b^*_{k}b_{k} + \sum_{k > 0}\hbar \omega_{\xi,k} {\tilde \xi}^*_{k}{\tilde \xi}_{k} \nonumber \\
&+ 2 \sum_{k>0}\hbar \omega_{k}{\tilde \xi}^*_{k}b_{k}b_{-k} + 2 \sum_{k>0}\hbar \omega_{k}{\tilde \xi}_{k}b^*_{k}b^*_{-k}.
\end{align}

\section{Perturbation theory for the imaginary-time correlation functions} \label{app: perturbation}

The aim of this appendix is to give an overview of our perturbative analysis of the multipoint correlation functions of the field variables arising in the correlation function $\chi'(k,\tau)$.

We first rewrite $\chi'(k,\tau)$ as 
\begin{align}
\frac{2\pi}{|k|}\chi'(k,\tau) = &-\langle f_{k}(\tau)b^*_{k}(\tau)f_{k}(0)b^*_{-k}(0) \rangle \nonumber \\
&-\langle f_{k}(\tau)b^*_{k}(\tau)f^*_{k}(0)b_{k}(0) \rangle \nonumber \\
&-\langle f^*_{k}(\tau)b_{-k}(\tau)f_{k}(0)b^*_{-k}(0) \rangle \nonumber \\
&-\langle f^*_{k}(\tau)b_{-k}(\tau)f^*_{k}(0)b_{k}(0) \rangle \nonumber \\
\equiv &\;\; C_1 + C_2 + C_3 + C_4,
\end{align}
and then expand $f_{k}(\tau)$ in $\delta \eta_{k}$ to the first order. 
For example, $C_{2}(k,\tau) = -\langle f_{k}(\tau)b^*_{k}(\tau)f^*_{k}(0)b_{k}(0) \rangle$ is approximated as 
\begin{align}
C_{2} \approx & \;\;\;\; c^{0000}_{2} \langle b^*_{k}(\tau)b_{k}(0) \rangle \nonumber \\
&+c^{1000}_{2} \langle \xi_{k}(\tau) b^*_{k}(\tau)b_{k}(0) \rangle + c^{0100}_{2} \langle \xi^*_{k}(\tau) b^*_{k}(\tau)b_{k}(0) \rangle \nonumber \\
&+c^{0010}_{2} \langle b^*_{k}(\tau)\xi_{k}(0)b_{k}(0) \rangle + c^{0001}_{2} \langle b^*_{k}(\tau) \xi^*_{k}(0) b_{k}(0) \rangle \nonumber \\
&+ c^{1010}_{2}\langle \xi_{k}(\tau) b^*_{k}(\tau) \xi_{k}(0) b_{k}(0) \rangle \nonumber \\
&+ c^{1001}_{2}\langle \xi_{k}(\tau) b^*_{k}(\tau) \xi^*_{k}(0) b_{k}(0) \rangle \\
&+ c^{0110}_{2}\langle \xi^*_{k}(\tau) b^*_{k}(\tau) \xi_{k}(0) b_{k}(0) \rangle \nonumber \\
&+ c^{0101}_{2}\langle \xi^*_{k}(\tau) b^*_{k}(\tau) \xi^*_{k}(0) b_{k}(0) \rangle. \nonumber
\end{align} 
The coefficients in front of the individual correlation functions are
\begin{align}
c^{0000}_{2} &= -(f_{k}[\eta_0])^2 = -K, \nonumber \\
c^{1000}_{2} &= -\frac{f_{k}[\eta_0]}{\sqrt{2}}\left[ {\tilde f}^{(0,1)}_{k}r^{-1} + f^{(1,0)}_{k}r \right] = c^{0001}_{2},  \nonumber \\
c^{0100}_{2} &= \frac{f_{k}[\eta_0]}{\sqrt{2}}\left[  {\tilde f}^{(0,1)}_{k}r^{-1} - f^{(1,0)}_{k}r \right] = c^{0010}_{2}, \nonumber \\
c^{1010}_{2} &= -\frac{1}{2}\left[(r f^{(1,0)}_{k})^2 - (r^{-1}  {\tilde f}^{(0,1)}_{k})^2 \right] = c^{0101}_{2}, \nonumber \\
c^{1001}_{2} &= -\frac{1}{2}\left[r^{-1}{\tilde f}^{(0,1)}_{k} + r f^{(1,0)}_{k} \right]^2, \nonumber \\
c^{0110}_{2} &= -\frac{1}{2}\left[ -r^{-1}{\tilde f}^{(0,1)}_{k} + r f^{(1,0)}_{k}\right]^2. \nonumber
\end{align} 
Here we used the notation 
\begin{align}
f_{k}(\tau) &= \sqrt{K} +f_{k}^{(1,0)}\delta \eta^{R}_{k}(\tau)+f_{k}^{(0,1)}\delta \eta^{I}_{k}(\tau),
\end{align}
where $f^{(1,0)}_{k} = {\rm cosh}(\eta_0) + {\rm sinh}(\eta_0) = \sqrt{K}$ and $f^{(0,1)}_{k} = \frac{i}{\eta_0}{\rm sinh}(\eta_0) \equiv i {\tilde f}^{(0,1)}_{k}$.

\subsection{Two-point correlation functions}

We begin with evaluating the two-point correlation functions in the fluctuation expansion.
In particular, we focus on a {\it diagonal} propagator $\langle b_{-k}(\tau)b^*_{-k}(0)\rangle$. 
The remaining two-point correlation functions may be obtained in the same way.
In terms of the squeezed-field path-integral representation, the diagonal propagator is
\begin{align}
\langle b_{-k}(\tau)b_{-k}^*(0)\rangle = \frac{\int [{\cal D}(b,\eta)] b_{-k}(\tau)b_{-k}^*(0)e^{-\frac{1}{\hbar}({\cal S}_{2}+{\cal S}_{3})}}{\int [{\cal D}(b,\eta)]e^{-\frac{1}{\hbar}({\cal S}_{2}+{\cal S}_{3})}}. \label{eq: functional_ave_b_bs}
\end{align}
The quadratic action ${\cal S}_{2}$ describes the Gaussian fluctuations of the system, while the perturbative term ${\cal S}_{3}=\sum_{k>0}{\cal S}^{k}_{\rm 3}$ does the cubic interactions among the TLL and the squeezing modes.

The interaction term of the action has the Matsubara-frequency representation as
\begin{align}
{\cal S}^{k}_{3} 
&= \frac{\hbar}{(\hbar \beta)^{\frac{1}{2}}}\sum_{n_1,n_2}V_{2}(n_1,n_2){\tilde \xi}^*_{k}(\omega_{n_1+n_2})b_{k}(\omega_{n_1})b_{-k}(\omega_{n_2}) \nonumber \\
&+ \frac{\hbar}{(\hbar \beta)^{\frac{1}{2}}}\sum_{n_1,n_2}V_{2}(n_1,n_2){\tilde \xi}_{k}(\omega_{n_1+n_2})b^*_{k}(\omega_{n_1})b^*_{-k}(\omega_{n_2}) \nonumber \\
&+ \frac{\hbar}{(\hbar \beta)^{\frac{1}{2}}}\sum_{n_1,n_2}V_{1}(n_1,n_2) b^*_{k}(\omega_{n_1})b_{k}(\omega_{n_2}){\tilde \xi}_{k}(\omega_{n_1-n_2}) \nonumber \\
&+ \frac{\hbar}{(\hbar \beta)^{\frac{1}{2}}}\sum_{n_1,n_2}V_{1}(n_1,n_2) b^*_{-k}(\omega_{n_1})b_{-k}(\omega_{n_2}){\tilde \xi}_{k}(\omega_{n_1-n_2}) \nonumber \\
&+ \frac{\hbar}{(\hbar \beta)^{\frac{1}{2}}}\sum_{n_1,n_2}{\overline V}_{1}(n_1,n_2) b^*_{k}(\omega_{n_1})b_{k}(\omega_{n_2}){\tilde \xi}^*_{k}(\omega_{n_2-n_1}) \nonumber \\
&+ \frac{\hbar}{(\hbar \beta)^{\frac{1}{2}}}\sum_{n_1,n_2}{\overline V}_{1}(n_1,n_2) b^*_{-k}(\omega_{n_1})b_{-k}(\omega_{n_2}){\tilde \xi}^*_{k}(\omega_{n_2-n_1}), \nonumber
\end{align}
where 
\begin{align}
V_{1}(n_1,n_2) &= w_{1}(-i\omega_{n_1}+i\omega_{n_2}) \\
&= -{\overline V}_{1}(n_1,n_2), \nonumber \\
V_{2}(n_1,n_2) &= 2 \omega_{k} -i\omega_{n_1} -i\omega_{n_2}. 
\end{align}
We note that the first vertex $V_{1}(n_1,n_2)$, which is associated with the process of annihilation and creation of TLL modes for a fixed $k$, is anti-symmetric under inversion between $n_1$ and $n_2$.
On the other hand, the second vertex $V_{2}(n_1,n_2)$, which is associated with creation or annihilation of two TLL modes with $\pm k$, is symmetric under the inversion.

\begin{figure*}
\includegraphics[width=150mm]{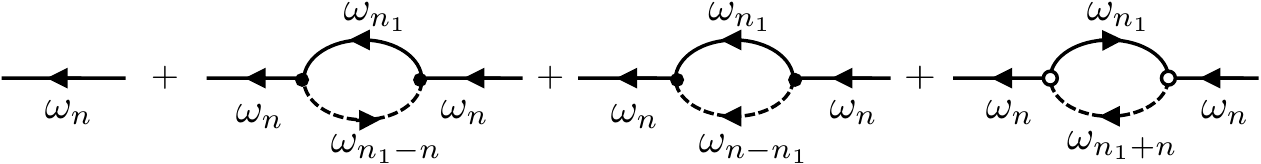}
\vspace{-1mm}
\caption{
Feynman diagrams for the perturbative terms in the two-point correlation function $\langle b_{-k}(\omega_{n})b^*_{-k}(\omega_{n}) \rangle$.
}
\label{fig: feynman_two_point}
\end{figure*}

A frequency component of the diagonal propagator, i.e., $\int d\tau e^{i\omega_{n}\tau}\langle b_{-k}(\tau)b^*_{-k}(0)\rangle = \langle b_{-k}(\omega_{n})b^*_{-k}(\omega_{n}) \rangle$ is obtained in the following way:
First, we expand Eq.~(\ref{eq: functional_ave_b_bs}) in ${\cal S}_{\rm int}$ to the second order.
Within this order, by performing a Wick contraction of the correction terms, there appear three relevant perturbative corrections in $\langle b_{-k}(\omega_{n})b^*_{-k}(\omega_{n}) \rangle$ such that
\begin{widetext}
\begin{align}
&\nonumber \\
\langle b_{-k}(\omega_{n})b^*_{-k}(\omega_{n}) \rangle 
=\frac{1}{-i\omega_{n}+\omega_{k}}
&+\frac{1}{(-i\omega_{n}+\omega_{k})^2}\frac{1}{\hbar \beta}\sum_{n_1}\frac{{\overline V}_{1}(n,n_1)V_{1}(n_1,n)}{(-i\omega_{n_1}+\omega_{k})(i\omega_{n}-i\omega_{n_1}+2\omega_{k})} \nonumber \\
&+\frac{1}{(-i\omega_{n}+\omega_{k})^2}\frac{1}{\hbar \beta}\sum_{n_1}\frac{{\overline V}_{1}(n_1,n)V_{1}(n,n_1)}{(-i\omega_{n_1}+\omega_{k})(i\omega_{n_1}-i\omega_{n}+2\omega_{k})} \nonumber \\
&+\frac{1}{(-i\omega_{n}+\omega_{k})^2}\frac{1}{\hbar \beta}\sum_{n_1}\frac{V_{2}(n_1,n)V_{2}(n,n_1)}{(-i\omega_{n_1}+\omega_{k})(-i\omega_{n}-i\omega_{n_1}+2\omega_{k})}. \label{eq: 2p correlation function with corrections}
\end{align}
\end{widetext}
The subscripted Matsubara frequency $\omega_{n_1}=2\pi n_{1}/(\hbar \beta)$ implies the summation of internal virtual processes among some unperturbed modes between two interaction vertices.

One can make a graphical representation for the individual terms in Eq.~(\ref{eq: 2p correlation function with corrections}) as drawn in Fig.~\ref{fig: feynman_two_point}.
In Fig.~\ref{fig: feynman_two_point}, the solid and dashed arrows stand for the unperturbed propagators of the TLL and squeezing modes, respectively.
Moreover, the tail and head of the arrows correspond to creation and annihilation of the modes.
Furthermore, for each interaction vertex, the filled bullet ($\bullet$) means $V_{1}(n_1,n_2)$ or ${\overline V}_{1}(n_1,n_2)$, whereas the empty circle ($\circ$) does $V_{2}(n_1,n_2)$.
The in-coming and out-going lines connected to the vertices must ensure the energy conservation law. 

Evaluation of each Matsubara sum in Eq.~(\ref{eq: 2p correlation function with corrections}) can be carried out by using the mathematical formula \cite{altland2010condensed,fetter2012quantum}
\begin{align}
\sum_{n=-\infty}^{\infty}h(i\omega_{n}) 
&= \frac{1}{2\pi i}\oint_{C} dz g_{\rm pole}(z)h(z), \label{eq: cauchy}
\end{align}
where $g_{\rm pole}(z)=\hbar\beta/(e^{\beta \hbar z}-1)$ and $C$ is an integration counter over a complex plane $z \in {\mathbb C}$, enveloping the singular points of $g_{\rm pole}(z)$ at $z=i\omega_{n}$.  
The following formulas would be very useful for later use:
\begin{align}
\frac{1}{\hbar \beta}\sum_{n \in {\mathbb Z}}\frac{e^{i\omega_{n}0^{+}}}{i\omega_{n}-{\overline \omega}} &= -n_{\rm B}({\overline \omega}), \label{eq: sum_formula1} \\
\frac{1}{\hbar \beta}\sum_{n \in {\mathbb Z}}\frac{e^{i\omega_{n}0^{+}}(-i\omega_{n})^{s}}{i\omega_{n}-{\overline \omega}} &= -(-{\overline \omega})^{s}n_{\rm B}({\overline \omega}), \label{eq: sum_formula2}
\end{align}
where ${\overline \omega} \in {\mathbb R}$ and $s$ ($\geq 0$) is an integer.
The phase factor $e^{i\omega_{n}0^{+}}$ is a convergence factor to regularize the series.

Let us calculate the first perturbative term in Eq.~(\ref{eq: 2p correlation function with corrections}), which is denoted by 
\begin{align}
{\cal G}_{2}(k,\omega_{n})&=\frac{1}{(-i\omega_{n}+\omega_{k})^2} \nonumber  \\
&\;\;\; \times \frac{1}{\hbar \beta}\sum_{n_1}\frac{{\overline V}_{1}(n,n_1)V_{1}(n_1,n)}{(-i\omega_{n_1}+\omega_{k})(i\omega_{n}-i\omega_{n_1}+2\omega_{k})}. \nonumber
\end{align}
To obtain a converged result, it is required to insert a convergence factor $e^{i\omega_{n}{\varepsilon}}$ such that $\sum_{n_1} \rightarrow \sum_{n_1}e^{i\omega_{n_1}{\varepsilon}}$ \cite{fetter2012quantum}.
Completing the summation of $n_1$ by using Eq.~(\ref{eq: cauchy}) or Eqs.~(\ref{eq: sum_formula1}) and (\ref{eq: sum_formula2}), and then taking a limit $\varepsilon \rightarrow +0$, one can verify that 
\begin{align}
{\cal G}_{2}(k,\omega_{n})&=\frac{w_{1}^2(n_{{\rm B},1}-n_{{\rm B},2})}{i\omega_{n}+\omega_{k}} \nonumber \\
&\;\; - \frac{w_1^2 n_{{\rm B},2}}{-i\omega_{n}+\omega_{k}} - \frac{2\omega_{k}w_{1}^2n_{{\rm B},2}}{(-i\omega_{n}+\omega_{k})^2}. 
\end{align}
For our convenience, we introduced $n_{{\rm B},s} \equiv n_{\rm B}(s\omega_{k})$.
Being parallel to this calculation, we also find that the second perturbative term denoted by ${\cal G}_{2}'(k,\omega_{n})$ becomes
\begin{align}
{\cal G}_{2}'(k,\omega_{n})
&=\frac{w_{1}^2(1+n_{{\rm B},1}+n_{{\rm B},2})}{-i\omega_{n}+3\omega_{k}} \nonumber \\
&\;\; - \frac{w_{1}^2(1+n_{{\rm B},2})}{-i\omega_{n}+\omega_{k}} + \frac{2\omega_{k}w_{1}^2(1+n_{{\rm B},2})}{(-i\omega_{n}+\omega_{k})^2}. 
\end{align}
We note that there is an important difference between the pole properties of ${\cal G}_{2}(k,\omega_{n})$ and ${\cal G}_{2}'(k,\omega_{n})$ reflecting their contraction structures of the diagrams.
Due to the diagrammatic property that two internal lines in ${\cal G}_{2}(k,\omega_{n})$ propagate oppositely to each other, ${\cal G}_{2}(k,\omega_{n})$ acquires a pole at $z = -\omega_{k}$, which can be regarded as a side shift from $z=\omega_{k}$ with difference $2\omega_{k}$.  
By contrast, the corresponding pole in ${\cal G}_{2}'(k,\omega_{n})$, whose internal lines have the same direction, appears in the opposite side, i.e., $z = 3\omega_{k} = \omega_{k}+2\omega_{k}$.

Moreover, ${\cal G}_{2}(k,\omega_{n})$ vanishes and ${\cal G}_{2}'(k,\omega_{n})$ remains finite when $T=0$.
These properties can also be attributed to their diagrammatic structures.
In general, similar 1-loop diagrams to those of ${\cal G}_{2}(k,\omega_{n})$ and ${\cal G}_{2}'(k,\omega_{n})$ emerge in perturbative analyses of quasiparticle Green's functions for generic superfluid systems of bosons.
Such diagrams lead to similar properties in, e.g., the damping rates of quasiparticles in symmetry-broken phases, i.e., vanishing and non-vanishing damping rates at $T=0$ in the Landau and Beliaev damping processes of quasiparticles (see, e.g., Refs.~\cite{nagao2016damping,nagao2018response} for the details). 

In the third perturbative term, for which we write ${\cal G}''_{2}(k,\omega_{n})$, the symmetric vertex $V_{2}(n,n_1)=2\omega_{k}-i\omega_{n}-i\omega_{n_1}$ exactly cancels the squeezing-mode propagator in the virtual process.
After direct calculations, it is shown that ${\cal G}''_{2}(k,\omega_{n})$ gives a modification to the weight of the zeroth-order propagator with dispersion $+\omega_{k}$
\begin{align}
{\cal G}_{2}''(k,\omega_{n})
&= \frac{n_{{\rm B},1}}{-i\omega_{n}+\omega_{k}}.
\end{align}
This contribution also vanishes at $T=0$.

Let us make some remarks on other types of perturbative term, which exist when drawing all the possible Feynman diagrams associated with $\langle b_{-k}(\omega_{n})b^*_{-k}(\omega_{n}) \rangle$, but will produce no contribution to it. 
One of those is the so-called tadpole graph \cite{altland2010condensed}, which contains a closed TLL-mode loop connected to a single squeezing-mode line.
Because $V_{1}(n_1,n_2)$ is anti-symmetric in its indices, such a graph is found to be exactly zero.
We further note that the perturbative terms from the denominator of the path integral (\ref{eq: functional_ave_b_bs}) can cancel the Feynman diagrams that are generated from the numerator of (\ref{eq: functional_ave_b_bs}) and have some parts disconnected from the external lines.

Evaluating other types of two-point correlation function is entirely parallel to the above.
Hence, we will not repeat it further.

\subsection{Three-point correlation functions}

\begin{figure}
\includegraphics[width=80mm]{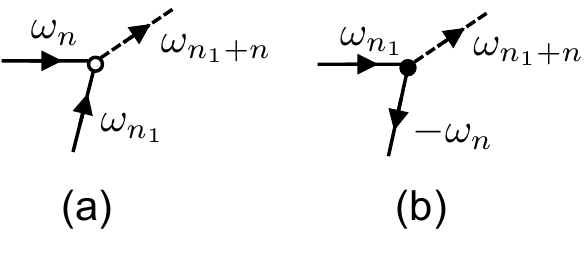}
\vspace{-1mm}
\caption{
Tree diagrams corresponding to $\langle {\tilde \xi}_{k}(\tau) b^*_{k}(\tau)b^*_{-k}(0) \rangle$ (a) and $\langle {\tilde \xi}_{k}(\tau) b^*_{k}(\tau)b_{k}(0) \rangle$ (b).
}
\label{fig: feynman_three_point}
\end{figure}

Next, we focus on the three-point correlation functions in $\chi'(k,\tau)$.
Within the second-order perturbation, multiple tree diagrams of first order give leading-order corrections to $\chi'(k,\tau)$ (see also Fig.~\ref{fig: feynman_three_point}).

We analyze ${\cal G}_{3}(k,\omega_{n}) \equiv \int d\tau e^{i\omega_{n}\tau} \langle {\tilde \xi}_{k}(\tau) b^*_{k}(\tau)b^*_{-k}(0) \rangle$ in $C_1$ as an example. 
It is easy to verify that
\begin{align}
{\cal G}_{3}(k,\omega_{n}) 
&= \frac{1}{(\beta\hbar)^{1/2}}\sum_{n_1}\langle {\tilde \xi}_{k}(\omega_{n_1+n}) b^*_{k}(\omega_{n_1})b^*_{-k}(\omega_{n}) \rangle \nonumber \\
&= \frac{1}{-i\omega_{n}+\omega_{k}}\frac{-1}{\beta\hbar}\sum_{n_1}\frac{V_{2}(n,n_1)}{(\omega_{k}-i\omega_{n_1})(2\omega_{k}-i\omega_{n+n_1})} \nonumber \\
&= -\frac{n_{{\rm B},1}}{-i\omega_{n}+\omega_{k}}. \nonumber
\end{align}
In the second line, $V_{2}(n,n_1)$ is cancelled by $(-i\omega_{n+n_1}+2\omega_{k})$ in the denominator.
In addition, the frequency summation was done after the regularization $\sum_{n_1} \rightarrow \sum_{n_1}e^{i\omega_{n_1}{\varepsilon}}$ (see also the previous subsection).
The diagrammatic representation of this term is given in Fig.~\ref{fig: feynman_three_point}(a).

As another example, let us calculate ${\cal G}_{3}'(k,\omega_{n}) \equiv \int d\tau e^{i\omega_{n}\tau} \langle {\tilde \xi}_{k}(\tau) b^*_{k}(\tau)b_{k}(0) \rangle$ in $C_2$.
Using Eqs.~(\ref{eq: sum_formula1}) and (\ref{eq: sum_formula2}), we obtain
\begin{align}
{\cal G}_{3}'(k,\omega_{n}) 
&= \frac{1}{(\beta\hbar)^{1/2}}\sum_{n_1}\langle {\tilde \xi}_{k}(\omega_{n_1+n}) b^*_{k}(\omega_{n_1})b_{k}(-\omega_{n}) \rangle \nonumber \\
&= \frac{1}{i\omega_{n}+\omega_{k}}\frac{-1}{\beta\hbar}\sum_{n_1}\frac{{\overline V}_{1}(-n,n_1)}{(\omega_{k}-i\omega_{n_1})(2\omega_{k}-i\omega_{n+n_1})} \nonumber \\
&= \frac{w_1 n_{{\rm B},1}}{-i\omega_{n}+\omega_{k}} - \frac{2\omega_{k} w_1 n_{{\rm B},2}}{(i\omega_{n}+\omega_{k})(-i\omega_{n}+\omega_{k})}. \nonumber
\end{align}
Its diagrammatic representation is Fig.~\ref{fig: feynman_three_point}(b).

\subsection{Four-point correlation functions}

We here evaluate the four-point functions in $\chi'(k,\tau)$.
As we will describe below, performing this requires more careful treatments than those for the two- and three-point correlation functions.

\begin{figure}
\includegraphics[width=85mm]{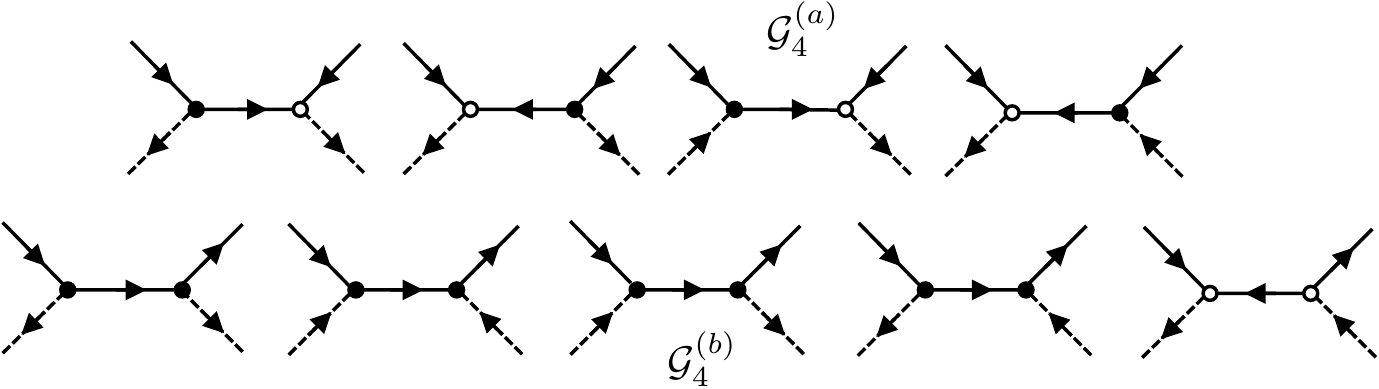}
\vspace{-1mm}
\caption{
Diagrammatic representation of the ladder-type contributions in the four-point functions.
The top and bottom diagrams are the ones in $C_1=-\langle f_{k}(\tau)b^*_{k}(\tau) f_{k}(0) b^*_{-k}(0) \rangle$ and $C_2=-\langle f_{k}(\tau)b^*_{k}(\tau) f^*_{k}(0) b_{k}(0) \rangle$, respectively.
The specific contributions of ${\cal G}^{(a)}_{4}$ and ${\cal G}^{(b)}_{4}$ are explicitly evaluated in Eqs.~(\ref{eq: g4a}) and (\ref{eq: g4b}).
}
\label{fig: feynman_four_point_1}
\end{figure}

The Wick contractions of the perturbative terms of the four-point correlation functions can be categorized into some subclasses associated with their topology of contraction.
Among of them, there is a class of ladder-type Feynman diagrams in $C_1+C_2=-\langle f_{k}(\tau)b^*_{k}(\tau) f_{k}(0) b^*_{-k}(0) \rangle-\langle f_{k}(\tau)b^*_{k}(\tau) f^*_{k}(0) b_{k}(0) \rangle$, which is drawn in Fig.~\ref{fig: feynman_four_point_1}.
Each ladder diagram has two Matsubara sums in its analytical expression.
Since the sums are completely independent of each other, one can carry out them by simply regularizing each sum with the convergence factor $e^{i\omega_{n}\varepsilon}$.
For instance, as for the contribution ${\cal G}^{(a)}_{4}(k,\omega_{n})$ indicated in Fig.~\ref{fig: feynman_four_point_1}, the calculation is done as follows:
\begin{align}
{\cal G}^{(a)}_{4}(k,\omega_{n}) 
&= \frac{1}{\beta \hbar}\sum_{n_1}\frac{e^{i\omega_{n_1}0^+} V_{1}(-n,n_1)}{(\omega_{k}-i\omega_{n_1})(i\omega_{n}+i\omega_{n_1}+2\omega_{k})} \nonumber \\
&\;\;\;\; \times \frac{1}{\beta \hbar}\sum_{n_2}\frac{e^{i\omega_{n_2}0^+}V_{2}(-n,n_2)}{(\omega_{k}-i\omega_{n_2})(i\omega_{n}-i\omega_{n_2}+2\omega_{k})} \nonumber \\
&\;\;\;\; \times \frac{1}{i\omega_{n}+\omega_{k}} \nonumber \\
&= \frac{w_{1}n_{{\rm B},1}(1+n_{{\rm B},1}+n_{{\rm B},2})}{i\omega_{n}+3\omega_{k}} \nonumber \\
&\;\;\;\;\;\;\;\;\;\; - \frac{w_{1}n_{{\rm B},1}(1+n_{{\rm B},2})}{i\omega_{n}+\omega_{k}} \nonumber \\
&\approx \frac{w_{1}n_{{\rm B},1}}{i\omega_{n}+3\omega_{k}} - \frac{w_{1}n_{{\rm B},1}}{i\omega_{n}+\omega_{k}}. \label{eq: g4a}
\end{align}
In the last line, we have approximated multiple products of the distribution function $n_{{\rm B},s}$, such as $n_{{\rm B},1}^2$ and $n_{{\rm B},1}n_{{\rm B},2}$, to zero because the thermal occupation of the excitation branches is supposed to be small.
Throughout this subsection, the symbol $\approx$ is used to imply some approximation in the same sense. 
Likewise, one can obtain the contribution ${\cal G}^{(b)}_{4}(k,\omega_{n})$ in Fig.~\ref{fig: feynman_four_point_1}, i.e., 
\begin{align}
{\cal G}^{(b)}_{4}(k,\omega_{n}) 
&= \frac{1}{\beta\hbar}\sum_{n_1}\frac{e^{i\omega_{n_1}0^+} V_{1}(-n,n_1)}{(\omega_{k}-i\omega_{n_1})(i\omega_{n_1}+i\omega_{n}+2\omega_{k})} \nonumber \\
&\;\;\; \times \frac{1}{\beta\hbar}\sum_{n_2}\frac{e^{i\omega_{n_2}0^+} {\overline V}_{1}(n_2,-n)}{(\omega_{k}-i\omega_{n_2})(i\omega_{n}+i\omega_{n_2}+2\omega_{k})} \nonumber \\
&\;\;\; \times \frac{1}{i\omega_{n}+\omega_{k}} \nonumber \\
&\approx \frac{w_{1}^2(1+2n_{{\rm B},2})}{i\omega_{n}+\omega_k} - \frac{w_{1}^2(1+2n_{{\rm B},2})}{i\omega_{n}+3\omega_k} \nonumber \\
&\;\;\;\;\;\; - \frac{w_{1}^2(2\omega_{k})(1+2n_{{\rm B},1}+2n_{{\rm B},2})}{(i\omega_{n}+3\omega_{k})^2}.  \label{eq: g4b}
\end{align}
The remaining contributions in the class can also be calculated in the same way.

\begin{figure}
\includegraphics[width=85mm]{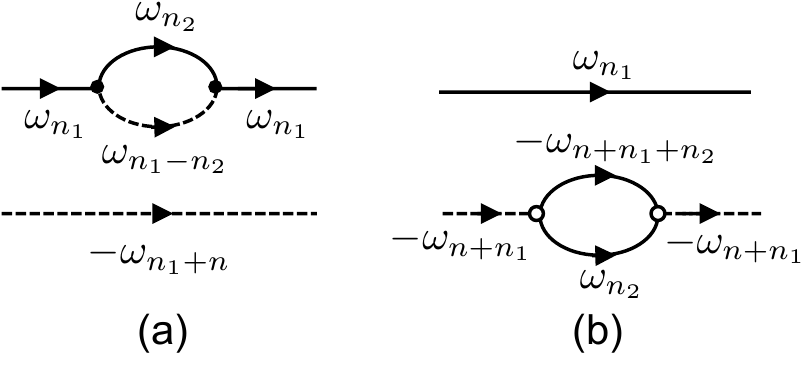}
\vspace{-1mm}
\caption{
Convolution of the free and perturbed propagators in $\langle {\tilde \xi}^*_{k}(\tau)b^*_{k}(\tau){\tilde \xi}_{k}(0)b_{k}(0) \rangle$. 
(a) The TLL-mode propagator is modified by virtual processes of emission of squeezing and TLL modes while the squeezing-mode one is unperturbed.
(b) The squeezing-mode propagator acquires a loop correction associated with virtual emission of two TLL modes while the TLL-mode one is unperturbed.
}
\label{fig: feynman_four_point_2}
\end{figure}

In addition to the ladder-type diagrams, there exist some topologically-disconnected diagrams in the same approximation order, as sketched in Fig.~\ref{fig: feynman_four_point_2}.
The disconnected diagrams are nothing but simple convolutions between the TLL and squeezing mode propagators, either of which is perturbed by the interaction vertices.
To evaluate such diagrams, we first calculate some perturbed single-mode propagators of the second order, and then insert them into the analytical expressions.
As for Fig.~\ref{fig: feynman_four_point_2}(a), in which the TLL-mode propagator is modified by the perturbations with $V_{1}$ and ${\overline V}_{1}$, the evaluation of its analytical expression, for which we write ${\cal G}^{(c)}_{4}(k,\omega_{n})$, is performed as follows:
\begin{align}
{\cal G}^{(c)}_{4}(k,\omega_{n})
&= \frac{1}{\beta\hbar}\sum_{n_1}\frac{e^{i\omega_{n_1}0^{+}}}{i\omega_{n+n_1}+2\omega_{k}}\left[ \frac{w_1^2\left(1+n_{{\rm B},1}+n_{{\rm B},2}\right)}{-i\omega_{n_1}+3\omega_{k}} \right] \nonumber \\
&\approx \frac{w_{1}^2(1+n_{{\rm B},1}+2n_{{\rm B},2}+n_{{\rm B},3})}{i\omega_{n}+5\omega_{k}} \nonumber \\
&\;\;\;\;\;\;\;\;\;\;\;\; - \frac{w_{1}^2(1+n_{{\rm B},1}+2n_{{\rm B},2})}{i\omega_{n}+3\omega_{k}} \nonumber \\
&\;\;\;\;\;\;\;\;\;\;\;\;\;\;\;\; + \frac{2\omega_{k}w_{1}^2(1+n_{{\rm B},1}+2n_{{\rm B},2})}{(i\omega_{n}+3\omega_{k})^2}. \nonumber 
\end{align}
We note that this contribution is a responsible part for the emergent pole at $z = 5\omega_{k}$.
Similarly, another contribution ${\cal G}^{(d)}_{4}(k,\omega_{n})$, which corresponds to Fig.~\ref{fig: feynman_four_point_2}(b), reads
\begin{align}
{\cal G}^{(d)}_{4}(k,\omega_{n})
&= \frac{1}{\beta\hbar}\sum_{n_1}\frac{e^{i\omega_{n_1}0^{+}}}{-i\omega_{n_1}+\omega_{k}}\left[ \frac{1+2n_{{\rm B},1}}{i\omega_{n+n_1}+2\omega_{k}}\right] \nonumber \\
&\approx \frac{1+3n_{{\rm B},1}+n_{{\rm B},2}}{i\omega_{n}+3\omega_{k}}. \nonumber 
\end{align}
As indicated by Fig.~\ref{fig: feynman_four_point_2}(b), this contribution emerges due to a process, under which propagation of a squeezing mode is perturbed by couplings to two TLL modes.

\begin{figure}
\includegraphics[width=85mm]{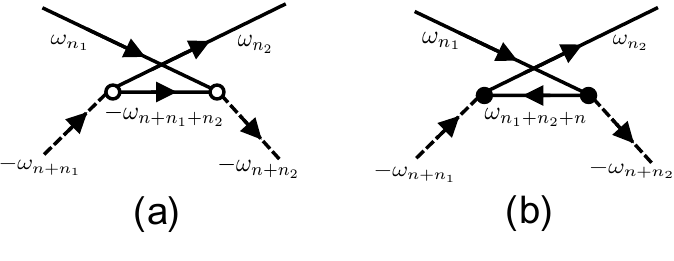}
\vspace{-1mm}
\caption{
(a) The cross-type Feynman diagram associated with $V_{2}$.
(b) The similar cross-type diagram associated with $V_{1}$ and ${\overline V}_{1}$.
}
\label{fig: feynman_four_point_3}
\end{figure}

As the rest of the contributions, we treat the so-called cross-type Feynman diagrams (Fig.~\ref{fig: feynman_four_point_3}).
In particular, let us consider contributions that appear in a normal four-point correlation function $\langle b^{*}_{k}(\tau) {\tilde \xi}^*_{k}(\tau) b_{k}(0) {\tilde \xi}_{k}(0) \rangle$.
The term as {\it normal} implies that it is nonzero even when the interaction is switched off.
There are two species of the diagrams in the normal correlation function:
\begin{widetext}
\begin{align}
{\cal G}^{(e)}_{4}(k,\omega_{n}) &= \frac{1}{(\beta\hbar)^2}\sum_{n_1,n_2}\frac{V_{2}(n_2,-n-n_1-n_2)V_{2}(n_1,-n-n_1-n_2)}{(\omega_{k}-i\omega_{n_1})(\omega_{k}-i\omega_{n_2})(i\omega_{n_1+n}+2\omega_{k})(i\omega_{n_2+n}+2\omega_{k})(i\omega_{n+n_1+n_2}+\omega_{k})}, \\
{\cal G}^{(f)}_{4}(k,\omega_{n}) &= \frac{1}{(\beta\hbar)^2}\sum_{n_1,n_2}\frac{V_{1}(n_2,n+n_1+n_2){\overline V}_{1}(n+n_1+n_2,n_1)}{(\omega_{k}-i\omega_{n_1})(\omega_{k}-i\omega_{n_2})(i\omega_{n_1+n}+2\omega_{k})(i\omega_{n_2+n}+2\omega_{k})(-i\omega_{n+n_1+n_2}+\omega_{k})}.
\end{align}
\end{widetext}
Here ${\cal G}^{(e)}_{4}(k,\omega_{n})$ and ${\cal G}^{(f)}_{4}(k,\omega_{n})$ correspond to Fig.~\ref{fig: feynman_four_point_3}(a) and (b), respectively.
Moreover, we make a remark on that the integrands of these functions remain unchanged under exchanges $n_{1} \leftrightarrow n_{2}$.
If one first integrates $n_2$ and then does $n_1$ by using the formula (\ref{eq: cauchy}), we obtain 
\begin{align} 
{\cal G}^{(e)}_{4}(k,\omega_{n}) &\approx \frac{1+3n_{{\rm B},1}+n_{{\rm B},2}}{i\omega_{n}+3\omega_{k}}, \nonumber \\
{\cal G}^{(f)}_{4}(k,\omega_{n}) &\approx -\frac{w_{1}^2(1+2n_{{\rm B},1}+2n_{{\rm B},2})}{i\omega_{n}+3\omega_{k}} \nonumber \\
&\;\;\;\; + 2\omega_{k} w_{1}^2 \frac{1+2n_{{\rm B},1}+2n_{{\rm B},2}}{(i\omega_{n}+3\omega_{k})^2} \nonumber \\
&\;\;\;\; + \frac{w_1^2(1+n_{{\rm B},1}+2n_{{\rm B},2}+n_{{\rm B},3})}{i\omega_{n}+5\omega_{k}}. \nonumber
\end{align}
We note that the latter one also yields the pole at $z=5\omega_{k}$.

Nonzero contributions that are similar to ${\cal G}^{(e)}_{4}(k,\omega_{n})$ and ${\cal G}^{(f)}_{4}(k,\omega_{n})$ also emerge in another normal four-point correlation function, i.e., $\langle b^{*}_{k}(\tau) {\tilde \xi}_{k}(\tau) b_{k}(0) {\tilde \xi}^*_{k}(0) \rangle$.
However, it can be directly checked that the calculated weights contain second-order terms of the mean occupation numbers of the thermal excitations.
Therefore, these are negligible compared to other contributions of order ${\cal O}(1)$ or ${\cal O}(n_{{\rm B},s})$.
 
As well as the normal ones, the {\it anomalous} four-point correlation functions, such as $\langle b^*_{k}(\tau) {\tilde \xi}_{k}(\tau) b^*_{-k}(0) {\tilde \xi}_{k}(0) \rangle$ and $\langle b^*_{k}(\tau) {\tilde \xi}^*_{k}(\tau) b_{k}(0) {\tilde \xi}^*_{k}(0) \rangle$, will also contain some cross-type Feynman diagrams as a result of the Wick contraction.
However, if one first integrates $n_2$ and then does $n_1$ for a certain diagram of them by using the formula (\ref{eq: cauchy}), the result will be different from the one obtained from the reversed procedure.
Such a non-commutativity, to our guess, could be related to some convergence problems of series with doubled infinite sums of the Matsubara frequencies.
To avoid some ambiguity in the final result of the spectral function itself, in this work, we just put a prescription, in which one neglects all the cross-type contributions in the anomalous correlators.
In some separate publications, we will address the issues as described above in detail in order to clarify the degree to which the excluded contributions may change the total contributions.

\subsection{Renormalized velocities and spectral weights}

Finally, to obtain the renormalized phase velocities and weights, we combine all the relevant contributions discussed in the above subsections.
Within the second-order perturbation, it turns out that one can write $\chi''(k,\omega_{n})$ as
\begin{align}
\frac{2\pi}{|k|}\chi''(k,\omega_{n}) 
&= \frac{{\cal A}}{-i\omega_{n}+\omega_{k}} + \frac{2\omega_{k}{\cal A}'}{(-i\omega_{n}+\omega_{k})^2} \nonumber \\
&+ \frac{{\cal B}}{-i\omega_{n}+3\omega_{k}} + \frac{2\omega_{k}{\cal B}'}{(-i\omega_{n}+3\omega_{k})^2} \nonumber \\
&+ \frac{{\cal C}}{-i\omega_{n}+5\omega_{k}} + (-i\omega_{n} \leftrightarrow i\omega_{n}). \label{eq: chi_matsubara}
\end{align}
Here the weights of the poles of first order are given as functions of $k$, $T$, and $K$:
\begin{widetext}
\begin{align}
{\cal A} 
&= -K\left[ 1 - w_{1}^2 + (1-w_{1})^2n_{{\rm B},1} - 3 w_{1}^2 n_{{\rm B},2} \right] +\frac{1}{\sqrt{2r^2}}\left[-2\gamma_{2}w_1 - 4 \gamma_{1}n_{{\rm B},1} + 2w_1\left(2\gamma_{1} - \gamma_2\right)n_{{\rm B},2} \right] \nonumber \\
&-\frac{\sigma_{3}}{r^2}\left[ - w_1n_{{\rm B},1} + \frac{w_1^2}{2}(n_{{\rm B},1}-n_{{\rm B},2}) - w_1^2 n_{{\rm B},2} \right] - \frac{\sigma_{1}}{2r^2}\left[ (1-w_1^2)(n_{{\rm B},1}-n_{{\rm B},2}) + w_1^2(n_{{\rm B},2}-n_{{\rm B},3}) \right] \nonumber \\
&-\frac{\sigma_{2}}{2r^2}w_1^2(1+2n_{{\rm B},2}), \\
{\cal B}
&= -K w_{1}^2 \left( 1+n_{{\rm B},1}+n_{{\rm B},2} \right) + \frac{2\gamma_{2}w_{1}}{\sqrt{2r^2}}\left(1+n_{{\rm B},1}+n_{{\rm B},2}\right) \nonumber \\
&-\frac{\sigma_{3}}{r^2}\left[\frac{w_1^2}{2}\left(n_{{\rm B},1}+n_{{\rm B},2}\right)\right] - \frac{\sigma_{1}}{2r^2}w_{1}^2\left(n_{{\rm B},1}-n_{{\rm B},2}\right) - \frac{\sigma_{2}}{2r^2}\left[ 2-3w_1^2 + (7-3w_1^2)n_{{\rm B},1} + (2-7w_1^2)n_{{\rm B},2} \right], \\
{\cal C}
&= -\frac{\sigma_{2}}{r^2}w_1^2(1+n_{{\rm B},1}+2n_{{\rm B},2}+n_{{\rm B},3}). 
\end{align}
\end{widetext}
In these expressions, the constants 
\begin{align}
\gamma_{1}&=\frac{f_{k}[\eta_0]}{\sqrt{2}}\left[f^{(1,0)}_{k}r + {\tilde f}^{(0,1)}_{k}r^{-1} \right], \nonumber \\
\gamma_{2}&=\frac{f_{k}[\eta_0]}{\sqrt{2}}\left[ f^{(1,0)}_{k}r - {\tilde f}^{(0,1)}_{k}r^{-1} \right], \nonumber
\end{align}
give the coefficients in front of the three-point correlation functions in the spectral function while 
\begin{align}
\sigma_{1}&=\frac{1}{2}\left(f^{(1,0)}_{k}r+{\tilde f}^{(0,1)}_{k}r^{-1}\right)^2, \nonumber \\
\sigma_{2}&=\frac{1}{2}\left(f^{(1,0)}_{k}r-{\tilde f}^{(0,1)}_{k}r^{-1}\right)^2, \nonumber \\
\sigma_{3}&=\frac{1}{2}\left[\left(rf^{(1,0)}_{k}\right)^2-\left({\tilde f}^{(0,1)}_{k}\right)^2\right], \nonumber 
\end{align}
do those of the four-point correlation functions. 

The poles of second order are characterized by 
\begin{align}
{\cal A}' &= -Kw_1^2 + \frac{\sigma_{1}}{2r^2}w_1^2(n_1-n_2), \\
{\cal B}' &= -\frac{\sigma_{2}}{2r^2}w_1^2(1+n_{{\rm B},1}+n_{{\rm B},2}).
\end{align}
We note that ${\cal A}'$ is composed of the contributions from the two-point and four-point correlation functions, whereas ${\cal B}'$ originates from the perturbative corrections to the four-point correlation functions. 
Moreover, all the contributions to the second-order poles are constructed from $V_{1}$ and ${\overline V}_{1}$, but they are not associated with $V_{2}$.

The emergent poles of second order imply that there is a shift of {\it chemical potential} to the bare band dispersion.
More precisely, Eq.~(\ref{eq: chi_matsubara}) is reminiscent of the following expansion of a propagator with respect to a chemical potential shift $\Sigma$:
\begin{align}
\frac{1}{i\omega_{n}-\omega-\Sigma} = \frac{1}{i\omega_{n}-\omega} +  \frac{1}{i\omega_{n}-\omega} \Sigma \frac{1}{i\omega_{n}-\omega} + \cdots. \nonumber 
\end{align}
For the present case, the corresponding chemical potentials are proportional to $\omega_{k}=v_0|k|$, so that they just give renormalization effects on the velocities of the dispersions.
In other words, the system still remains gapless.
Thus, within the second-order validity of the expansion, $\chi''(k,\omega_{n})$ has the form 
\begin{align}
\frac{2\pi}{|k|}\chi''(k,\omega_{n}) 
&\approx \frac{{\cal A}}{-i\omega_{n}+{\tilde v}_1 |k|} + \frac{{\cal B}}{-i\omega_{n}+{\tilde v}_3 |k|} \nonumber \\
&+ \frac{{\cal C}}{-i\omega_{n}+5\omega_{k}} + (-i\omega_{n} \leftrightarrow i\omega_{n}), 
\end{align}
and one can obtain the renormalized velocities ${\tilde v}_1$ and ${\tilde v}_3$ in Eq.~(\ref{eq: vel}).

\bibliography{reference}

\begin{thebibliography}{67}%
\makeatletter
\providecommand \@ifxundefined [1]{%
 \@ifx{#1\undefined}
}%
\providecommand \@ifnum [1]{%
 \ifnum #1\expandafter \@firstoftwo
 \else \expandafter \@secondoftwo
 \fi
}%
\providecommand \@ifx [1]{%
 \ifx #1\expandafter \@firstoftwo
 \else \expandafter \@secondoftwo
 \fi
}%
\providecommand \natexlab [1]{#1}%
\providecommand \enquote  [1]{``#1''}%
\providecommand \bibnamefont  [1]{#1}%
\providecommand \bibfnamefont [1]{#1}%
\providecommand \citenamefont [1]{#1}%
\providecommand \href@noop [0]{\@secondoftwo}%
\providecommand \href [0]{\begingroup \@sanitize@url \@href}%
\providecommand \@href[1]{\@@startlink{#1}\@@href}%
\providecommand \@@href[1]{\endgroup#1\@@endlink}%
\providecommand \@sanitize@url [0]{\catcode `\\12\catcode `\$12\catcode
  `\&12\catcode `\#12\catcode `\^12\catcode `\_12\catcode `\%12\relax}%
\providecommand \@@startlink[1]{}%
\providecommand \@@endlink[0]{}%
\providecommand \url  [0]{\begingroup\@sanitize@url \@url }%
\providecommand \@url [1]{\endgroup\@href {#1}{\urlprefix }}%
\providecommand \urlprefix  [0]{URL }%
\providecommand \Eprint [0]{\href }%
\providecommand \doibase [0]{http://dx.doi.org/}%
\providecommand \selectlanguage [0]{\@gobble}%
\providecommand \bibinfo  [0]{\@secondoftwo}%
\providecommand \bibfield  [0]{\@secondoftwo}%
\providecommand \translation [1]{[#1]}%
\providecommand \BibitemOpen [0]{}%
\providecommand \bibitemStop [0]{}%
\providecommand \bibitemNoStop [0]{.\EOS\space}%
\providecommand \EOS [0]{\spacefactor3000\relax}%
\providecommand \BibitemShut  [1]{\csname bibitem#1\endcsname}%
\let\auto@bib@innerbib\@empty
\bibitem [{\citenamefont {Nagaosa}(1999)}]{nagaosa1999quantum}%
  \BibitemOpen
  \bibfield  {author} {\bibinfo {author} {\bibfnamefont {N.}~\bibnamefont
  {Nagaosa}},\ }\href@noop {} {\emph {\bibinfo {title} {Quantum field theory in
  strongly correlated electronic systems}}}\ (\bibinfo  {publisher} {Springer
  Science \& Business Media},\ \bibinfo {year} {1999})\BibitemShut {NoStop}%
\bibitem [{\citenamefont {Giamarchi}(2004)}]{giamarchi2004quantum}%
  \BibitemOpen
  \bibfield  {author} {\bibinfo {author} {\bibfnamefont {T.}~\bibnamefont
  {Giamarchi}},\ }\href@noop {} {\emph {\bibinfo {title} {Quantum Physics in
  One Dimension}}}\ (\bibinfo  {publisher} {Oxford University Press},\ \bibinfo
  {year} {2004})\BibitemShut {NoStop}%
\bibitem [{\citenamefont {Fradkin}(2013)}]{fradkin2013field}%
  \BibitemOpen
  \bibfield  {author} {\bibinfo {author} {\bibfnamefont {E.}~\bibnamefont
  {Fradkin}},\ }\href@noop {} {\emph {\bibinfo {title} {Field theories of
  condensed matter physics}}}\ (\bibinfo  {publisher} {Cambridge University
  Press},\ \bibinfo {year} {2013})\BibitemShut {NoStop}%
\bibitem [{\citenamefont {Avouris}\ \emph {et~al.}(2008)\citenamefont
  {Avouris}, \citenamefont {Freitag},\ and\ \citenamefont
  {Perebeinos}}]{avouris2008carbon}%
  \BibitemOpen
  \bibfield  {author} {\bibinfo {author} {\bibfnamefont {P.}~\bibnamefont
  {Avouris}}, \bibinfo {author} {\bibfnamefont {M.}~\bibnamefont {Freitag}}, \
  and\ \bibinfo {author} {\bibfnamefont {V.}~\bibnamefont {Perebeinos}},\
  }\href {https://www.nature.com/articles/nphoton.2008.94} {\bibfield
  {journal} {\bibinfo  {journal} {Nature Photonics}\ }\textbf {\bibinfo
  {volume} {2}},\ \bibinfo {pages} {341} (\bibinfo {year} {2008})}\BibitemShut
  {NoStop}%
\bibitem [{\citenamefont {Ishii}\ \emph {et~al.}(2003)\citenamefont {Ishii},
  \citenamefont {Kataura}, \citenamefont {Shiozawa}, \citenamefont {Yoshioka},
  \citenamefont {Otsubo}, \citenamefont {Takayama}, \citenamefont {Miyahara},
  \citenamefont {Suzuki}, \citenamefont {Achiba}, \citenamefont {Nakatake},
  \citenamefont {Narimura}, \citenamefont {Higashiguchi}, \citenamefont
  {Shimada}, \citenamefont {Namatame},\ and\ \citenamefont
  {Taniguchi}}]{ishii2003direct}%
  \BibitemOpen
  \bibfield  {author} {\bibinfo {author} {\bibfnamefont {H.}~\bibnamefont
  {Ishii}}, \bibinfo {author} {\bibfnamefont {H.}~\bibnamefont {Kataura}},
  \bibinfo {author} {\bibfnamefont {H.}~\bibnamefont {Shiozawa}}, \bibinfo
  {author} {\bibfnamefont {H.}~\bibnamefont {Yoshioka}}, \bibinfo {author}
  {\bibfnamefont {H.}~\bibnamefont {Otsubo}}, \bibinfo {author} {\bibfnamefont
  {Y.}~\bibnamefont {Takayama}}, \bibinfo {author} {\bibfnamefont
  {T.}~\bibnamefont {Miyahara}}, \bibinfo {author} {\bibfnamefont
  {S.}~\bibnamefont {Suzuki}}, \bibinfo {author} {\bibfnamefont
  {Y.}~\bibnamefont {Achiba}}, \bibinfo {author} {\bibfnamefont
  {M.}~\bibnamefont {Nakatake}}, \bibinfo {author} {\bibfnamefont
  {T.}~\bibnamefont {Narimura}}, \bibinfo {author} {\bibfnamefont
  {M.}~\bibnamefont {Higashiguchi}}, \bibinfo {author} {\bibfnamefont
  {K.}~\bibnamefont {Shimada}}, \bibinfo {author} {\bibfnamefont
  {H.}~\bibnamefont {Namatame}}, \ and\ \bibinfo {author} {\bibfnamefont
  {M.}~\bibnamefont {Taniguchi}},\ }\href {\doibase 10.1038/nature02074}
  {\bibfield  {journal} {\bibinfo  {journal} {Nature}\ }\textbf {\bibinfo
  {volume} {426}},\ \bibinfo {pages} {540} (\bibinfo {year}
  {2003})}\BibitemShut {NoStop}%
\bibitem [{\citenamefont {Shi}\ \emph {et~al.}(2015)\citenamefont {Shi},
  \citenamefont {Hong}, \citenamefont {Bechtel}, \citenamefont {Zeng},
  \citenamefont {Martin}, \citenamefont {Watanabe}, \citenamefont {Taniguchi},
  \citenamefont {Shen},\ and\ \citenamefont {Wang}}]{shi2015observation}%
  \BibitemOpen
  \bibfield  {author} {\bibinfo {author} {\bibfnamefont {Z.}~\bibnamefont
  {Shi}}, \bibinfo {author} {\bibfnamefont {X.}~\bibnamefont {Hong}}, \bibinfo
  {author} {\bibfnamefont {H.~A.}\ \bibnamefont {Bechtel}}, \bibinfo {author}
  {\bibfnamefont {B.}~\bibnamefont {Zeng}}, \bibinfo {author} {\bibfnamefont
  {M.~C.}\ \bibnamefont {Martin}}, \bibinfo {author} {\bibfnamefont
  {K.}~\bibnamefont {Watanabe}}, \bibinfo {author} {\bibfnamefont
  {T.}~\bibnamefont {Taniguchi}}, \bibinfo {author} {\bibfnamefont {Y.-R.}\
  \bibnamefont {Shen}}, \ and\ \bibinfo {author} {\bibfnamefont
  {F.}~\bibnamefont {Wang}},\ }\href {\doibase 10.1038/nphoton.2015.123}
  {\bibfield  {journal} {\bibinfo  {journal} {Nature Photonics}\ }\textbf
  {\bibinfo {volume} {9}},\ \bibinfo {pages} {515} (\bibinfo {year}
  {2015})}\BibitemShut {NoStop}%
\bibitem [{\citenamefont {Sato}\ \emph {et~al.}(2019)\citenamefont {Sato},
  \citenamefont {Matsuo}, \citenamefont {Hsu}, \citenamefont {Stano},
  \citenamefont {Ueda}, \citenamefont {Takeshige}, \citenamefont {Kamata},
  \citenamefont {Lee}, \citenamefont {Shojaei}, \citenamefont {Wickramasinghe},
  \citenamefont {Shabani}, \citenamefont {Palmstr\o{}m}, \citenamefont
  {Tokura}, \citenamefont {Loss},\ and\ \citenamefont
  {Tarucha}}]{sato2019strong}%
  \BibitemOpen
  \bibfield  {author} {\bibinfo {author} {\bibfnamefont {Y.}~\bibnamefont
  {Sato}}, \bibinfo {author} {\bibfnamefont {S.}~\bibnamefont {Matsuo}},
  \bibinfo {author} {\bibfnamefont {C.-H.}\ \bibnamefont {Hsu}}, \bibinfo
  {author} {\bibfnamefont {P.}~\bibnamefont {Stano}}, \bibinfo {author}
  {\bibfnamefont {K.}~\bibnamefont {Ueda}}, \bibinfo {author} {\bibfnamefont
  {Y.}~\bibnamefont {Takeshige}}, \bibinfo {author} {\bibfnamefont
  {H.}~\bibnamefont {Kamata}}, \bibinfo {author} {\bibfnamefont {J.~S.}\
  \bibnamefont {Lee}}, \bibinfo {author} {\bibfnamefont {B.}~\bibnamefont
  {Shojaei}}, \bibinfo {author} {\bibfnamefont {K.}~\bibnamefont
  {Wickramasinghe}}, \bibinfo {author} {\bibfnamefont {J.}~\bibnamefont
  {Shabani}}, \bibinfo {author} {\bibfnamefont {C.}~\bibnamefont
  {Palmstr\o{}m}}, \bibinfo {author} {\bibfnamefont {Y.}~\bibnamefont
  {Tokura}}, \bibinfo {author} {\bibfnamefont {D.}~\bibnamefont {Loss}}, \ and\
  \bibinfo {author} {\bibfnamefont {S.}~\bibnamefont {Tarucha}},\ }\href
  {\doibase 10.1103/PhysRevB.99.155304} {\bibfield  {journal} {\bibinfo
  {journal} {Phys. Rev. B}\ }\textbf {\bibinfo {volume} {99}},\ \bibinfo
  {pages} {155304} (\bibinfo {year} {2019})}\BibitemShut {NoStop}%
\bibitem [{\citenamefont {Wang}\ \emph {et~al.}(2020)\citenamefont {Wang},
  \citenamefont {Zhao}, \citenamefont {Shi}, \citenamefont {Wu}, \citenamefont
  {Zhao}, \citenamefont {Jiang}, \citenamefont {Watanabe}, \citenamefont
  {Taniguchi}, \citenamefont {Zettl}, \citenamefont {Zhou},\ and\ \citenamefont
  {Wang}}]{wang2020nonlinear}%
  \BibitemOpen
  \bibfield  {author} {\bibinfo {author} {\bibfnamefont {S.}~\bibnamefont
  {Wang}}, \bibinfo {author} {\bibfnamefont {S.}~\bibnamefont {Zhao}}, \bibinfo
  {author} {\bibfnamefont {Z.}~\bibnamefont {Shi}}, \bibinfo {author}
  {\bibfnamefont {F.}~\bibnamefont {Wu}}, \bibinfo {author} {\bibfnamefont
  {Z.}~\bibnamefont {Zhao}}, \bibinfo {author} {\bibfnamefont {L.}~\bibnamefont
  {Jiang}}, \bibinfo {author} {\bibfnamefont {K.}~\bibnamefont {Watanabe}},
  \bibinfo {author} {\bibfnamefont {T.}~\bibnamefont {Taniguchi}}, \bibinfo
  {author} {\bibfnamefont {A.}~\bibnamefont {Zettl}}, \bibinfo {author}
  {\bibfnamefont {C.}~\bibnamefont {Zhou}}, \ and\ \bibinfo {author}
  {\bibfnamefont {F.}~\bibnamefont {Wang}},\ }\href {\doibase
  10.1038/s41563-020-0652-5} {\bibfield  {journal} {\bibinfo  {journal} {Nature
  Materials}\ }\textbf {\bibinfo {volume} {19}},\ \bibinfo {pages} {986}
  (\bibinfo {year} {2020})}\BibitemShut {NoStop}%
\bibitem [{\citenamefont {Wada}\ \emph {et~al.}(2001)\citenamefont {Wada},
  \citenamefont {Taniguchi}, \citenamefont {Ikegami}, \citenamefont {Inagaki},\
  and\ \citenamefont {Fukushima}}]{wada2001helium}%
  \BibitemOpen
  \bibfield  {author} {\bibinfo {author} {\bibfnamefont {N.}~\bibnamefont
  {Wada}}, \bibinfo {author} {\bibfnamefont {J.}~\bibnamefont {Taniguchi}},
  \bibinfo {author} {\bibfnamefont {H.}~\bibnamefont {Ikegami}}, \bibinfo
  {author} {\bibfnamefont {S.}~\bibnamefont {Inagaki}}, \ and\ \bibinfo
  {author} {\bibfnamefont {Y.}~\bibnamefont {Fukushima}},\ }\href {\doibase
  10.1103/PhysRevLett.86.4322} {\bibfield  {journal} {\bibinfo  {journal}
  {Phys. Rev. Lett.}\ }\textbf {\bibinfo {volume} {86}},\ \bibinfo {pages}
  {4322} (\bibinfo {year} {2001})}\BibitemShut {NoStop}%
\bibitem [{\citenamefont {Taniguchi}\ \emph {et~al.}(2005)\citenamefont
  {Taniguchi}, \citenamefont {Yamaguchi}, \citenamefont {Ishimoto},
  \citenamefont {Ikegami}, \citenamefont {Matsushita}, \citenamefont {Wada},
  \citenamefont {Gatica}, \citenamefont {Cole}, \citenamefont {Ancilotto},
  \citenamefont {Inagaki},\ and\ \citenamefont
  {Fukushima}}]{taniguchi2005possible}%
  \BibitemOpen
  \bibfield  {author} {\bibinfo {author} {\bibfnamefont {J.}~\bibnamefont
  {Taniguchi}}, \bibinfo {author} {\bibfnamefont {A.}~\bibnamefont
  {Yamaguchi}}, \bibinfo {author} {\bibfnamefont {H.}~\bibnamefont {Ishimoto}},
  \bibinfo {author} {\bibfnamefont {H.}~\bibnamefont {Ikegami}}, \bibinfo
  {author} {\bibfnamefont {T.}~\bibnamefont {Matsushita}}, \bibinfo {author}
  {\bibfnamefont {N.}~\bibnamefont {Wada}}, \bibinfo {author} {\bibfnamefont
  {S.~M.}\ \bibnamefont {Gatica}}, \bibinfo {author} {\bibfnamefont {M.~W.}\
  \bibnamefont {Cole}}, \bibinfo {author} {\bibfnamefont {F.}~\bibnamefont
  {Ancilotto}}, \bibinfo {author} {\bibfnamefont {S.}~\bibnamefont {Inagaki}},
  \ and\ \bibinfo {author} {\bibfnamefont {Y.}~\bibnamefont {Fukushima}},\
  }\href {\doibase 10.1103/PhysRevLett.94.065301} {\bibfield  {journal}
  {\bibinfo  {journal} {Phys. Rev. Lett.}\ }\textbf {\bibinfo {volume} {94}},\
  \bibinfo {pages} {065301} (\bibinfo {year} {2005})}\BibitemShut {NoStop}%
\bibitem [{\citenamefont {Savard}\ \emph {et~al.}(2011)\citenamefont {Savard},
  \citenamefont {Dauphinais},\ and\ \citenamefont
  {Gervais}}]{savard2011hydrodynamics}%
  \BibitemOpen
  \bibfield  {author} {\bibinfo {author} {\bibfnamefont {M.}~\bibnamefont
  {Savard}}, \bibinfo {author} {\bibfnamefont {G.}~\bibnamefont {Dauphinais}},
  \ and\ \bibinfo {author} {\bibfnamefont {G.}~\bibnamefont {Gervais}},\ }\href
  {\doibase 10.1103/PhysRevLett.107.254501} {\bibfield  {journal} {\bibinfo
  {journal} {Phys. Rev. Lett.}\ }\textbf {\bibinfo {volume} {107}},\ \bibinfo
  {pages} {254501} (\bibinfo {year} {2011})}\BibitemShut {NoStop}%
\bibitem [{\citenamefont {Bertaina}\ \emph {et~al.}(2016)\citenamefont
  {Bertaina}, \citenamefont {Motta}, \citenamefont {Rossi}, \citenamefont
  {Vitali},\ and\ \citenamefont {Galli}}]{bertaina2016one}%
  \BibitemOpen
  \bibfield  {author} {\bibinfo {author} {\bibfnamefont {G.}~\bibnamefont
  {Bertaina}}, \bibinfo {author} {\bibfnamefont {M.}~\bibnamefont {Motta}},
  \bibinfo {author} {\bibfnamefont {M.}~\bibnamefont {Rossi}}, \bibinfo
  {author} {\bibfnamefont {E.}~\bibnamefont {Vitali}}, \ and\ \bibinfo {author}
  {\bibfnamefont {D.~E.}\ \bibnamefont {Galli}},\ }\href {\doibase
  10.1103/PhysRevLett.116.135302} {\bibfield  {journal} {\bibinfo  {journal}
  {Phys. Rev. Lett.}\ }\textbf {\bibinfo {volume} {116}},\ \bibinfo {pages}
  {135302} (\bibinfo {year} {2016})}\BibitemShut {NoStop}%
\bibitem [{\citenamefont {Cazalilla}\ \emph {et~al.}(2011)\citenamefont
  {Cazalilla}, \citenamefont {Citro}, \citenamefont {Giamarchi}, \citenamefont
  {Orignac},\ and\ \citenamefont {Rigol}}]{cazalilla2011one}%
  \BibitemOpen
  \bibfield  {author} {\bibinfo {author} {\bibfnamefont {M.~A.}\ \bibnamefont
  {Cazalilla}}, \bibinfo {author} {\bibfnamefont {R.}~\bibnamefont {Citro}},
  \bibinfo {author} {\bibfnamefont {T.}~\bibnamefont {Giamarchi}}, \bibinfo
  {author} {\bibfnamefont {E.}~\bibnamefont {Orignac}}, \ and\ \bibinfo
  {author} {\bibfnamefont {M.}~\bibnamefont {Rigol}},\ }\href {\doibase
  10.1103/RevModPhys.83.1405} {\bibfield  {journal} {\bibinfo  {journal} {Rev.
  Mod. Phys.}\ }\textbf {\bibinfo {volume} {83}},\ \bibinfo {pages} {1405}
  (\bibinfo {year} {2011})}\BibitemShut {NoStop}%
\bibitem [{\citenamefont {Guan}\ \emph {et~al.}(2013)\citenamefont {Guan},
  \citenamefont {Batchelor},\ and\ \citenamefont {Lee}}]{wen2013fermi}%
  \BibitemOpen
  \bibfield  {author} {\bibinfo {author} {\bibfnamefont {X.-W.}\ \bibnamefont
  {Guan}}, \bibinfo {author} {\bibfnamefont {M.~T.}\ \bibnamefont {Batchelor}},
  \ and\ \bibinfo {author} {\bibfnamefont {C.}~\bibnamefont {Lee}},\ }\href
  {\doibase 10.1103/RevModPhys.85.1633} {\bibfield  {journal} {\bibinfo
  {journal} {Rev. Mod. Phys.}\ }\textbf {\bibinfo {volume} {85}},\ \bibinfo
  {pages} {1633} (\bibinfo {year} {2013})}\BibitemShut {NoStop}%
\bibitem [{\citenamefont {Capponi}\ \emph {et~al.}(2016)\citenamefont
  {Capponi}, \citenamefont {Lecheminant},\ and\ \citenamefont
  {Totsuka}}]{capponi2016phases}%
  \BibitemOpen
  \bibfield  {author} {\bibinfo {author} {\bibfnamefont {S.}~\bibnamefont
  {Capponi}}, \bibinfo {author} {\bibfnamefont {P.}~\bibnamefont
  {Lecheminant}}, \ and\ \bibinfo {author} {\bibfnamefont {K.}~\bibnamefont
  {Totsuka}},\ }\href@noop {} {\bibfield  {journal} {\bibinfo  {journal}
  {Annals of Physics}\ }\textbf {\bibinfo {volume} {367}},\ \bibinfo {pages}
  {50} (\bibinfo {year} {2016})}\BibitemShut {NoStop}%
\bibitem [{\citenamefont {St\"oferle}\ \emph {et~al.}(2004)\citenamefont
  {St\"oferle}, \citenamefont {Moritz}, \citenamefont {Schori}, \citenamefont
  {K\"ohl},\ and\ \citenamefont {Esslinger}}]{stoferle2004transition}%
  \BibitemOpen
  \bibfield  {author} {\bibinfo {author} {\bibfnamefont {T.}~\bibnamefont
  {St\"oferle}}, \bibinfo {author} {\bibfnamefont {H.}~\bibnamefont {Moritz}},
  \bibinfo {author} {\bibfnamefont {C.}~\bibnamefont {Schori}}, \bibinfo
  {author} {\bibfnamefont {M.}~\bibnamefont {K\"ohl}}, \ and\ \bibinfo {author}
  {\bibfnamefont {T.}~\bibnamefont {Esslinger}},\ }\href {\doibase
  10.1103/PhysRevLett.92.130403} {\bibfield  {journal} {\bibinfo  {journal}
  {Phys. Rev. Lett.}\ }\textbf {\bibinfo {volume} {92}},\ \bibinfo {pages}
  {130403} (\bibinfo {year} {2004})}\BibitemShut {NoStop}%
\bibitem [{\citenamefont {Kinoshita}\ \emph {et~al.}(2006)\citenamefont
  {Kinoshita}, \citenamefont {Wenger},\ and\ \citenamefont
  {Weiss}}]{kinoshita2006a}%
  \BibitemOpen
  \bibfield  {author} {\bibinfo {author} {\bibfnamefont {T.}~\bibnamefont
  {Kinoshita}}, \bibinfo {author} {\bibfnamefont {T.}~\bibnamefont {Wenger}}, \
  and\ \bibinfo {author} {\bibfnamefont {D.~S.}\ \bibnamefont {Weiss}},\ }\href
  {\doibase 10.1038/nature04693} {\bibfield  {journal} {\bibinfo  {journal}
  {Nature}\ }\textbf {\bibinfo {volume} {440}},\ \bibinfo {pages} {900}
  (\bibinfo {year} {2006})}\BibitemShut {NoStop}%
\bibitem [{\citenamefont {Citro}\ \emph {et~al.}(2007)\citenamefont {Citro},
  \citenamefont {Orignac}, \citenamefont {De~Palo},\ and\ \citenamefont
  {Chiofalo}}]{citro2007evidence}%
  \BibitemOpen
  \bibfield  {author} {\bibinfo {author} {\bibfnamefont {R.}~\bibnamefont
  {Citro}}, \bibinfo {author} {\bibfnamefont {E.}~\bibnamefont {Orignac}},
  \bibinfo {author} {\bibfnamefont {S.}~\bibnamefont {De~Palo}}, \ and\
  \bibinfo {author} {\bibfnamefont {M.~L.}\ \bibnamefont {Chiofalo}},\ }\href
  {\doibase 10.1103/PhysRevA.75.051602} {\bibfield  {journal} {\bibinfo
  {journal} {Phys. Rev. A}\ }\textbf {\bibinfo {volume} {75}},\ \bibinfo
  {pages} {051602} (\bibinfo {year} {2007})}\BibitemShut {NoStop}%
\bibitem [{\citenamefont {Widera}\ \emph {et~al.}(2008)\citenamefont {Widera},
  \citenamefont {Trotzky}, \citenamefont {Cheinet}, \citenamefont {F\"olling},
  \citenamefont {Gerbier}, \citenamefont {Bloch}, \citenamefont {Gritsev},
  \citenamefont {Lukin},\ and\ \citenamefont {Demler}}]{widera2008quantum}%
  \BibitemOpen
  \bibfield  {author} {\bibinfo {author} {\bibfnamefont {A.}~\bibnamefont
  {Widera}}, \bibinfo {author} {\bibfnamefont {S.}~\bibnamefont {Trotzky}},
  \bibinfo {author} {\bibfnamefont {P.}~\bibnamefont {Cheinet}}, \bibinfo
  {author} {\bibfnamefont {S.}~\bibnamefont {F\"olling}}, \bibinfo {author}
  {\bibfnamefont {F.}~\bibnamefont {Gerbier}}, \bibinfo {author} {\bibfnamefont
  {I.}~\bibnamefont {Bloch}}, \bibinfo {author} {\bibfnamefont
  {V.}~\bibnamefont {Gritsev}}, \bibinfo {author} {\bibfnamefont {M.~D.}\
  \bibnamefont {Lukin}}, \ and\ \bibinfo {author} {\bibfnamefont
  {E.}~\bibnamefont {Demler}},\ }\href {\doibase
  10.1103/PhysRevLett.100.140401} {\bibfield  {journal} {\bibinfo  {journal}
  {Phys. Rev. Lett.}\ }\textbf {\bibinfo {volume} {100}},\ \bibinfo {pages}
  {140401} (\bibinfo {year} {2008})}\BibitemShut {NoStop}%
\bibitem [{\citenamefont {Haller}\ \emph {et~al.}(2010)\citenamefont {Haller},
  \citenamefont {Hart}, \citenamefont {Mark}, \citenamefont {Danzl},
  \citenamefont {Reichs{\"o}llner}, \citenamefont {Gustavsson}, \citenamefont
  {Dalmonte}, \citenamefont {Pupillo},\ and\ \citenamefont
  {N{\"a}gerl}}]{haller2010pinning}%
  \BibitemOpen
  \bibfield  {author} {\bibinfo {author} {\bibfnamefont {E.}~\bibnamefont
  {Haller}}, \bibinfo {author} {\bibfnamefont {R.}~\bibnamefont {Hart}},
  \bibinfo {author} {\bibfnamefont {M.~J.}\ \bibnamefont {Mark}}, \bibinfo
  {author} {\bibfnamefont {J.~G.}\ \bibnamefont {Danzl}}, \bibinfo {author}
  {\bibfnamefont {L.}~\bibnamefont {Reichs{\"o}llner}}, \bibinfo {author}
  {\bibfnamefont {M.}~\bibnamefont {Gustavsson}}, \bibinfo {author}
  {\bibfnamefont {M.}~\bibnamefont {Dalmonte}}, \bibinfo {author}
  {\bibfnamefont {G.}~\bibnamefont {Pupillo}}, \ and\ \bibinfo {author}
  {\bibfnamefont {H.-C.}\ \bibnamefont {N{\"a}gerl}},\ }\href
  {https://www.nature.com/articles/nature09259/} {\bibfield  {journal}
  {\bibinfo  {journal} {Nature}\ }\textbf {\bibinfo {volume} {466}},\ \bibinfo
  {pages} {597} (\bibinfo {year} {2010})}\BibitemShut {NoStop}%
\bibitem [{\citenamefont {Hu}\ \emph {et~al.}(2011)\citenamefont {Hu},
  \citenamefont {Mathey}, \citenamefont {Tiesinga}, \citenamefont {Danshita},
  \citenamefont {Williams},\ and\ \citenamefont {Clark}}]{hu2011detecting}%
  \BibitemOpen
  \bibfield  {author} {\bibinfo {author} {\bibfnamefont {A.}~\bibnamefont
  {Hu}}, \bibinfo {author} {\bibfnamefont {L.}~\bibnamefont {Mathey}}, \bibinfo
  {author} {\bibfnamefont {E.}~\bibnamefont {Tiesinga}}, \bibinfo {author}
  {\bibfnamefont {I.}~\bibnamefont {Danshita}}, \bibinfo {author}
  {\bibfnamefont {C.~J.}\ \bibnamefont {Williams}}, \ and\ \bibinfo {author}
  {\bibfnamefont {C.~W.}\ \bibnamefont {Clark}},\ }\href {\doibase
  10.1103/PhysRevA.84.041609} {\bibfield  {journal} {\bibinfo  {journal} {Phys.
  Rev. A}\ }\textbf {\bibinfo {volume} {84}},\ \bibinfo {pages} {041609}
  (\bibinfo {year} {2011})}\BibitemShut {NoStop}%
\bibitem [{\citenamefont {Gring}\ \emph {et~al.}(2012)\citenamefont {Gring},
  \citenamefont {Kuhnert}, \citenamefont {Langen}, \citenamefont {Kitagawa},
  \citenamefont {Rauer}, \citenamefont {Schreitl}, \citenamefont {Mazets},
  \citenamefont {Smith}, \citenamefont {Demler},\ and\ \citenamefont
  {Schmiedmayer}}]{gring2012relaxation}%
  \BibitemOpen
  \bibfield  {author} {\bibinfo {author} {\bibfnamefont {M.}~\bibnamefont
  {Gring}}, \bibinfo {author} {\bibfnamefont {M.}~\bibnamefont {Kuhnert}},
  \bibinfo {author} {\bibfnamefont {T.}~\bibnamefont {Langen}}, \bibinfo
  {author} {\bibfnamefont {T.}~\bibnamefont {Kitagawa}}, \bibinfo {author}
  {\bibfnamefont {B.}~\bibnamefont {Rauer}}, \bibinfo {author} {\bibfnamefont
  {M.}~\bibnamefont {Schreitl}}, \bibinfo {author} {\bibfnamefont
  {I.}~\bibnamefont {Mazets}}, \bibinfo {author} {\bibfnamefont {D.~A.}\
  \bibnamefont {Smith}}, \bibinfo {author} {\bibfnamefont {E.}~\bibnamefont
  {Demler}}, \ and\ \bibinfo {author} {\bibfnamefont {J.}~\bibnamefont
  {Schmiedmayer}},\ }\href {\doibase 10.1126/science.1224953} {\bibfield
  {journal} {\bibinfo  {journal} {Science}\ }\textbf {\bibinfo {volume}
  {337}},\ \bibinfo {pages} {1318} (\bibinfo {year} {2012})}\BibitemShut
  {NoStop}%
\bibitem [{\citenamefont {Cheneau}\ \emph {et~al.}(2012)\citenamefont
  {Cheneau}, \citenamefont {Barmettler}, \citenamefont {Poletti}, \citenamefont
  {Endres}, \citenamefont {Schau{\ss}}, \citenamefont {Fukuhara}, \citenamefont
  {Gross}, \citenamefont {Bloch}, \citenamefont {Kollath},\ and\ \citenamefont
  {Kuhr}}]{cheneau2012light}%
  \BibitemOpen
  \bibfield  {author} {\bibinfo {author} {\bibfnamefont {M.}~\bibnamefont
  {Cheneau}}, \bibinfo {author} {\bibfnamefont {P.}~\bibnamefont {Barmettler}},
  \bibinfo {author} {\bibfnamefont {D.}~\bibnamefont {Poletti}}, \bibinfo
  {author} {\bibfnamefont {M.}~\bibnamefont {Endres}}, \bibinfo {author}
  {\bibfnamefont {P.}~\bibnamefont {Schau{\ss}}}, \bibinfo {author}
  {\bibfnamefont {T.}~\bibnamefont {Fukuhara}}, \bibinfo {author}
  {\bibfnamefont {C.}~\bibnamefont {Gross}}, \bibinfo {author} {\bibfnamefont
  {I.}~\bibnamefont {Bloch}}, \bibinfo {author} {\bibfnamefont
  {C.}~\bibnamefont {Kollath}}, \ and\ \bibinfo {author} {\bibfnamefont
  {S.}~\bibnamefont {Kuhr}},\ }\href {\doibase 10.1038/nature10748} {\bibfield
  {journal} {\bibinfo  {journal} {Nature}\ }\textbf {\bibinfo {volume} {481}},\
  \bibinfo {pages} {484} (\bibinfo {year} {2012})}\BibitemShut {NoStop}%
\bibitem [{\citenamefont {Danshita}(2013)}]{danshita2013universal}%
  \BibitemOpen
  \bibfield  {author} {\bibinfo {author} {\bibfnamefont {I.}~\bibnamefont
  {Danshita}},\ }\href {\doibase 10.1103/PhysRevLett.111.025303} {\bibfield
  {journal} {\bibinfo  {journal} {Phys. Rev. Lett.}\ }\textbf {\bibinfo
  {volume} {111}},\ \bibinfo {pages} {025303} (\bibinfo {year}
  {2013})}\BibitemShut {NoStop}%
\bibitem [{\citenamefont {Fabbri}\ \emph {et~al.}(2015)\citenamefont {Fabbri},
  \citenamefont {Panfil}, \citenamefont {Cl\'ement}, \citenamefont {Fallani},
  \citenamefont {Inguscio}, \citenamefont {Fort},\ and\ \citenamefont
  {Caux}}]{fabbri2015dynamical}%
  \BibitemOpen
  \bibfield  {author} {\bibinfo {author} {\bibfnamefont {N.}~\bibnamefont
  {Fabbri}}, \bibinfo {author} {\bibfnamefont {M.}~\bibnamefont {Panfil}},
  \bibinfo {author} {\bibfnamefont {D.}~\bibnamefont {Cl\'ement}}, \bibinfo
  {author} {\bibfnamefont {L.}~\bibnamefont {Fallani}}, \bibinfo {author}
  {\bibfnamefont {M.}~\bibnamefont {Inguscio}}, \bibinfo {author}
  {\bibfnamefont {C.}~\bibnamefont {Fort}}, \ and\ \bibinfo {author}
  {\bibfnamefont {J.-S.}\ \bibnamefont {Caux}},\ }\href {\doibase
  10.1103/PhysRevA.91.043617} {\bibfield  {journal} {\bibinfo  {journal} {Phys.
  Rev. A}\ }\textbf {\bibinfo {volume} {91}},\ \bibinfo {pages} {043617}
  (\bibinfo {year} {2015})}\BibitemShut {NoStop}%
\bibitem [{\citenamefont {Kunimi}\ and\ \citenamefont
  {Danshita}(2017)}]{kunimi2017thermally}%
  \BibitemOpen
  \bibfield  {author} {\bibinfo {author} {\bibfnamefont {M.}~\bibnamefont
  {Kunimi}}\ and\ \bibinfo {author} {\bibfnamefont {I.}~\bibnamefont
  {Danshita}},\ }\href {\doibase 10.1103/PhysRevA.95.033637} {\bibfield
  {journal} {\bibinfo  {journal} {Phys. Rev. A}\ }\textbf {\bibinfo {volume}
  {95}},\ \bibinfo {pages} {033637} (\bibinfo {year} {2017})}\BibitemShut
  {NoStop}%
\bibitem [{\citenamefont {Yang}\ \emph {et~al.}(2017)\citenamefont {Yang},
  \citenamefont {Chen}, \citenamefont {Zheng}, \citenamefont {Sun},
  \citenamefont {Dai}, \citenamefont {Guan}, \citenamefont {Yuan},\ and\
  \citenamefont {Pan}}]{yang2017quantum}%
  \BibitemOpen
  \bibfield  {author} {\bibinfo {author} {\bibfnamefont {B.}~\bibnamefont
  {Yang}}, \bibinfo {author} {\bibfnamefont {Y.-Y.}\ \bibnamefont {Chen}},
  \bibinfo {author} {\bibfnamefont {Y.-G.}\ \bibnamefont {Zheng}}, \bibinfo
  {author} {\bibfnamefont {H.}~\bibnamefont {Sun}}, \bibinfo {author}
  {\bibfnamefont {H.-N.}\ \bibnamefont {Dai}}, \bibinfo {author} {\bibfnamefont
  {X.-W.}\ \bibnamefont {Guan}}, \bibinfo {author} {\bibfnamefont {Z.-S.}\
  \bibnamefont {Yuan}}, \ and\ \bibinfo {author} {\bibfnamefont {J.-W.}\
  \bibnamefont {Pan}},\ }\href {\doibase 10.1103/PhysRevLett.119.165701}
  {\bibfield  {journal} {\bibinfo  {journal} {Phys. Rev. Lett.}\ }\textbf
  {\bibinfo {volume} {119}},\ \bibinfo {pages} {165701} (\bibinfo {year}
  {2017})}\BibitemShut {NoStop}%
\bibitem [{\citenamefont {Ozaki}\ \emph {et~al.}(2020)\citenamefont {Ozaki},
  \citenamefont {Nagao}, \citenamefont {Danshita},\ and\ \citenamefont
  {Kasamatsu}}]{ozaki2020semiclassical}%
  \BibitemOpen
  \bibfield  {author} {\bibinfo {author} {\bibfnamefont {Y.}~\bibnamefont
  {Ozaki}}, \bibinfo {author} {\bibfnamefont {K.}~\bibnamefont {Nagao}},
  \bibinfo {author} {\bibfnamefont {I.}~\bibnamefont {Danshita}}, \ and\
  \bibinfo {author} {\bibfnamefont {K.}~\bibnamefont {Kasamatsu}},\ }\href
  {\doibase 10.1103/PhysRevResearch.2.033272} {\bibfield  {journal} {\bibinfo
  {journal} {Phys. Rev. Research}\ }\textbf {\bibinfo {volume} {2}},\ \bibinfo
  {pages} {033272} (\bibinfo {year} {2020})}\BibitemShut {NoStop}%
\bibitem [{\citenamefont {Tomonaga}(1950)}]{tomonaga1950remarks}%
  \BibitemOpen
  \bibfield  {author} {\bibinfo {author} {\bibfnamefont {S.}~\bibnamefont
  {Tomonaga}},\ }\href {\doibase 10.1143/ptp/5.4.544} {\bibfield  {journal}
  {\bibinfo  {journal} {Prog. Theor. Phys.}\ }\textbf {\bibinfo {volume} {5}},\
  \bibinfo {pages} {544} (\bibinfo {year} {1950})}\BibitemShut {NoStop}%
\bibitem [{\citenamefont {Luttinger}(1963)}]{luttinger1963exactly}%
  \BibitemOpen
  \bibfield  {author} {\bibinfo {author} {\bibfnamefont {J.}~\bibnamefont
  {Luttinger}},\ }\href@noop {} {\bibfield  {journal} {\bibinfo  {journal} {J.
  Math. Phys.}\ }\textbf {\bibinfo {volume} {4}},\ \bibinfo {pages} {1154}
  (\bibinfo {year} {1963})}\BibitemShut {NoStop}%
\bibitem [{\citenamefont {Mattis}\ and\ \citenamefont
  {Lieb}(1965)}]{mattis1965exact}%
  \BibitemOpen
  \bibfield  {author} {\bibinfo {author} {\bibfnamefont {D.~C.}\ \bibnamefont
  {Mattis}}\ and\ \bibinfo {author} {\bibfnamefont {E.~H.}\ \bibnamefont
  {Lieb}},\ }\href@noop {} {\bibfield  {journal} {\bibinfo  {journal} {J. Math.
  Phys.}\ }\textbf {\bibinfo {volume} {6}},\ \bibinfo {pages} {304} (\bibinfo
  {year} {1965})}\BibitemShut {NoStop}%
\bibitem [{\citenamefont {Cazalilla}(2004)}]{cazalilla2004bosonizing}%
  \BibitemOpen
  \bibfield  {author} {\bibinfo {author} {\bibfnamefont {M.~A.}\ \bibnamefont
  {Cazalilla}},\ }\href {\doibase 10.1088/0953-4075/37/7/051} {\bibfield
  {journal} {\bibinfo  {journal} {Journal of Physics B: Atomic, Molecular and
  Optical Physics}\ }\textbf {\bibinfo {volume} {37}},\ \bibinfo {pages} {S1}
  (\bibinfo {year} {2004})}\BibitemShut {NoStop}%
\bibitem [{\citenamefont {Gogolin}\ \emph {et~al.}(2004)\citenamefont
  {Gogolin}, \citenamefont {Nersesyan},\ and\ \citenamefont
  {Tsvelik}}]{gogolin2004bosonization}%
  \BibitemOpen
  \bibfield  {author} {\bibinfo {author} {\bibfnamefont {A.~O.}\ \bibnamefont
  {Gogolin}}, \bibinfo {author} {\bibfnamefont {A.~A.}\ \bibnamefont
  {Nersesyan}}, \ and\ \bibinfo {author} {\bibfnamefont {A.~M.}\ \bibnamefont
  {Tsvelik}},\ }\href@noop {} {\emph {\bibinfo {title} {Bosonization and
  strongly correlated systems}}}\ (\bibinfo  {publisher} {Cambridge university
  press},\ \bibinfo {year} {2004})\BibitemShut {NoStop}%
\bibitem [{\citenamefont {Haldane}(1981)}]{haldane1981effective}%
  \BibitemOpen
  \bibfield  {author} {\bibinfo {author} {\bibfnamefont {F.~D.~M.}\
  \bibnamefont {Haldane}},\ }\href {\doibase 10.1103/PhysRevLett.47.1840}
  {\bibfield  {journal} {\bibinfo  {journal} {Phys. Rev. Lett.}\ }\textbf
  {\bibinfo {volume} {47}},\ \bibinfo {pages} {1840} (\bibinfo {year}
  {1981})}\BibitemShut {NoStop}%
\bibitem [{\citenamefont {Furusaki}\ and\ \citenamefont
  {Nagaosa}(1994)}]{furusaki1994kondo}%
  \BibitemOpen
  \bibfield  {author} {\bibinfo {author} {\bibfnamefont {A.}~\bibnamefont
  {Furusaki}}\ and\ \bibinfo {author} {\bibfnamefont {N.}~\bibnamefont
  {Nagaosa}},\ }\href {\doibase 10.1103/PhysRevLett.72.892} {\bibfield
  {journal} {\bibinfo  {journal} {Phys. Rev. Lett.}\ }\textbf {\bibinfo
  {volume} {72}},\ \bibinfo {pages} {892} (\bibinfo {year} {1994})}\BibitemShut
  {NoStop}%
\bibitem [{\citenamefont {Mathey}\ \emph {et~al.}(2004)\citenamefont {Mathey},
  \citenamefont {Wang}, \citenamefont {Hofstetter}, \citenamefont {Lukin},\
  and\ \citenamefont {Demler}}]{mathey2004luttinger}%
  \BibitemOpen
  \bibfield  {author} {\bibinfo {author} {\bibfnamefont {L.}~\bibnamefont
  {Mathey}}, \bibinfo {author} {\bibfnamefont {D.-W.}\ \bibnamefont {Wang}},
  \bibinfo {author} {\bibfnamefont {W.}~\bibnamefont {Hofstetter}}, \bibinfo
  {author} {\bibfnamefont {M.~D.}\ \bibnamefont {Lukin}}, \ and\ \bibinfo
  {author} {\bibfnamefont {E.}~\bibnamefont {Demler}},\ }\href {\doibase
  10.1103/PhysRevLett.93.120404} {\bibfield  {journal} {\bibinfo  {journal}
  {Phys. Rev. Lett.}\ }\textbf {\bibinfo {volume} {93}},\ \bibinfo {pages}
  {120404} (\bibinfo {year} {2004})}\BibitemShut {NoStop}%
\bibitem [{\citenamefont {Mathey}(2007)}]{mathey2007commensurate}%
  \BibitemOpen
  \bibfield  {author} {\bibinfo {author} {\bibfnamefont {L.}~\bibnamefont
  {Mathey}},\ }\href {\doibase 10.1103/PhysRevB.75.144510} {\bibfield
  {journal} {\bibinfo  {journal} {Phys. Rev. B}\ }\textbf {\bibinfo {volume}
  {75}},\ \bibinfo {pages} {144510} (\bibinfo {year} {2007})}\BibitemShut
  {NoStop}%
\bibitem [{\citenamefont {Tokuno}\ \emph {et~al.}(2008)\citenamefont {Tokuno},
  \citenamefont {Oshikawa},\ and\ \citenamefont {Demler}}]{tokuno2008dynamics}%
  \BibitemOpen
  \bibfield  {author} {\bibinfo {author} {\bibfnamefont {A.}~\bibnamefont
  {Tokuno}}, \bibinfo {author} {\bibfnamefont {M.}~\bibnamefont {Oshikawa}}, \
  and\ \bibinfo {author} {\bibfnamefont {E.}~\bibnamefont {Demler}},\ }\href
  {\doibase 10.1103/PhysRevLett.100.140402} {\bibfield  {journal} {\bibinfo
  {journal} {Phys. Rev. Lett.}\ }\textbf {\bibinfo {volume} {100}},\ \bibinfo
  {pages} {140402} (\bibinfo {year} {2008})}\BibitemShut {NoStop}%
\bibitem [{\citenamefont {Kitagawa}\ \emph {et~al.}(2010)\citenamefont
  {Kitagawa}, \citenamefont {Pielawa}, \citenamefont {Imambekov}, \citenamefont
  {Schmiedmayer}, \citenamefont {Gritsev},\ and\ \citenamefont
  {Demler}}]{kitagawa2010ramsey}%
  \BibitemOpen
  \bibfield  {author} {\bibinfo {author} {\bibfnamefont {T.}~\bibnamefont
  {Kitagawa}}, \bibinfo {author} {\bibfnamefont {S.}~\bibnamefont {Pielawa}},
  \bibinfo {author} {\bibfnamefont {A.}~\bibnamefont {Imambekov}}, \bibinfo
  {author} {\bibfnamefont {J.}~\bibnamefont {Schmiedmayer}}, \bibinfo {author}
  {\bibfnamefont {V.}~\bibnamefont {Gritsev}}, \ and\ \bibinfo {author}
  {\bibfnamefont {E.}~\bibnamefont {Demler}},\ }\href {\doibase
  10.1103/PhysRevLett.104.255302} {\bibfield  {journal} {\bibinfo  {journal}
  {Phys. Rev. Lett.}\ }\textbf {\bibinfo {volume} {104}},\ \bibinfo {pages}
  {255302} (\bibinfo {year} {2010})}\BibitemShut {NoStop}%
\bibitem [{\citenamefont {Eggel}\ \emph {et~al.}(2011)\citenamefont {Eggel},
  \citenamefont {Cazalilla},\ and\ \citenamefont
  {Oshikawa}}]{eggel2011dynamical}%
  \BibitemOpen
  \bibfield  {author} {\bibinfo {author} {\bibfnamefont {T.}~\bibnamefont
  {Eggel}}, \bibinfo {author} {\bibfnamefont {M.~A.}\ \bibnamefont
  {Cazalilla}}, \ and\ \bibinfo {author} {\bibfnamefont {M.}~\bibnamefont
  {Oshikawa}},\ }\href {\doibase 10.1103/PhysRevLett.107.275302} {\bibfield
  {journal} {\bibinfo  {journal} {Phys. Rev. Lett.}\ }\textbf {\bibinfo
  {volume} {107}},\ \bibinfo {pages} {275302} (\bibinfo {year}
  {2011})}\BibitemShut {NoStop}%
\bibitem [{\citenamefont {Ruhman}\ \emph {et~al.}(2015)\citenamefont {Ruhman},
  \citenamefont {Berg},\ and\ \citenamefont {Altman}}]{ruhman2015topological}%
  \BibitemOpen
  \bibfield  {author} {\bibinfo {author} {\bibfnamefont {J.}~\bibnamefont
  {Ruhman}}, \bibinfo {author} {\bibfnamefont {E.}~\bibnamefont {Berg}}, \ and\
  \bibinfo {author} {\bibfnamefont {E.}~\bibnamefont {Altman}},\ }\href
  {\doibase 10.1103/PhysRevLett.114.100401} {\bibfield  {journal} {\bibinfo
  {journal} {Phys. Rev. Lett.}\ }\textbf {\bibinfo {volume} {114}},\ \bibinfo
  {pages} {100401} (\bibinfo {year} {2015})}\BibitemShut {NoStop}%
\bibitem [{\citenamefont {Okamoto}\ \emph {et~al.}(2016)\citenamefont
  {Okamoto}, \citenamefont {Mathey},\ and\ \citenamefont
  {H\"artle}}]{okamoto2016hierarchical}%
  \BibitemOpen
  \bibfield  {author} {\bibinfo {author} {\bibfnamefont {J.-i.}\ \bibnamefont
  {Okamoto}}, \bibinfo {author} {\bibfnamefont {L.}~\bibnamefont {Mathey}}, \
  and\ \bibinfo {author} {\bibfnamefont {R.}~\bibnamefont {H\"artle}},\ }\href
  {\doibase 10.1103/PhysRevB.94.235411} {\bibfield  {journal} {\bibinfo
  {journal} {Phys. Rev. B}\ }\textbf {\bibinfo {volume} {94}},\ \bibinfo
  {pages} {235411} (\bibinfo {year} {2016})}\BibitemShut {NoStop}%
\bibitem [{\citenamefont {Ashida}\ \emph {et~al.}(2016)\citenamefont {Ashida},
  \citenamefont {Furukawa},\ and\ \citenamefont {Ueda}}]{ashida2016quantum}%
  \BibitemOpen
  \bibfield  {author} {\bibinfo {author} {\bibfnamefont {Y.}~\bibnamefont
  {Ashida}}, \bibinfo {author} {\bibfnamefont {S.}~\bibnamefont {Furukawa}}, \
  and\ \bibinfo {author} {\bibfnamefont {M.}~\bibnamefont {Ueda}},\ }\href
  {\doibase 10.1103/PhysRevA.94.053615} {\bibfield  {journal} {\bibinfo
  {journal} {Phys. Rev. A}\ }\textbf {\bibinfo {volume} {94}},\ \bibinfo
  {pages} {053615} (\bibinfo {year} {2016})}\BibitemShut {NoStop}%
\bibitem [{\citenamefont {Matveev}\ and\ \citenamefont
  {Andreev}(2017)}]{matveev2017second}%
  \BibitemOpen
  \bibfield  {author} {\bibinfo {author} {\bibfnamefont {K.~A.}\ \bibnamefont
  {Matveev}}\ and\ \bibinfo {author} {\bibfnamefont {A.~V.}\ \bibnamefont
  {Andreev}},\ }\href {\doibase 10.1103/PhysRevLett.119.266801} {\bibfield
  {journal} {\bibinfo  {journal} {Phys. Rev. Lett.}\ }\textbf {\bibinfo
  {volume} {119}},\ \bibinfo {pages} {266801} (\bibinfo {year}
  {2017})}\BibitemShut {NoStop}%
\bibitem [{\citenamefont {Matveev}\ and\ \citenamefont
  {Andreev}(2018)}]{matveev2018hybrid}%
  \BibitemOpen
  \bibfield  {author} {\bibinfo {author} {\bibfnamefont {K.~A.}\ \bibnamefont
  {Matveev}}\ and\ \bibinfo {author} {\bibfnamefont {A.~V.}\ \bibnamefont
  {Andreev}},\ }\href {\doibase 10.1103/PhysRevLett.121.026803} {\bibfield
  {journal} {\bibinfo  {journal} {Phys. Rev. Lett.}\ }\textbf {\bibinfo
  {volume} {121}},\ \bibinfo {pages} {026803} (\bibinfo {year}
  {2018})}\BibitemShut {NoStop}%
\bibitem [{\citenamefont {Imamura}\ \emph {et~al.}(2019)\citenamefont
  {Imamura}, \citenamefont {Totsuka},\ and\ \citenamefont
  {Hansson}}]{imamura2019from}%
  \BibitemOpen
  \bibfield  {author} {\bibinfo {author} {\bibfnamefont {Y.}~\bibnamefont
  {Imamura}}, \bibinfo {author} {\bibfnamefont {K.}~\bibnamefont {Totsuka}}, \
  and\ \bibinfo {author} {\bibfnamefont {T.~H.}\ \bibnamefont {Hansson}},\
  }\href {\doibase 10.1103/PhysRevB.100.125148} {\bibfield  {journal} {\bibinfo
   {journal} {Phys. Rev. B}\ }\textbf {\bibinfo {volume} {100}},\ \bibinfo
  {pages} {125148} (\bibinfo {year} {2019})}\BibitemShut {NoStop}%
\bibitem [{\citenamefont {Sachdev}(2011)}]{sachdev2011quantum}%
  \BibitemOpen
  \bibfield  {author} {\bibinfo {author} {\bibfnamefont {S.}~\bibnamefont
  {Sachdev}},\ }\href@noop {} {\emph {\bibinfo {title} {Quantum Phase
  Transitions}}}\ (\bibinfo  {publisher} {Cambridge University Press},\
  \bibinfo {year} {2011})\BibitemShut {NoStop}%
\bibitem [{\citenamefont {Barak}\ \emph {et~al.}(2010)\citenamefont {Barak},
  \citenamefont {Steinberg}, \citenamefont {Pfeiffer}, \citenamefont {West},
  \citenamefont {Glazman}, \citenamefont {Von~Oppen},\ and\ \citenamefont
  {Yacoby}}]{barak2010interacting}%
  \BibitemOpen
  \bibfield  {author} {\bibinfo {author} {\bibfnamefont {G.}~\bibnamefont
  {Barak}}, \bibinfo {author} {\bibfnamefont {H.}~\bibnamefont {Steinberg}},
  \bibinfo {author} {\bibfnamefont {L.~N.}\ \bibnamefont {Pfeiffer}}, \bibinfo
  {author} {\bibfnamefont {K.~W.}\ \bibnamefont {West}}, \bibinfo {author}
  {\bibfnamefont {L.}~\bibnamefont {Glazman}}, \bibinfo {author} {\bibfnamefont
  {F.}~\bibnamefont {Von~Oppen}}, \ and\ \bibinfo {author} {\bibfnamefont
  {A.}~\bibnamefont {Yacoby}},\ }\href
  {https://www.nature.com/articles/nphys1678?page=3} {\bibfield  {journal}
  {\bibinfo  {journal} {Nature Physics}\ }\textbf {\bibinfo {volume} {6}},\
  \bibinfo {pages} {489} (\bibinfo {year} {2010})}\BibitemShut {NoStop}%
\bibitem [{\citenamefont {Pustilnik}\ \emph {et~al.}(2006)\citenamefont
  {Pustilnik}, \citenamefont {Khodas}, \citenamefont {Kamenev},\ and\
  \citenamefont {Glazman}}]{pustilnik2006dynamic}%
  \BibitemOpen
  \bibfield  {author} {\bibinfo {author} {\bibfnamefont {M.}~\bibnamefont
  {Pustilnik}}, \bibinfo {author} {\bibfnamefont {M.}~\bibnamefont {Khodas}},
  \bibinfo {author} {\bibfnamefont {A.}~\bibnamefont {Kamenev}}, \ and\
  \bibinfo {author} {\bibfnamefont {L.~I.}\ \bibnamefont {Glazman}},\ }\href
  {\doibase 10.1103/PhysRevLett.96.196405} {\bibfield  {journal} {\bibinfo
  {journal} {Phys. Rev. Lett.}\ }\textbf {\bibinfo {volume} {96}},\ \bibinfo
  {pages} {196405} (\bibinfo {year} {2006})}\BibitemShut {NoStop}%
\bibitem [{\citenamefont {Imambekov}\ and\ \citenamefont
  {Glazman}(2009)}]{imambekov2009universal}%
  \BibitemOpen
  \bibfield  {author} {\bibinfo {author} {\bibfnamefont {A.}~\bibnamefont
  {Imambekov}}\ and\ \bibinfo {author} {\bibfnamefont {L.~I.}\ \bibnamefont
  {Glazman}},\ }\href {\doibase 10.1126/science.1165403} {\bibfield  {journal}
  {\bibinfo  {journal} {Science}\ }\textbf {\bibinfo {volume} {323}},\ \bibinfo
  {pages} {228} (\bibinfo {year} {2009})}\BibitemShut {NoStop}%
\bibitem [{\citenamefont {Schmidt}\ \emph {et~al.}(2010)\citenamefont
  {Schmidt}, \citenamefont {Imambekov},\ and\ \citenamefont
  {Glazman}}]{schmidt2010spin}%
  \BibitemOpen
  \bibfield  {author} {\bibinfo {author} {\bibfnamefont {T.~L.}\ \bibnamefont
  {Schmidt}}, \bibinfo {author} {\bibfnamefont {A.}~\bibnamefont {Imambekov}},
  \ and\ \bibinfo {author} {\bibfnamefont {L.~I.}\ \bibnamefont {Glazman}},\
  }\href {\doibase 10.1103/PhysRevB.82.245104} {\bibfield  {journal} {\bibinfo
  {journal} {Phys. Rev. B}\ }\textbf {\bibinfo {volume} {82}},\ \bibinfo
  {pages} {245104} (\bibinfo {year} {2010})}\BibitemShut {NoStop}%
\bibitem [{\citenamefont {Imambekov}\ \emph {et~al.}(2012)\citenamefont
  {Imambekov}, \citenamefont {Schmidt},\ and\ \citenamefont
  {Glazman}}]{imambekov2012one}%
  \BibitemOpen
  \bibfield  {author} {\bibinfo {author} {\bibfnamefont {A.}~\bibnamefont
  {Imambekov}}, \bibinfo {author} {\bibfnamefont {T.~L.}\ \bibnamefont
  {Schmidt}}, \ and\ \bibinfo {author} {\bibfnamefont {L.~I.}\ \bibnamefont
  {Glazman}},\ }\href {\doibase 10.1103/RevModPhys.84.1253} {\bibfield
  {journal} {\bibinfo  {journal} {Rev. Mod. Phys.}\ }\textbf {\bibinfo {volume}
  {84}},\ \bibinfo {pages} {1253} (\bibinfo {year} {2012})}\BibitemShut
  {NoStop}%
\bibitem [{\citenamefont {Matveev}\ and\ \citenamefont
  {Furusaki}(2013)}]{matveev2013decay}%
  \BibitemOpen
  \bibfield  {author} {\bibinfo {author} {\bibfnamefont {K.~A.}\ \bibnamefont
  {Matveev}}\ and\ \bibinfo {author} {\bibfnamefont {A.}~\bibnamefont
  {Furusaki}},\ }\href {\doibase 10.1103/PhysRevLett.111.256401} {\bibfield
  {journal} {\bibinfo  {journal} {Phys. Rev. Lett.}\ }\textbf {\bibinfo
  {volume} {111}},\ \bibinfo {pages} {256401} (\bibinfo {year}
  {2013})}\BibitemShut {NoStop}%
\bibitem [{\citenamefont {Protopopov}\ \emph {et~al.}(2014)\citenamefont
  {Protopopov}, \citenamefont {Gutman},\ and\ \citenamefont
  {Mirlin}}]{protopopov2014relaxation}%
  \BibitemOpen
  \bibfield  {author} {\bibinfo {author} {\bibfnamefont {I.~V.}\ \bibnamefont
  {Protopopov}}, \bibinfo {author} {\bibfnamefont {D.~B.}\ \bibnamefont
  {Gutman}}, \ and\ \bibinfo {author} {\bibfnamefont {A.~D.}\ \bibnamefont
  {Mirlin}},\ }\href {\doibase 10.1103/PhysRevB.90.125113} {\bibfield
  {journal} {\bibinfo  {journal} {Phys. Rev. B}\ }\textbf {\bibinfo {volume}
  {90}},\ \bibinfo {pages} {125113} (\bibinfo {year} {2014})}\BibitemShut
  {NoStop}%
\bibitem [{\citenamefont {Apostolov}\ \emph {et~al.}(2013)\citenamefont
  {Apostolov}, \citenamefont {Liu}, \citenamefont {Maizelis},\ and\
  \citenamefont {Levchenko}}]{apostolov2013thermal}%
  \BibitemOpen
  \bibfield  {author} {\bibinfo {author} {\bibfnamefont {S.}~\bibnamefont
  {Apostolov}}, \bibinfo {author} {\bibfnamefont {D.~E.}\ \bibnamefont {Liu}},
  \bibinfo {author} {\bibfnamefont {Z.}~\bibnamefont {Maizelis}}, \ and\
  \bibinfo {author} {\bibfnamefont {A.}~\bibnamefont {Levchenko}},\ }\href
  {\doibase 10.1103/PhysRevB.88.045435} {\bibfield  {journal} {\bibinfo
  {journal} {Phys. Rev. B}\ }\textbf {\bibinfo {volume} {88}},\ \bibinfo
  {pages} {045435} (\bibinfo {year} {2013})}\BibitemShut {NoStop}%
\bibitem [{\citenamefont {Buchhold}\ and\ \citenamefont
  {Diehl}(2015)}]{buchhold2015nonequilibrium}%
  \BibitemOpen
  \bibfield  {author} {\bibinfo {author} {\bibfnamefont {M.}~\bibnamefont
  {Buchhold}}\ and\ \bibinfo {author} {\bibfnamefont {S.}~\bibnamefont
  {Diehl}},\ }\href {\doibase 10.1103/PhysRevA.92.013603} {\bibfield  {journal}
  {\bibinfo  {journal} {Phys. Rev. A}\ }\textbf {\bibinfo {volume} {92}},\
  \bibinfo {pages} {013603} (\bibinfo {year} {2015})}\BibitemShut {NoStop}%
\bibitem [{\citenamefont {Samanta}\ \emph {et~al.}(2019)\citenamefont
  {Samanta}, \citenamefont {Protopopov}, \citenamefont {Mirlin},\ and\
  \citenamefont {Gutman}}]{samanta2019thermal}%
  \BibitemOpen
  \bibfield  {author} {\bibinfo {author} {\bibfnamefont {R.}~\bibnamefont
  {Samanta}}, \bibinfo {author} {\bibfnamefont {I.~V.}\ \bibnamefont
  {Protopopov}}, \bibinfo {author} {\bibfnamefont {A.~D.}\ \bibnamefont
  {Mirlin}}, \ and\ \bibinfo {author} {\bibfnamefont {D.~B.}\ \bibnamefont
  {Gutman}},\ }\href {\doibase 10.1103/PhysRevLett.122.206801} {\bibfield
  {journal} {\bibinfo  {journal} {Phys. Rev. Lett.}\ }\textbf {\bibinfo
  {volume} {122}},\ \bibinfo {pages} {206801} (\bibinfo {year}
  {2019})}\BibitemShut {NoStop}%
\bibitem [{\citenamefont {Seifie}\ \emph {et~al.}(2019)\citenamefont {Seifie},
  \citenamefont {Singh},\ and\ \citenamefont {Mathey}}]{seifie2019squeezed}%
  \BibitemOpen
  \bibfield  {author} {\bibinfo {author} {\bibfnamefont {I.~M.~H.}\
  \bibnamefont {Seifie}}, \bibinfo {author} {\bibfnamefont {V.~P.}\
  \bibnamefont {Singh}}, \ and\ \bibinfo {author} {\bibfnamefont
  {L.}~\bibnamefont {Mathey}},\ }\href {\doibase 10.1103/PhysRevA.100.013602}
  {\bibfield  {journal} {\bibinfo  {journal} {Phys. Rev. A}\ }\textbf {\bibinfo
  {volume} {100}},\ \bibinfo {pages} {013602} (\bibinfo {year}
  {2019})}\BibitemShut {NoStop}%
\bibitem [{\citenamefont {Altland}\ and\ \citenamefont
  {Simons}(2010)}]{altland2010condensed}%
  \BibitemOpen
  \bibfield  {author} {\bibinfo {author} {\bibfnamefont {A.}~\bibnamefont
  {Altland}}\ and\ \bibinfo {author} {\bibfnamefont {B.~D.}\ \bibnamefont
  {Simons}},\ }\href@noop {} {\emph {\bibinfo {title} {Condensed matter field
  theory}}}\ (\bibinfo  {publisher} {Cambridge university press},\ \bibinfo
  {year} {2010})\BibitemShut {NoStop}%
\bibitem [{\citenamefont {Tsue}\ and\ \citenamefont
  {Fujiwara}(1991)}]{tsue1991time}%
  \BibitemOpen
  \bibfield  {author} {\bibinfo {author} {\bibfnamefont {Y.}~\bibnamefont
  {Tsue}}\ and\ \bibinfo {author} {\bibfnamefont {Y.}~\bibnamefont
  {Fujiwara}},\ }\href {https://academic.oup.com/ptp/article/86/2/443/1853690}
  {\bibfield  {journal} {\bibinfo  {journal} {Prog. Theor. Phys.}\ }\textbf
  {\bibinfo {volume} {86}},\ \bibinfo {pages} {443} (\bibinfo {year}
  {1991})}\BibitemShut {NoStop}%
\bibitem [{\citenamefont {Guaita}\ \emph {et~al.}(2019)\citenamefont {Guaita},
  \citenamefont {Hackl}, \citenamefont {Shi}, \citenamefont {Hubig},
  \citenamefont {Demler},\ and\ \citenamefont {Cirac}}]{guaita2019gaussian}%
  \BibitemOpen
  \bibfield  {author} {\bibinfo {author} {\bibfnamefont {T.}~\bibnamefont
  {Guaita}}, \bibinfo {author} {\bibfnamefont {L.}~\bibnamefont {Hackl}},
  \bibinfo {author} {\bibfnamefont {T.}~\bibnamefont {Shi}}, \bibinfo {author}
  {\bibfnamefont {C.}~\bibnamefont {Hubig}}, \bibinfo {author} {\bibfnamefont
  {E.}~\bibnamefont {Demler}}, \ and\ \bibinfo {author} {\bibfnamefont {J.~I.}\
  \bibnamefont {Cirac}},\ }\href {\doibase 10.1103/PhysRevB.100.094529}
  {\bibfield  {journal} {\bibinfo  {journal} {Phys. Rev. B}\ }\textbf {\bibinfo
  {volume} {100}},\ \bibinfo {pages} {094529} (\bibinfo {year}
  {2019})}\BibitemShut {NoStop}%
\bibitem [{\citenamefont {Schollw{\"o}ck}(2011)}]{schollwock2011the}%
  \BibitemOpen
  \bibfield  {author} {\bibinfo {author} {\bibfnamefont {U.}~\bibnamefont
  {Schollw{\"o}ck}},\ }\href {\doibase
  https://doi.org/10.1016/j.aop.2010.09.012} {\bibfield  {journal} {\bibinfo
  {journal} {Ann. Phys.}\ }\textbf {\bibinfo {volume} {326}},\ \bibinfo {pages}
  {96 } (\bibinfo {year} {2011})}\BibitemShut {NoStop}%
\bibitem [{\citenamefont {Vidal}(2007)}]{vidal2007classical}%
  \BibitemOpen
  \bibfield  {author} {\bibinfo {author} {\bibfnamefont {G.}~\bibnamefont
  {Vidal}},\ }\href {\doibase 10.1103/PhysRevLett.98.070201} {\bibfield
  {journal} {\bibinfo  {journal} {Phys. Rev. Lett.}\ }\textbf {\bibinfo
  {volume} {98}},\ \bibinfo {pages} {070201} (\bibinfo {year}
  {2007})}\BibitemShut {NoStop}%
\bibitem [{\citenamefont {Walls}\ and\ \citenamefont
  {Milburn}(2007)}]{walls2007quantum}%
  \BibitemOpen
  \bibfield  {author} {\bibinfo {author} {\bibfnamefont {D.~F.}\ \bibnamefont
  {Walls}}\ and\ \bibinfo {author} {\bibfnamefont {G.~J.}\ \bibnamefont
  {Milburn}},\ }\href@noop {} {\emph {\bibinfo {title} {Quantum optics}}}\
  (\bibinfo  {publisher} {Springer Science \& Business Media},\ \bibinfo {year}
  {2007})\BibitemShut {NoStop}%
\bibitem [{\citenamefont {Fetter}\ and\ \citenamefont
  {Walecka}(2012)}]{fetter2012quantum}%
  \BibitemOpen
  \bibfield  {author} {\bibinfo {author} {\bibfnamefont {A.~L.}\ \bibnamefont
  {Fetter}}\ and\ \bibinfo {author} {\bibfnamefont {J.~D.}\ \bibnamefont
  {Walecka}},\ }\href@noop {} {\emph {\bibinfo {title} {Quantum theory of
  many-particle systems}}}\ (\bibinfo  {publisher} {Courier Corporation},\
  \bibinfo {year} {2012})\BibitemShut {NoStop}%
\bibitem [{\citenamefont {Nagao}\ and\ \citenamefont
  {Danshita}(2016)}]{nagao2016damping}%
  \BibitemOpen
  \bibfield  {author} {\bibinfo {author} {\bibfnamefont {K.}~\bibnamefont
  {Nagao}}\ and\ \bibinfo {author} {\bibfnamefont {I.}~\bibnamefont
  {Danshita}},\ }\href {https://doi.org/10.1093/ptep/ptw061} {\bibfield
  {journal} {\bibinfo  {journal} {Prog. Theor. Exp. Phys.}\ }\textbf {\bibinfo
  {volume} {2016}} (\bibinfo {year} {2016})}\BibitemShut {NoStop}%
\bibitem [{\citenamefont {Nagao}\ \emph {et~al.}(2018)\citenamefont {Nagao},
  \citenamefont {Takahashi},\ and\ \citenamefont
  {Danshita}}]{nagao2018response}%
  \BibitemOpen
  \bibfield  {author} {\bibinfo {author} {\bibfnamefont {K.}~\bibnamefont
  {Nagao}}, \bibinfo {author} {\bibfnamefont {Y.}~\bibnamefont {Takahashi}}, \
  and\ \bibinfo {author} {\bibfnamefont {I.}~\bibnamefont {Danshita}},\ }\href
  {\doibase 10.1103/PhysRevA.97.043628} {\bibfield  {journal} {\bibinfo
  {journal} {Phys. Rev. A}\ }\textbf {\bibinfo {volume} {97}},\ \bibinfo
  {pages} {043628} (\bibinfo {year} {2018})}\BibitemShut {NoStop}%
\end{thebibliography}%

\end{document}